\RequirePackage[2020-02-02]{latexrelease}

\documentclass{article}
\usepackage{preprint}

\usepackage{enumerate}
\usepackage{import}
\usepackage{tabularx}
\usepackage{graphicx}
\usepackage{multirow, makecell}
\usepackage{multicol}

\usepackage{amsmath, amsthm, amssymb, amsfonts}
\makeatletter
\g@addto@macro\normalsize{%
  \setlength\abovedisplayskip{10pt}
  \setlength\belowdisplayskip{10pt}
  \setlength\abovedisplayshortskip{10pt}
  \setlength\belowdisplayshortskip{10pt}
}
\usepackage[normalem]{ulem}
\usepackage{algorithm}
\usepackage{algpseudocode}
\usepackage{MnSymbol}%
\usepackage{wasysym}%
\usepackage{mathtools}
\newcommand{\Z}{\mathbb{Z}}
\newcommand{\R}{\mathbb{R}}

\DeclareMathOperator\Arg{Arg}

\usepackage{setspace}
\usepackage[authoryear,sort]{natbib}
\usepackage{xcolor}		
\usepackage[colorlinks = true,
            linkcolor = purple,
            urlcolor  = blue,
            citecolor = cyan,
            anchorcolor = black]{hyperref}	
\usepackage{booktabs} 		
\usepackage{nicefrac}		
\usepackage{microtype}		
\usepackage{lineno}		
\usepackage{float}			
\usepackage{calc}

\usepackage{lipsum}		
\usepackage{comment}
\usepackage{newfloat}
\DeclareFloatingEnvironment[name={Supplementary Figure}]{suppfigure}
\usepackage{sidecap}
\sidecaptionvpos{figure}{c}

\newenvironment{myitemize}
{ \begin{itemize}
    \setlength{\itemsep}{3pt}
    \setlength{\parskip}{3pt}
    \setlength{\parsep}{3pt}     }
{ \end{itemize}                  } 

\newenvironment{myenumerate}
{ \begin{enumerate}
    \setlength{\itemsep}{3pt}
    \setlength{\parskip}{3pt}
    \setlength{\parsep}{3pt}     }
{ \end{enumerate}                  } 

\usepackage{titlesec}
\titlespacing\section{0pt}{12pt plus 3pt minus 3pt}{1pt plus 1pt minus 1pt}
\titlespacing\subsection{0pt}{10pt plus 3pt minus 3pt}{1pt plus 1pt minus 1pt}
\titlespacing\subsubsection{0pt}{8pt plus 3pt minus 3pt}{1pt plus 1pt minus 1pt}

\makeatletter

\renewcommand\tagform@[1]{\maketag@@@{\ignorespaces#1\unskip\@@italiccorr}}
\renewcommand\theequation{(\oldtheequation)}
\makeatother

\newcommand\SecNum[1]{%
  \href{#1}{\getrefnumber{#1}}%
}

\usepackage{placeins}

\usepackage{tikz,xcolor,hyperref}

\definecolor{lime}{HTML}{A6CE39}
\DeclareRobustCommand{\orcidicon}{
	\begin{tikzpicture}
	\draw[lime, fill=lime] (0,0) 
	circle [radius=0.16] 
	node[white] {{\fontfamily{qag}\selectfont \tiny ID}};
	\draw[white, fill=white] (-0.0625,0.095) 
	circle [radius=0.007];
	\end{tikzpicture}
	\hspace{-2mm}
}
\foreach \x in {A, ..., Z}{\expandafter\xdef\csname orcid\x\endcsname{\noexpand\href{https://orcid.org/\csname orcidauthor\x\endcsname}
			{\noexpand\orcidicon}}
}

\newcommand{\orcidauthorD}{0000-0002-1825-6763}

\usepackage{xwatermark}

\usepackage{authblk}

\author[1\thanks{\tt{rvawo@dtu.dk}}]{Rebekka V. Woldseth\orcidA{}}
\author[2]{J. Andreas B\ae rentzen\orcidB{}}
\author[1]{Ole Sigmund\orcidC{}}
\affil[1]{Department of Civil and Mechanical Engineering, Technical University of Denmark. Koppels All\'{e}, B.404, 2800 Kgs. Lyngby, Denmark.}
\affil[2]{Department of Applied Mathematics and Computer Science, Technical University of Denmark. Richard Petersens Plads, B.321, 2800 Kgs. Lyngby, Denmark.}

\title{Phasor Noise for Dehomogenisation in 2D Multiscale Topology Optimisation}

\begin{document}

\maketitle

\begin{abstract}
This paper presents an alternative approach to dehomogenisation of elastic Rank-N laminate structures based on the computer graphics discipline of phasor noise. The proposed methodology offers an improvement of existing methods, where high-quality single-scale designs can be obtained efficiently without the utilisation of any least-squares problem or pre-trained models. By utilising a continuous and periodic representation of the translation at each intermediate step, appropriate length-scale and thicknesses can be obtained. Numerical tests verifies the performance of the proposed methodology compared to state-of-the-art alternatives, and the dehomogenised designs achieve structural performance within a few percentages of the optimised homogenised solution. The nature of the phasor-based dehomogenisation is inherently mesh-independent and highly parallelisable, allowing for further efficient implementations and future extensions to 3D problems on unstructured meshes.
\end{abstract}

\vspace{0.35cm}


\section{Introduction}\label{sec:intro_main}

Homogenisation-based topology optimisation stems from the original works on topology optimisation exploiting the notion that theoretically optimal topologies consist of multilayered periodic composites oriented in local principal stress directions (\citealt{Allaire2002}, \citealt{BendsoeKikuchi1988}, \citealt{BendsoeSigmund2004}). This approach allowed for a more compact design representation than the later more prevailing density-based methods, and can be performed at a much lower cost. As a consequence of the assumption of infinitesimally small features, the optimised homogenisation-based solutions are not suited for obtaining a manufacturable design directly, which is why the density-based methods were preferred in the further development of topology optimisation as computational capacity increased. The homogenisation-based approach was revived by the introduction of projection-based approaches aimed at mapping the infinitesimal microstructures to solid-void designs at a finite length scale (\citealt{PantzTrabelsi2008}, \citealt{PantzTrabelsi2010}). This idea has inspired several improvements in terms of computational efficiency and length scale control to obtain high-resolution, manufacturable designs efficiently (\citealt{DondersAllaire2020}, \citealt{DondersOlivi2020}, \citealt{AllairePantz2019}, \citealt{Groen18}, \citealt{GroenFork2019}). The framework has been further extended to incorporate mechanisms to handle singularities in the orientation field by identifying seamlines (\citealt{Stutzetal2020}) or regularisation (\citealt{DondersAllairePantz2017}), multiple loading cases (\citealt{JensenGroen2022}) and 3D optimised designs (\citealt{DondersAllaire2020}, \citealt{GroenStutz2020}).

The above-mentioned projection-based approaches rely on solving a global least-squares problem to obtain a conformal mapping field (\citealt{Groen18}). The globality of this approach may cause instabilities in the presence of singularities in the orientation field and makes solving the least-squares problem computationally demanding when problem size or complexity grows. Further, the strict orientation coherence requirement incorporated in the least-squares formulation is prone to cause large variation in the projected unit-cell sizes. This may cause the realised structure to not adhere to constant periodicity, an artefact which diminishes the homogenisation assumption of perfect periodicity, and may also cause disconnections or instabilities in thin structural members. The introduction of branching to increase the periodicity coherence as a post processing step to the projection based methods has proven to improve on this (\citealt{GroenFork2019}). 

\citealt{Senhoraetal2022} and \citealt{ZhengKochmann2021} perform dehomogenisation utilising spinodal microstructures, which build on many of the same mathematical concepts as the phasor noise utilised in this article, but the nature of the spinodal microstructures is inherently suboptimal. For 3D-dehomogenisation, heuristics based on stream-surfaces have been proposed in replacement of solving the least-squares problem (\citealt{Stutzetal2022}, \citealt{jensenogdeandre2023}). This approach reduces the sensitivity with respect to singularities compared to the projection-based methods, but gives rise to other challenges and is still computationally demanding. 

Motivated by these instabilities and the limited scalability of the global projection-based methods, more locally based methods relying on different computer graphics disciplines have been proposed. \citealt{Elingaardetal2022} trained a Convolutional Neural Network (CNN) to directly predict an intermediate mapping field, omitting the expensive solution of the least-squares problem. This procedure benefits from its locality which increases stability in the presence of singularities and the gain in efficiency by the direct mapping to the intermediate field, but due to the CNN being trained with the aim of maintaining constant periodicity local disconnections occur in this intermediate field at locations where branching is needed to maintain connectivity. An image-processing based procedure for locating and connecting these branches was proposed, at the expense of the overall dehomogenisation time and the mechanical soundness of the branch connections. The procedure is still more efficient than that of \citealt{GroenFork2019}, but this comes at a loss in solution quality. Additionally, this approach is prone to imperfections in terms of failing to discover all disconnections and adding additional unloaded artefacts in the branching regions. \citealt{Garnieretal2022} proposed an alternative pattern-generation approach to dehomogenisation based on the dynamical solution to a reaction-diffusion partial differential equation system. This procedure effectively grows the structure in a local way by iterating through the structural domain on some intermediate mesh. The main contribution is the observation that certain parameter choices allow for obtaining connected branches directly when building the pattern. These branches become similar to those of \citealt{Elingaardetal2022} in shape without the potential artefacts. Still, the obtained dehomogenised designs have similar structural performances to those obtained by the CNN.



Procedural phasor noise offers real-time rendering of smoothly varying wave patterns with local control of periodicity and orientation (\citealt{Tricard_phasornoise_2019}, \citealt{Tricard_orientable_2020}) making it a suitable basis for efficient dehomogenisation. The formulation of and manipulation of the textures is performed on the coarse mesh and the intermediate field is obtained based on sampling functional values at specified points in space. This makes the representation of the textures compact and the nature of the pattern continuous and mesh-independent. Exploiting these attributes allows for creating a time and memory efficient framework for generating orthotropic infill oriented microstructures.

To extend the phasor noise methodology for generating pseudo-randomly oriented microstructures (\citealt{Tricard_orientable_2020}), to one suited for structural dehomogenisation several novelties are introduced. Firstly, anisotropy is introduced both in the kernel phase-alignments and the phasor field sampling to reduce curvature and the impact of local singularities in this phasor field. Secondly, a direct approach for determining the location of the singularities as well as how the disconnection resulting from each individual singularity is best treated for branch closure is proposed. Lastly, an image morphology-based procedure is introduced to close these branches and ensure structural connectivity with minimal deviation in orientations, periodicity and material prescribed.

The proposed methodology allows for circumventing the training of a CNN and several of the more expensive post-processing steps at fine resolution, currently required for the neural network approach (\citealt{Elingaardetal2022}). Direct access to disconnected locations in the intermediate field allows for defining an inherently continuous approach for connecting the branches based on interpolation. Further, the intermediate field is described by a periodic wave field readily translated to a triangular wave field for direct thresholding, which makes the post-processing procedure for the phasor method significantly more efficient than that proposed for the CNN. Also, the procedural nature of sampling the intermediate field in the phasor-based method is inherently efficient, allowing for an overall dehomogenisation procedure with run times competitive with those of the neural network. The directly controllable nature of the procedure makes it flexible and reliable beyond what is realistically attainable by the CNN. This is exemplified by a better and more stable performance across different test instances.

The principles of the phasor-based method is further utilised to develop a procedure for adding a smooth, variable-thickness structural boundary. This procedure includes an approach for smoothing the fine-scale indicator field to reduce the impact of staircase artefacts such as structural appendices originating from direct upscaling of the coarser underlying grid. This procedure adopts the computational efficiency of the phasor-based technique by constructing and adapting the boundary on the coarse resolution only, before sampling along the structural boundary at a finer resolution. The underlying idea of the approach is only reliant upon the access to a coarse resolution indicator field outlining the structural domain, and can thus also be utilised within other optimisation frameworks, not otherwise reliant on phasor-based components. 

This paper introduces the concepts involved in utilising phasor noise techniques from computer graphics in structural design. The descriptions of the methodology, and its fundamentals, are presented in a tutorial-style manner to make the intuitive understanding of the different building blocks more accessible, by bridging the gap between computer graphics and mechanics, which will hopefully contribute to new ideas. To this end, the paper is organised as follows; a brief introduction to the nature of the homogenised optimised solution and the task of dehomogenisation is given in \autoref{sec:Dehom_main}. Phasor noise and how it is modified to perform dehomogenisation is detailed in \autoref{sec:phasor_main}, and implementational details are summarised in \autoref{sec:Implementation_main}. \autoref{sec:reuslts_main} presents a series of numerical tests and comparisons to existing methods to validate the performance of the proposed method, while the overall findings and future outlooks are presented in \autoref{sec:conclusion_main}.
\section{Dehomogenisation} \label{sec:Dehom_main}
\begin{figure}[!htb]
    \centering
    \includegraphics[width=0.9\linewidth]{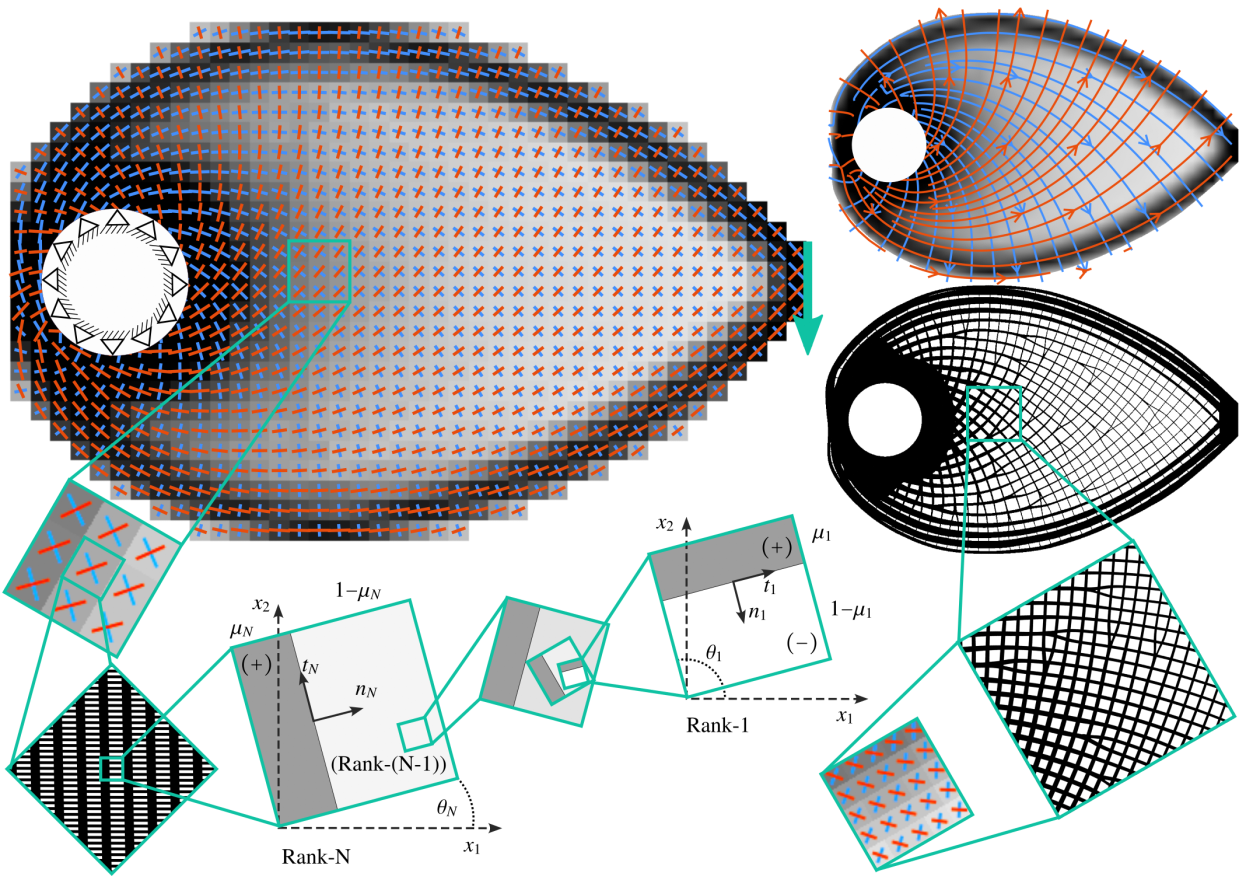}
    \caption{Illustration of the multi-scale Rank-N homogenisation-based formulation and how this translates to a single-scale dehomogenised design for a single load Rank-2 (N=2) example.}
    \label{fig:homogenised_sol_to_dehom}
\end{figure}

Homogenisation-based topology optimisation is a multi-scale design method utilising a design representation consisting of composite materials assumed to be periodic on a microscopic length-scale (\citealt{Bendsoe1989}). It has been proven that two-phase composites consisting of sequential laminates with a finite number of layers can achieve the theoretical maximum strain energy (\citealt{Avellaneda1987}, \citealt{FrancfortMurat1986}). These composites are referred to as Rank-N laminates for which the effective elastic properties can be derived analytically. For 2D single- and multi-load case problems Rank-2 and Rank-3 laminates are stiffness optimal, respectively. In 3D, the corresponding layer numbers ensuring optimal local stiffness are N=3 and N=6. The properties of the composites, in terms of lamination directions and layer thicknesses, are controlled only on the larger macroscopic length-scale, reducing the computational cost associated with the optimisation and analysis. For more detail on the derivation of the homogenised design representation and the corresponding optimisation, the reader is referred to (\citealt{WuSigGro21}).

The homogenisation-based TO approach allows for optimising the design on a coarse mesh, where the optimised result is a grey-scale solution representing the underlying assumption of infinite-periodic microstructures. This representation is crucial for efficiency and solution quality of the optimisation procedure, but the optimised result is not directly manufacturable. Dehomogenisation denotes the process of translating the multi-scale optimised design representation to a finite-periodic, single-scale physically realisable design with minimal loss in structural performance compared to the homogenised solution. \autoref{fig:homogenised_sol_to_dehom} illustrates the relation between the homogenised design, with underlying microstructure representation, and the single-scale dehomogenised design for a Michel-type beam. Consider the homogenised optimised solution of a Rank-N laminate where the solution is representated by a set of N layers each described by an orientation field and a relative thickness field. The dehomogenisation-goal is then to translate each of these layers to a smoothly varying constant-periodic lamination field with continuous transitions between the relative thicknesses on a fine resolution, where the final single-scale design is the union of these layers.

The dehomogenisation procedure proposed in this article considers these homogenised Rank-N optimisation results as input, meaning that it is assumed that N layers of thicknesses and orientations are provided for each element in the design domain (referred to as kernels in the context of phasors), and that there exists a minimal relative thickness $\mu_{min}>0$ such that any thickness value $\mu<\mu_{min}$ indicates void. For generating homogenised optimisation results to provide input datasets several approaches are available, for instance \citealt{WuSigGro21} with accompanying Matlab implementation. For details on how to avoid very small, but positive, relative thicknesses in the lamination layers, which otherwise could be challenging for single-scale realisations, the reader is referred to \citealt{Gieleetal2021} and \citealt{JensenGroen2022}. 

\section{Phasor noise for de-homogenisation} \label{sec:phasor_main}

This section introduces the concept of phasor noise and explains in detail how this methodology can be utilised for dehomogenisation. Procedural noise functions are pattern generating algorithms describing the characteristics of the pattern in an implicit manner through its parameterisation (\citealt{Lagaeetal2010ProceduralSurvey}). The procedural definition of a pattern allows for a compact representation of an inherently continuous and multi-resolution function that can be sampled at any point in space in constant time. This makes the pattern mesh-independent and the procedure of sampling the noise at different resolutions automated in a very efficient algorithmic manner.

\begin{figure}[!htb]
    \centering
    \includegraphics[width=1\textwidth]{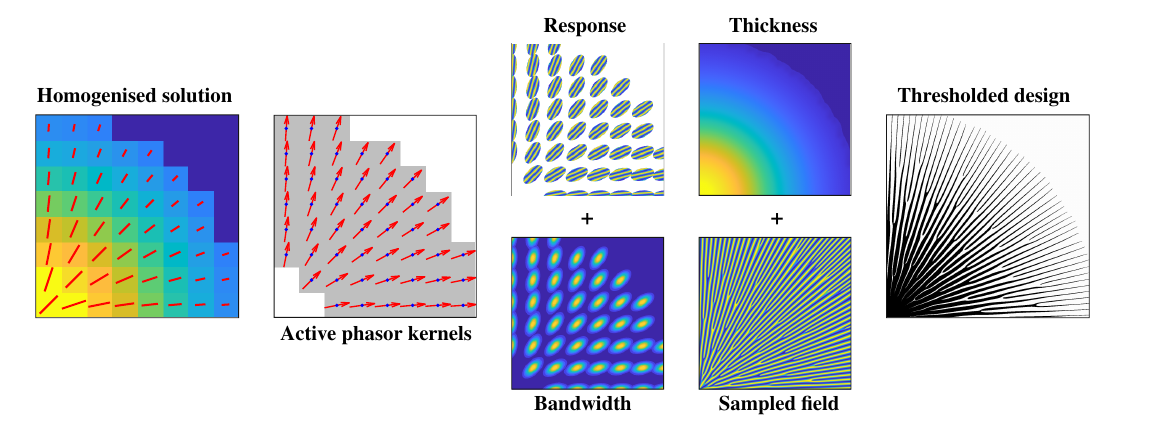}
    \caption{A homogenised solution is translated to a set of active kernels based on the element densities and orientations. With a prescribed periodicity and bandwidth each kernel transmits a complex wave signal with local response and the combined signal on the sampled domain provides a smoothly transitioning wave-field. The sampled field can be translated and directly thresholded based on the local thicknesses on the sample domain obtained by linear interpolation.}
    \label{fig:translate_sample_threshold}
\end{figure}

A phasor field is an instantaneous phase field obtained by summing a set of complex valued Gabor kernels (\citealt{Tricard_phasornoise_2019}). Computing the argument of the phasor field, a real valued phase is obtained. Applying a periodic function (e.g. sine) to this phase yields a striped pattern with frequency and orientation specified by the underlying Gabor kernels (\autoref{fig:translate_sample_threshold}).

Gabor noise represents a well-established sparse convolution procedural noise function (\citealt{Lagaeetal2009Gabor}). The kernels defining this noise, the Gabor kernels, are defined as the product between a Gaussian envelope and a harmonic (\autoref{fig:Isotropic_kernel_description}). Each kernel emits a signal describing a wave in space and the intensity of the signal is at any point determined by the magnitude of the Gaussian about its centre. Phasor noise is a sparse convolution noise defined by a set of weighted signal-emitting kernels in space. It is constructed by a reformulation of Gabor noise to a phasor field, allowing for a clear separation between intensity and phase.

 A key insight regarding phasor fields is that the local loss of contrast in the complex wave-field caused by destructive inference does not influence the frequency content of the field. Thus, by separating the phase from the intensity, highly contrasted periodic patterns with blended transitions between orientations, frequencies and thicknesses can be obtained. Only at singular points, where the phasor field is exactly zero, will there be discontinuities in the phase. These singular points occur at locations where the phase-field cannot both adhere to the constant periodicity and follow the specified orientations and therefore represent the exact locations of bifurcations in the phase field.

 \autoref{sec:phasor_basic_def} describes the basic definitions needed to formalise the noise and \autoref{sec:phasor_intro_sampling} utilises this formalisation to obtain the desired pattern. The details of the proposed dehomogenisation method are covered in \autoref{sec:phase_align}-\SecNum{subseq:branches}. Compared to \citealt{Tricard_phasornoise_2019} a more elaborate phasor alignment procedure is needed, which is described in \autoref{sec:phase_align}. Since the aim of the proposed procedure is to dehomogenise at a constant length scale, the phasor field will include bifurcations that must be handled appropriately. The phasor field formulation simplifies this process due to the relation between the bifurcations in the dehomogenised field and singular points in the phasor field. How the bifurcations are located and reshaped is described in \autoref{subseq:branches}.


\subsection{Basic definitions}\label{sec:phasor_basic_def}  
\begin{figure}[!htb]
\begin{minipage}{0.5\textwidth}
    \centering
    \includegraphics[width=0.9\textwidth]{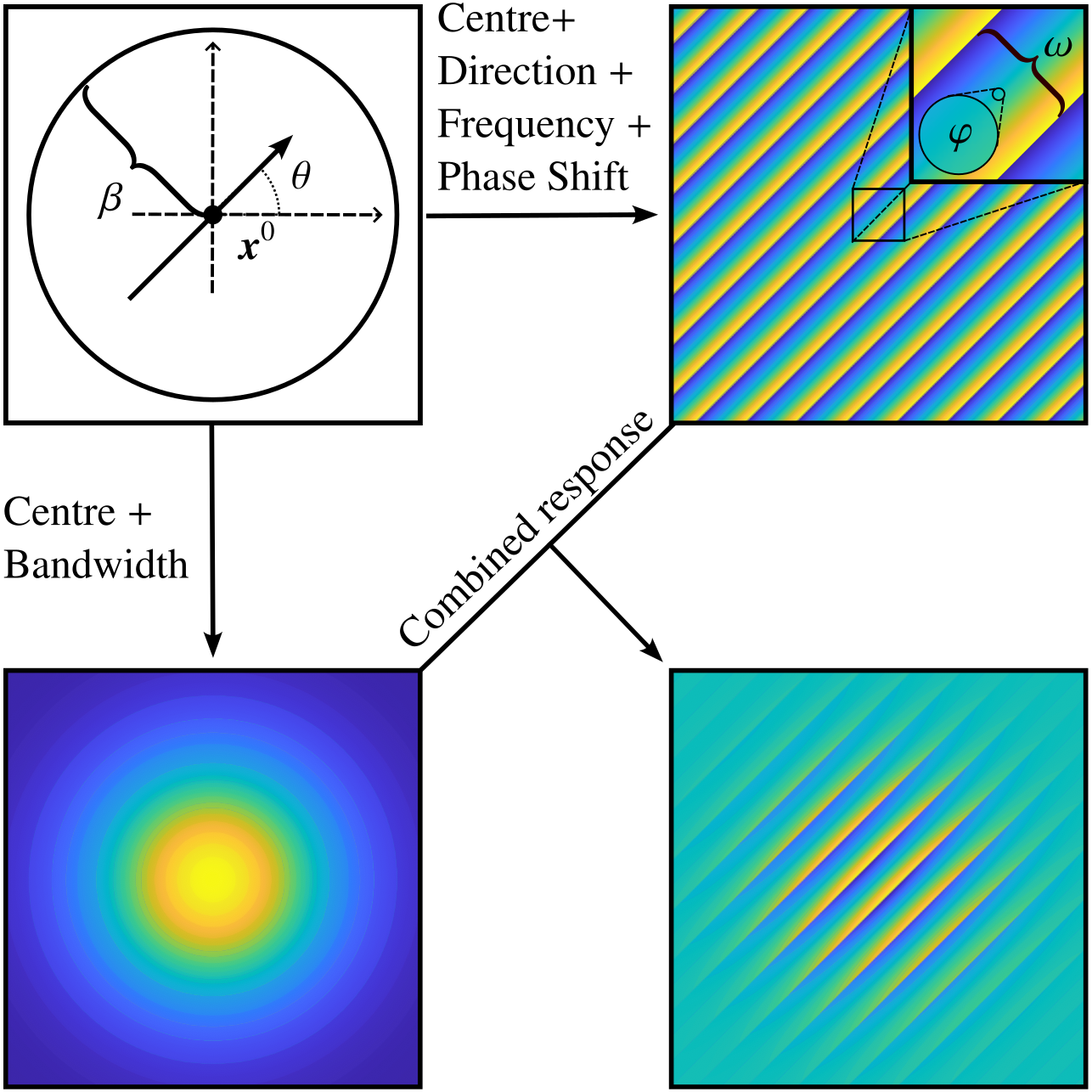}
\end{minipage}
\hfill
\fbox{
\begin{minipage}{0.45\textwidth}
\vspace{2mm}

\textbf{Notation}\\[2ex]
\begin{minipage}{0.95\textwidth}
\begin{flushleft}
    \begin{myitemize}
    \setlength{\itemindent}{-.1in}
    \item $\mathcal{K}$ : set of kernels
    \item $\boldsymbol{x_j^0}{=}\left(x^0_j,y^0_j\right)^T\in \R^2$: coordinate location of the origin of kernel $j\in \mathcal{K}$ (i.e. the centre)
    \item $\omega\in \R^+$ : fixed periodicity/frequency of the phasor field and dehomogenised design
    \item $\beta_j\in \R^+$ : the bandwidth of kernel $j\in \mathcal{K}$ controlling the radius of impact of the kernel
    \item $\theta_j$ : orientation of kernel $j\in \mathcal{K}$
    \item $\boldsymbol{d_j}{=}\left(d^x_j,d^y_j\right)^T$ : the direction of kernel $j\in \mathcal{K}$ in vector form $\left(d^x_j,d^y_j\right){=}\left(\cos(\theta_j),{-}\sin(\theta_j)\right)$
    \item $\varphi_j$ : the phase shift of kernel $j\in \mathcal{K}$
    \item $\mathcal{G}_j$ : The local response of kernel $j\in \mathcal{K}$
\end{myitemize}
\end{flushleft}
\end{minipage}
\vspace{2mm}
\end{minipage}}
    \caption{A basic isotropic phasor kernel with centre at $\boldsymbol{x}^0$, orientation $\theta$, bandwidth $\beta$, frequency $\omega$ and phase $\varphi$. The phasor kernel response (bottom right), $\mathcal{G}_j(\boldsymbol{x})$, is given by the product between a $\omega$-periodic sawtooth wave of phase shift $\varphi$ about the centre-line (top right) and a Gaussian of bandwidth $\beta$ about the kernel centre (bottom left).}
    \label{fig:Isotropic_kernel_description}
\end{figure}

Let $\mathcal{K}$ denote a set of 2D phasor kernels. Each kernel $j\in \mathcal{K}$ is uniquely defined by the coordinate location of its origin $\boldsymbol{x}_j^0\in\mathbb{R}^2$, its orientation $\theta_j\in [-\pi,\;\pi]$, its frequency $\omega\in \R^+$, its bandwidth $\beta_j<\omega$, and its phase shift $\varphi_j\in [-\pi,\;\pi]$. Based on these defining parameters the kernel $j\in \mathcal{K}$ is the product of two complex functions; a Gaussian of bandwidth $\beta_j$ centred at $\boldsymbol{x}_j^0$ and an oscillator of frequency $\omega$, phase $\varphi_j$ and direction $\boldsymbol{d_j}{=}\left(\cos(\theta_j),\sin(\theta_j)\right)^T$. The local response of kernel $j\in \mathcal{K}$ at the location $\boldsymbol{x}\in \mathbb{R}^2$ is dependant upon the distance between the phasor centre and the sampling point, weighted by the bandwidth, and is given;
\begin{equation}
    \mathcal{G}_j(\boldsymbol{x})=\mathrm{e}^{- \beta_j\|\boldsymbol{x}-\boldsymbol{x^0_j}\|^2} \mathrm{e}^{2i\pi \omega\boldsymbol{d}_j\cdot(\boldsymbol{x}-\boldsymbol{x^0_j})+i\varphi_j}
\end{equation}
Given a set of sampling points $\mathcal{T}\subset \mathbb{R}^2$ the phasor field value at sampling point $\boldsymbol{x}\in \mathcal{T}$, $\phi(\boldsymbol{x})$, is given the instantaneous phase, corresponding to the real-valued argument, of the complex Gabor-noise resulting from summing the phasor kernel contributions at the sampling point;
\begin{equation}
    \mathcal{G}(\boldsymbol{x}){=}\sum_{j\in \mathcal{K}}\mathcal{G}_j(\boldsymbol{x}),\quad \phi(\boldsymbol{x}){=}\Arg (\mathcal{G}(\boldsymbol{x}))
\end{equation}
The resulting phasor field $\phi(\boldsymbol{x})\in [-\pi,\pi]$ is a sawtooth wave-field with varying orientations and periodicities blended as defined by the phasor kernels. This periodic field can be subjected to any $2\pi$-periodic function to change its profile but maintain the direction and periodicity of the pattern. The phasor sine-wave is obtained by applying the $2\pi$-periodic function $\sin(\phi(\boldsymbol{x}))$ and forms the foundation for translating the phasor field to a de-homogenised design. This translation ensures a smoothly oscillating field, later to be translated to local densities and thresholded.

\autoref{fig:Isotropic_kernel_description} illustrates how the defining components of a single isotropic phasor kernel combines to the local response sampled near the kernel centre. The bandwidth of a kernel determines the intensity of its signal at any given point in  $\mathbb{R}^2$ based on the distance to the kernel centre. Effectively the bandwidth can be viewed as a parameter controlling the radius of the main area of impact of the given kernel. The bandwidth is directly related to the standard deviation of a Gaussian filter $\sigma={1}/{\sqrt{2\pi\beta_j}}$, a crucial consideration when later choosing the magnitude of the bandwidths. The frequency of the kernel determines the periodicity of the oscillations locally about the kernel centre, and its phase the shift in periodic behaviour about the centre.

\begin{figure}[!htb]
    \centering
   \includegraphics[width=1\linewidth]{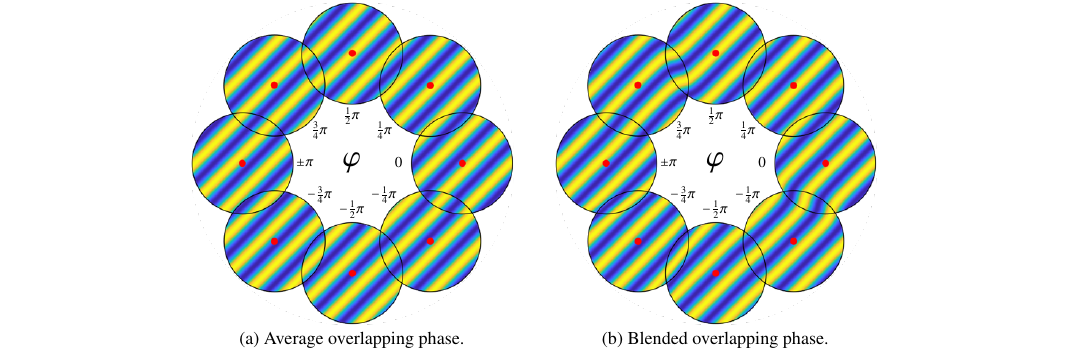}
    \caption{Illustrating how the kernel phase shift $\varphi$ shifts the phasor sine-wave, $\sin\phi(\boldsymbol{x})$, and how the phasor-sampling process provides smooth transitions between the shifted phasor kernels. The phase shift is $2\pi$-periodic.}
    \label{fig:circular_phase_shift_illustration}
\end{figure}

The phase shift $\varphi_j$ determines the value $\phi(\boldsymbol{x}_j^0)$. For the base case $\varphi_j=0$ the phasor field $\phi(\boldsymbol{x})$ has a zero-contour line along the kernels orientation passing through the kernel origin. Due to periodic behaviour this is true for any $\varphi_j=2k\pi,\; k\in \Z$. Given a kernel phase $\varphi_j=k\pi+a,\; a\in \R,\; k\in \Z$ the phasor field profile is shifted such that the contour-line passing through $\boldsymbol{x}_j^0$ has the value $a$. The effect of the phase shift of the same kernel is illustrated in \autoref{fig:circular_phase_shift_illustration}.

The combined response of the single phasor kernel in \autoref{fig:Isotropic_kernel_description} is weighted by the signal magnitude to indicate the strength of the kernel's influence when moving away from its centre. Defining the phasor field $\varphi$ as the argument of the complex signal reduces the occurrence of vanishing waves in regions of low intensity significantly. In the case of a single kernel, $\varphi(\boldsymbol{x})$ corresponds to the top right sawtooth wave in \autoref{fig:Isotropic_kernel_description}. In the case of blending multiple kernels the signal intensity determines to what degree the orientation and frequency of a kernel influences the local direction and periodicity at a sampling point compared to the remaining kernels.

\begin{figure}[!htb]
    \centering
\includegraphics[width=1\linewidth]{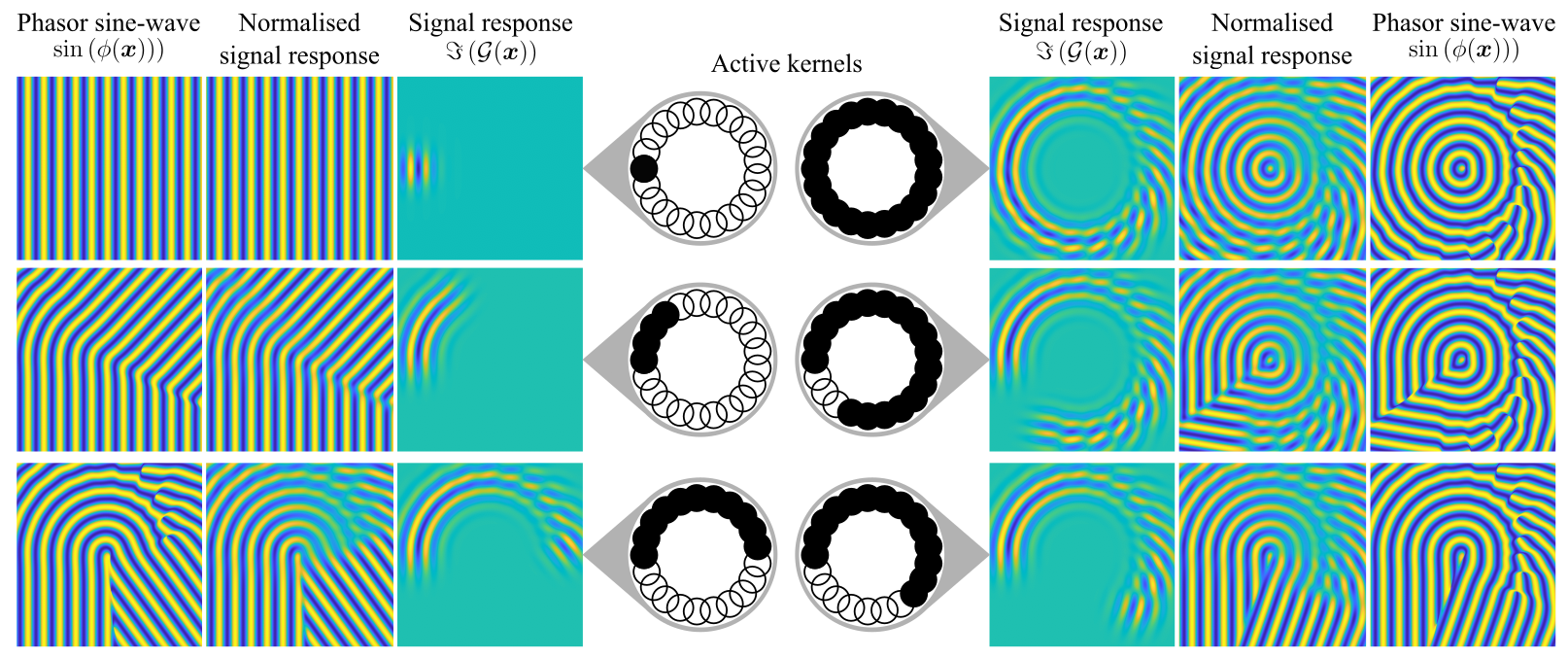}
    \caption{Illustrating the effect of sampling a phasor field from an increasing set of underlying active phasor kernels on a triplet consisting of the intensity-weighted signal response, the normalised response and the phasor sine-wave obtained by the instantaneous phase of the complex signal response.}
    \label{fig:sampling_with_different_no_kernels}
\end{figure}

\autoref{fig:sampling_with_different_no_kernels} illustrates the effect of blending the response from an increasing number of phasor kernels along a circle. The imaginary part of the complex Gabor noise $\Im{\mathcal{G}(\boldsymbol{x})}$ illustrates the intensity of the signal in the sampled domain. Effectively, $\Im{\mathcal{G}(\boldsymbol{x})}$ is a weighted sum of real sine waves defined by taking the imaginary part of the phasor field, corresponding to the sine-version of a Gabor noise (\citealt{Lagaeetal2009Gabor}). Normalising the signal response, $\Im{\mathcal{G}}$, by the sum of the Gaussian weights increases the intensity of the overall response field. However, due to the equivalence with summing real-valued wave-functions, destructive inference leads to regions with local loss of contrast as more kernels are added. The phasor sine wave, $\sin(\phi(\boldsymbol{x}))$, is obtained by applying a sine-transform to the instantaneous phase of the complex phasor field. By construction, phasor noise omits the intensity of the Gabor noise, such that the local loss of contrast is circumvented to form a perfectly oscillating wave-field.

\subsection{Translation of optimised solution}\label{sec:phasor_intro_sampling}
In dehomogenisation, as discussed in \autoref{sec:intro_main} and \autoref{sec:Dehom_main}, the aim is to obtain a constant periodic infill of specified relative thicknesses following the varying lamination directions (\autoref{fig:homogenised_sol_to_dehom}). Procedural pattern synthesis by phasor noise allows for obtaining such patterns in a computationally efficient and flexible manner. Given a homogenised topology optimised solution on a coarse mesh a set of element centres and lamination widths and directions are provided. Provided a finite periodicity and appropriate kernel bandwidth the coarse-mesh information can be translated directly to a set of uniquely defined phasor kernels. Summing these kernels one can obtain the phase field value at any location in the domain. The real-valued field obtained from taking the argument of the complex-valued phasor field corresponds to a spatially varying sawtooth wave-field controlled by the desired periodicity and local orientation. This allows for applying $2\pi$-periodic functions to change the profile of the sawtooth-waves. Therefore, a direct translation to a periodic triangle field with periodic linearly piecewise transitions in $[0, 1]$ is available. The resulting triangular wave-field can then be directly thresholded to satisfy the local relative thickness.

Considering each lamination layer separately, let each element of intermediate density, $\rho\in]\mu_{min},1[$, in the homogenised optimised solution represent a phasor kernel with origin at the element centre and orientation specified by the lamination orientation. Note that solid or void elements are excluded both to circumvent the issue of local singularities in the lamination orientations at these densities as well as for limiting the computational cost of later aligning and sampling of the phasor field. When performing thresholding to obtain the final dehomogenised design these regions will become fully solid or void, meaning the local orientation of the phasor field is irrelevant. The desired periodicity of the de-homogenised design defines the phasor-kernel frequencies to be constant across all kernels. Considering structured meshes, all kernels are assigned the same bandwidth ensuring a radius of impact with a small overlap between the coarse-scale elements (\autoref{tab:Parameter_choices}). In this way, the homogenised optimised solution constitutes a set of phasor kernels to be used for dehomogenisation. The choice of kernel phase shifts will be discussed in \autoref{sec:phase_align}.

\subsection{Phase-alignment}\label{sec:phase_align}
One of the main challenges associated with using phasor noise as a basis for generating dehomogenised structures is how the waves exhibit singularities where they diverge from an ideal regular oscillation. These defects occur where the field vanishes or the phase field has large local variation. It is found that even for perceivable smooth orientation fields such defects may cause local curvatures detrimental to structural performance. To counteract these negative effects, phase alignment is introduced to regularise the oscillations of the phasor noise by an iterative procedure adjusting the phase of a kernel based on the phase of its neighbours, similar to that presented by \citealt{Tricard_orientable_2020}. 

\autoref{alg:phase_alignment} outlines one iteration of the phase alignment procedure starting from a set of initial phase shifts $\varphi_j\in [-\pi,\;\pi]$ for all kernels $j\in \mathcal{K}$. For each kernel $j\in \mathcal{K}$ a predefined neighbourhood of nearby kernels $\mathcal{N}(j)$ is considered for the alignment of kernel $j\in \mathcal{K}$. Conflicting signals, causing destructive inference in the phasor field, can only occur between kernels with overlapping impact radius in the sampling-domain and thus, it is only necessary to align a kernel with a smaller set of its nearest neighbours. 

\begin{algorithm}[h]\onehalfspacing
\caption{One iteration of phase alignment}\label{alg:phase_alignment}
\begin{algorithmic}
\Require $\varphi_j$ initialised
\For{$j \in \mathcal{K}$}
\State $\upsilon_j=\sum_{i\in \mathcal{N}(j)} |\boldsymbol{d}_j\cdot \boldsymbol{d}_i|\mathrm{e}^{2i\pi \omega_i \boldsymbol{\tilde{d}}_{ij}\cdot(\boldsymbol{x}^0_j-\boldsymbol{x}^0_i)+i\tilde{\varphi}_{ij}}$ \Comment{Sum the complex-valued contribution of each kernel in the neighbourhood}
\State $\varphi_j=\Arg(\upsilon_j)$ \Comment{Obtain real-valued weighted average phase shift}
\EndFor
\end{algorithmic}
\end{algorithm}

The original phase-alignment considered $\boldsymbol{\tilde{d}}_{ij}=\boldsymbol{d}_i$ and $\tilde{\varphi}_{ij}=\varphi_i$. The updated phase of a kernel is then achieved by evaluating the value of the complex oscillator of its neighbours $i\in \mathcal{N}(j)$ at the centre $\boldsymbol{x}_j^0$ and performing a weighted average based on the absolute value of the dot-product between the directions of kernel $j$ and $i$. This weight ensures that the alignment effort is stronger with kernels of similar orientations and diminishes completely for orthogonally oriented kernels, as alignment of kernels with large deviation in orientation is invalid. Such jumps in lamination layer orientations may happen in homogenised solutions with angular singularities, as discussed in \citealt{Stutzetal2020}.

\begin{figure}[!htb]
\begin{minipage}{1\textwidth}
\centering
    \includegraphics[width=\textwidth]{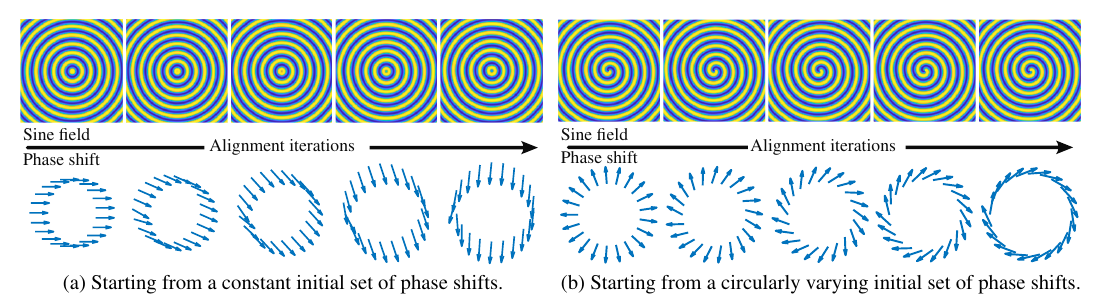}
\end{minipage}
\vfill
\begin{minipage}{\textwidth}
\centering
    \includegraphics[width=0.5\textwidth]{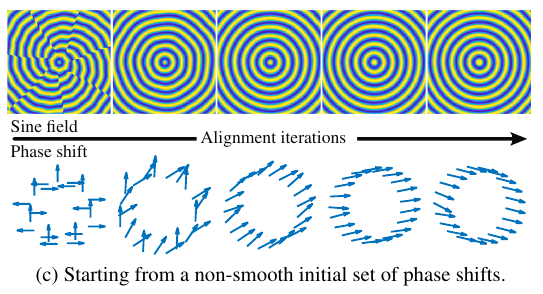}
\end{minipage}
\caption{Illustrating the effects of phase alignment on the sampled sine-field for iteration 0-4 of phase alignment starting from a constant initial set of phase shifts (a), a circularly varying initial set of phase shifts (b) and a non-smooth initial set of phase shifts (c). Note that the phase shift is a scalar but as it is applied by multiplication with $\mathrm{e}^{i\varphi}$ it effectively acts as a $2\pi$-periodic orientation which is utilised here to illustrate the phase-shifts by vectors during the alignment iterations. This representation clearly reflects both the effect of the alignment steps but also clarifies why a fluctuating iterative pattern with equally good alignemnts may occur.}
\label{fig:circle_phase_align_perfect_from_zero}
\end{figure}

\autoref{fig:circle_phase_align_perfect_from_zero}(a)-(c) illustrate different effects and aspects of the phase alignment procedure. The set of phasor kernels considered forms a circle with orientations along the circumference. Each figure shows the changes induced by performing four iterations of phase alignment, starting from different initial sets of kernel phase shifts. The neighbourhood of a kernel is in each case considered to be the previous and following kernels along the circle of kernels.

Due to the perfect circular definition of the phasor kernels starting from a constant zero phase shift in \autoref{fig:circle_phase_align_perfect_from_zero}(a) corresponds to starting from perfectly aligned phases. In this case the result of the phase alignment is a constant shift of all the kernel phase shifts resulting in an overall phase shift of the phasor field. As such, continued iteration will circulate through a periodic pattern of overall phase shifts, all equally valid as a perfect alignment. Starting from circularly varying phase shifts in \autoref{fig:circle_phase_align_perfect_from_zero}(b) further indicates that there are multiple optimal phase alignments for a set of kernels. Here the resulting phasor field is a perfect spiral instead of nested circles as before, but the solution is equally smooth and further phase alignment simply shifts the phase of the entire field simultaneously maintaining the perfect alignment. Starting from a non-perfect alignment in \autoref{fig:circle_phase_align_perfect_from_zero}(c) it is illustrated how the procedure efficiently restores the perfect alignment pattern from \autoref{fig:circle_phase_align_perfect_from_zero}(a).

As such, there may be multiple optimal phase-alignments for a set of phasor kernels. In the case of the presented circular case these optimal alignments correspond to a perfectly smooth alignment. However, not all sets of phasor kernels will have a perfect alignment, for example if the underlying directional field is divergent. There is therefore no guarantee that continued alignment iterations will ensure a better solution, and the alignment may even stagnate into an oscillating iterative pattern. Thus, it is considered sufficient to apply a fixed number of alignment iterations for later use.

To utilise phasor noise for dehomogenisation there are several additional challenges that must be accounted for, not considered in the original phase alignment procedure (\citealt{Tricard_orientable_2020}): matching the phase of opposing vectors, the negative impact of local lamination curvature and the occurrence of passive void or solid regions.
To this end, three extensions to the phase alignment formulation are introduced. The first is related to the rotational invariance in the homogenised representation requiring phase-matching for neighbouring kernels with opposite directional vectors. Secondly, an imbalance in how strongly a kernel is aligned with certain neighbours can induce local lamination curvature, especially near a singular point, in the resulting field, which should be reduced to improve structural integrity. Lastly, optimised structures may contain small solid or void regions within the structural body, where there are no active phasor kernels, complicating the construction of appropriate neighbourhoods for alignment.


\subsubsection{Matching the phase of opposing kernels}
The first challenge arises from the kernel directions spanning the entire unit circle, $\theta_j\in [-\pi,\;\pi[$. The originally presented phase alignment was developed for pseudo-randomly sampled directional fields with prescribed angular variation within the unit half circle $[0,\pi[$ (\citealt{Tricard_phasornoise_2019}). If two kernels are counter-oriented, their aligned phases will cause a shift in their sine-profiles causing abrupt field disconnections. To account for this artefact the phase-alignment update for kernel $j\in \mathcal{K}$ is modified, similarly to the phase-alignment procedure of \citealt{Lichtenbergetal2018}, such that the contribution of kernel $i\in \mathcal{K}$ is determined by its local candidate phase shift $\tilde{\varphi}_{ij}$ and similarly its local direction $\boldsymbol{\tilde{d}}_{ij}$.
\begin{equation}
   \tilde{\varphi}_{ij}=\begin{cases}
        \pi-\varphi_i & \textit{if } \boldsymbol{d}_j\cdot \boldsymbol{d}_i<0 \\
        \varphi_i & \textit{otherwise}
    \end{cases}\quad \land \quad \boldsymbol{\tilde{d}}_{ij}=\begin{cases}
        -\boldsymbol{{d}}_{i} & \textit{if } \boldsymbol{d}_j\cdot \boldsymbol{d}_i<0 \\
        \boldsymbol{{d}}_{i} & \textit{otherwise}
    \end{cases}
\end{equation}
These adjustments encourage coherence between the wave functions emitted from the two kernels also when they are of opposing directions. Note that $\sin(\pi-\varphi)=\sin(\varphi)$, meaning that for a real-valued periodic transformation of the phase field, the candidate phase shift $\tilde{\varphi}$ is consistent with $\varphi$ even when there are discontinuities in the directions of the waves in the complex phasor field. To illustrate the effect of this adjustment the circluar phasor kernel from \autoref{fig:sampling_with_different_no_kernels} is revisited, but now the directions of the kernels in the first and third quadrants are flipped to maintain the circular orientation but force a scenario where neighbouring kernels have opposite directions. A perfect alignment exists for this case, as for the original circle, but can only be found if the $2\pi$-periodic behaviour is accounted for.
\begin{figure}[!htb]
    \centering
    \includegraphics[width=\linewidth]{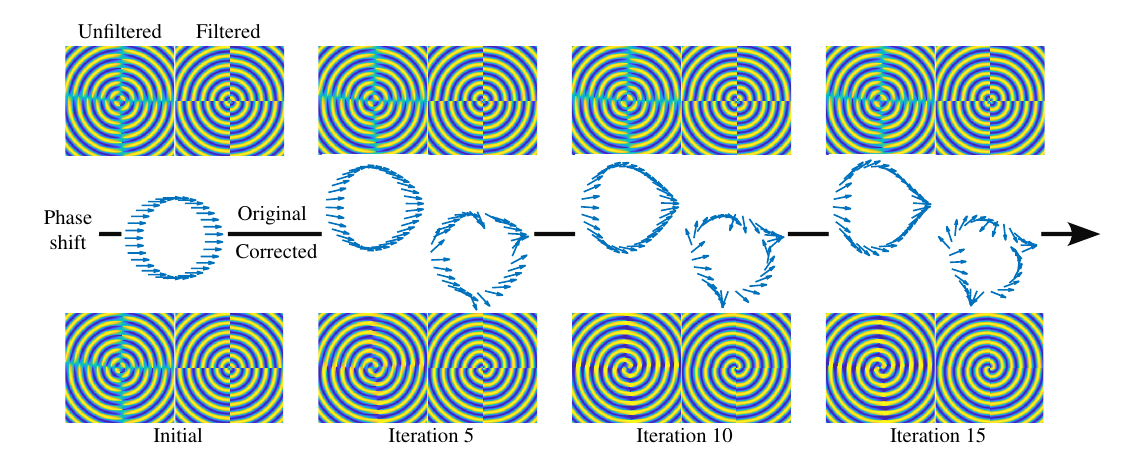}
    \caption{Ilustrating the phase alignment procedure for cases of neighbouring kernels with opposing directions utilising the original and the updated phase alignment procedure. Further, the effect of the sampling filter to be introduced in \autoref{sec:sampling_and_filter} is illustrated.}
    \label{fig:opposing_angle_alignment_exemplified_with_filter}
\end{figure}

\begin{figure}[!htb]
    \centering
    \includegraphics[width=0.9\linewidth]{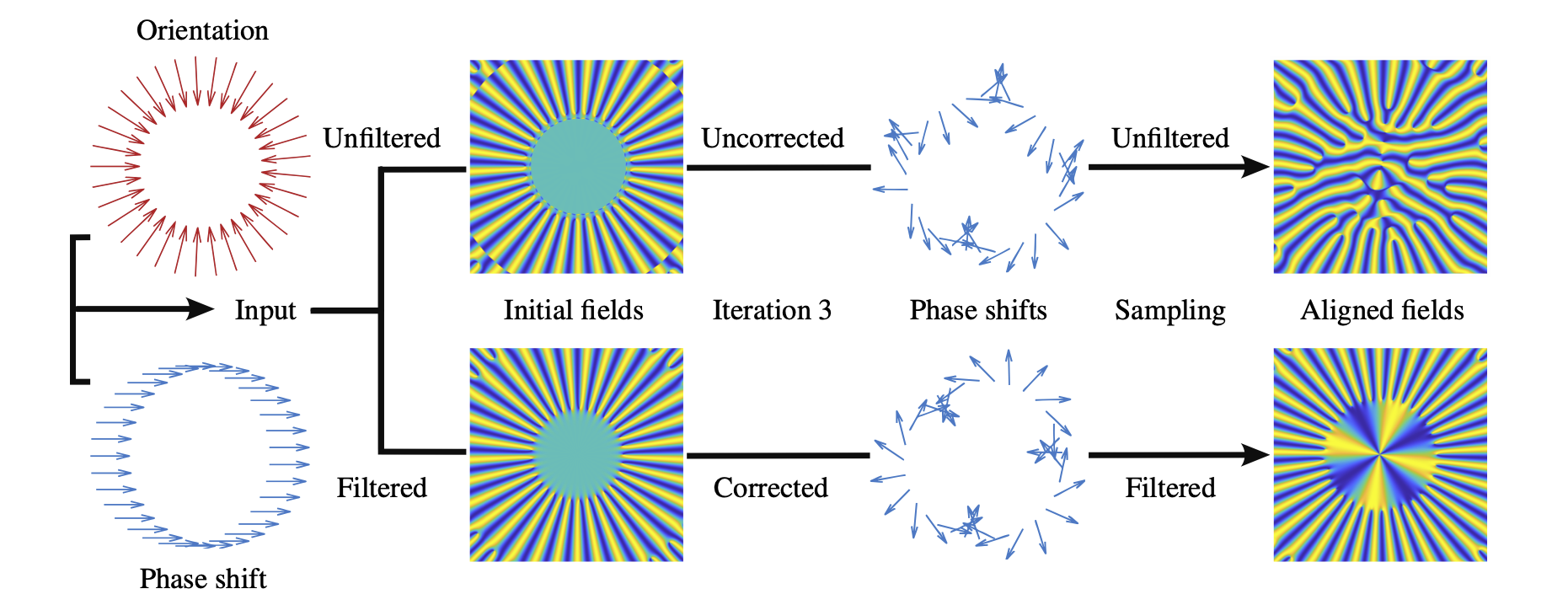}
    \caption{Illustrating the effect of having a central singularity in the orientation field by considering the case where all the phasor kernels along the circle have directions pointing towards the centre of the domain. The top sequence illustrates the case if no sampling filter or alignment correction is imposed, while the bottom illustrates the effect of incorporating both these features.}
    \label{fig:flipped_perfect_star_singularity}
\end{figure}

\autoref{fig:opposing_angle_alignment_exemplified_with_filter} provides a visual comparison to exemplify this concept. Starting from zero kernel phase shifts the original and extended phase alignment procedures are run for 15 iterations. The top row illustrates how the basic procedure fails to recognise the effect of opposing angles and thus at each iteration produces a phasor field where opposing directions results in a phase shift of $\pi$ in relation to each other. The bottom row conversely illustrates how the extended version aligns the phases of these opposing regions and approaches the perfectly aligned spiral similarly to \autoref{fig:circle_phase_align_perfect_from_zero}. Further, the periodic wave behaviour of the phasor field vanishes when sampling near the boundary between opposing regions. This effect can be relieved by introducing a filter in the sampling procedure, which will be covered in more detail in \autoref{sec:sampling_and_filter}. \autoref{fig:flipped_perfect_star_singularity} similarly illustrates the improvement in robustness against singularities in the orientation field, achieved by adding the described phase-alignment correction and sampling filter.

\subsubsection{Singularities and local curvature}
The second challenge to account for is related to an inherent artefact arising from the phasor field attempting to maintain constant periodicity even when branching across singular points in the phase field. The introduction of a branch may cause local curvature in the surrounding field to better fit the branching into the desired constant periodicity, which results in a small phase shift orthogonal to the lamination direction. \citealt{Tricard_orientable_2020} utilised phasor noise for designing microstructures where this was not considered a challenge, but for dehomogenisation of rank-N structures this may deteriorate the structural performance significantly. To best approximate constant periodicity globally, it is necessary to allow for local deviation about a branching point. Reducing the local curvature around a branching point can be achieved by increasing the region where deviation in periodicity is allowed. Done correctly, this will aid in removing the local curvature as well as improve the shape of the branching regions. This improvement in shape will be of great importance when later connecting disconnected branches in Section 3.5, which is crucial for mechanical performance. 

\begin{figure}[!htb]
    \centering
     \includegraphics[width=\linewidth]{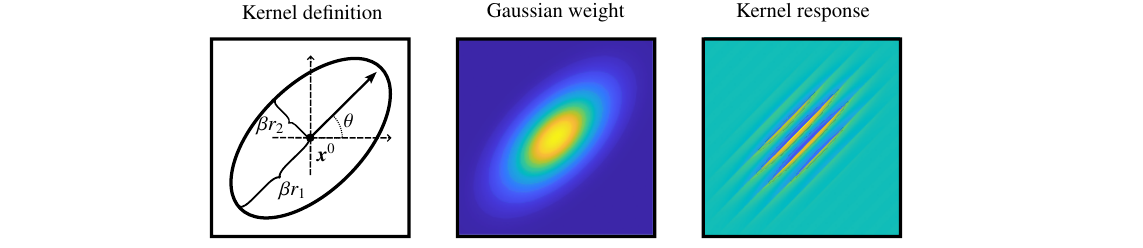}
    \caption{Illustrating how the introduction of anisotropy in the phasor sampling modifies the phasor description components compared to the isotropic kernel in \autoref{fig:Isotropic_kernel_description}. The changes are all driven by the modification of the shape of the Gaussian kernel determining the signal strength, where anisotropy here is introduced with the main axis along the kernel orientation.}
    \label{fig:anisotropic_kernel}
\end{figure}

One of the novelties of the phase alignment procedure presented in this work is an extended anisotropic definition of a kernel's neighbourhood, with major axis along the lamination direction. By reducing the degree to which a kernel is aligned with its neighbours in the orthogonal direction, and increasing to what extent it is aligned with its neighbours along the lamination direction, it is possible to reduce the curvature about singular points and encourage straighter curves along the lamination direction. Formally this is obtained by defining an oriented and directionally-weighted distance measure $\|\boldsymbol{\cdot}\|_{\theta}^{(r_1,\;r_2)}$ leading to the neighbourhood definition
\begin{equation}\label{eq:align_neighbourhood}
    \mathcal{N}(j)=\left\{i\in \mathcal{K}:\; \left\|(\boldsymbol{x_j^0}-\boldsymbol{x_i^0})\right\|_{\theta}^{(r,\;r^{-1})}<R^2_j\right\},\;\|\boldsymbol{\cdot}\|_{\theta_j}^{(r_1,\;r_2)}=\left\|\begin{pmatrix}r_1\sin\theta & {r_1}\cos\theta\\ -{r_2}\cos\theta & {r_2}\sin\theta\end{pmatrix}(\boldsymbol{\cdot})\right\|_2,\; r_1,r_2\in\mathbb{R}^{+}
\end{equation}
The desired anisotropy is achieved by selecting $r<1$ in \autoref{eq:align_neighbourhood}. As $r\rightarrow 0$ the alignment in the orthogonal direction vanishes almost completely, depending on the underlying kernel orientations. It is still prudent to maintain some alignment in the orthogonal direction for maintaining periodic behaviour in areas away from branching points. Thus, care should be taken as to not choose $r$ so small that the only kernels in the neighbourhood are those along the lamination direction. For further detail on the choice of these parameters the reader is referred to \autoref{app:parameter_select}. The Gaussian kernel in \autoref{fig:anisotropic_kernel} illustrates the anisotropic shape of the new neighbourhood definition.

\subsubsection{Alignment of partially isolated kernels}
The third potential challenge of utilising phasor noise for dehomogenisation is caused by the occurrence of thin structural members or smaller passive (solid or void) regions present within the structural body. These features are represented by only a few coarse mesh elements, fully contained within the desired phase-alignment neighbourhood. Such occurrences may induce erroneous behaviour in the phase-alignment procedure either by causing unwanted kernel-alignment across passive regions or insufficient alignment within thin structural members. The result of this is a finalised structure with unwanted curvature, larger variation in periodicity and potential structural disconnections.

To reduce unwanted alignment across passive regions, the kernel neighbourhoods considered for alignment can be chosen as to not significantly exceed the filter radius from the optimisation. As the minimal size of a passive void region is determined by the optimisation filter radius the kernel neighbourhood would have to be larger than the filter neighbourhood to span across a void region in any substantial manner. It is a requirement that the alignment neighbourhood spans more than one coarse element in any direction, and a larger span in the direction of anisotropy is beneficial, but it is also recommended to keep the neighbourhood relatively small both for computational performance and solution quality. Thus, depending on the relation between the filter radius imposed during optimisation and the radius of the alignment neighbourhood the occurrence of a void cut-off of a neighbourhood $\mathcal{N}(j)$ may not pose a significant challenge.

The problem occurring when a neighbourhood $\mathcal{N}(j)$ is drastically smaller than the theoretical maximum size due to cut-off by solid or void regions is more demanding to correct for. Especially challenging is the alignment of longer narrow regions where the lamination direction is orthogonal to the prolonged direction of the region. Here it is proposed to increase the size of $\mathcal{N}(j)$ by decreasing the degree of anisotropy for affected kernels. 

This is implemented for each lamination layer $l\in \mathcal{L}$ by considering the indicator field of active phasor kernels $\boldsymbol{z}^l$.
\begin{equation}
    \mathcal{K}^l=\left\{j\in \mathcal{K}\;:\;\mu_j^l\geq \mu_{min}\; \land \; \displaystyle\sum_{l\in\mathcal{L}}\mu_j^l\leq0.99\right\},\; {z}_j^l=\begin{cases}1 & \text{if }j\in\mathcal{K}^l\\0 & \text{otherwise} \end{cases}
\end{equation}
The boundary of this indicator field indicates what active kernels have solid or void neighbours. To obtain a measure for how much each active kernel is affected by these inactive kernels, in terms of its alignment neighbourhood becoming smaller, the indicator field $\boldsymbol{z}^l$ is first filtered using a 3x3 mean filter with replication padding to obtain $\boldsymbol{\tilde{z}}^l$. The Sobel-operator (\citealt{sobel1973}), a discrete gradient-filter often used for edge detection in images, is then used to obtain the gradient directions $\kappa^l_e,\; e\in \left\{j\in \mathcal{K}^l\;:\;{\tilde{z}_j}^l\in]0,\;1[\right\}$ along the filtered boundary of $\boldsymbol{\tilde{z}}^l$. Gradients along the boundary on opposing sides of thin members will have opposing directions. Thus, given the set $\mathcal{B}^l=\left\{j\in \mathcal{K}^l\;:\;{\tilde{z}_j}^l\in]0,\;1[\right\}$ of near boundary kernels with corresponding boundary gradients, given $\kappa^l_e$, thin members can be identified as the set of elements in a neighbourhood where the minimal dot product between the boundary gradients in the neighbourhood approaches -1. 

To account for the anisotropic nature of the alignment neighbourhood $\mathcal{N}(j)$ a larger 5x5 neighbourhood is used for determining $\Delta\kappa^l_e$ indicating how close a boundary region element $e\in \mathcal{B}^l$ is to a narrow structural member. 
\begin{equation}
    \Delta\kappa^l_e=\max_{m\in \mathcal{M}^{5\times 5}_{e}\cap \mathcal{B}^l}\left\{\dfrac{1-\left(\cos\kappa_e^l\cos\kappa_m^l+\sin\kappa_e^l\sin\kappa_m^l\right)}{2}\right\},\; \mathcal{M}^{(n+2)\times (n+2)}_e=\{j\in\mathcal{K}^l\;:\; \max\{|\boldsymbol{x}_e-\boldsymbol{x}_m|\}\leq n\},\; n\in \mathbb{N}
\end{equation}
Further, a reduction in the degree of anisotropy of a kernel neighbourhood in a narrow structural member will only increase the size of the alignment neighbourhood in any substantial or beneficial manner if the kernel orientation is not aligned with the main direction of the structural member. Therefore, $\Delta\kappa^l_e$ is adjusted for kernels where the boundary orientation follows the lamination orientation in the current layer, 
\begin{equation}\label{eq:phasealignrad}
     \Delta\kappa^l_e:= \Delta\kappa^l_e\left(1-\mathcal{H}\left(\left|\cos\kappa_e^l\cos\theta_e^l+\sin\kappa_e^l\sin\theta_e^l\right|,\;0.99\right)\right),\; \mathcal{H}(\delta,\;\eta)=\begin{cases}1 & \text{if }\delta\geq \eta\\ 0 & \text{otherwise}\end{cases}
\end{equation}
before a 3x3 pill-box filter, i.e. a circular averaging filter with radius 1 (\citealt{HornSjoberg1974}), is applied to ensure smooth transitions and obtain the measure $\Delta\tilde{\kappa}^l_e$. The alignment neighbourhood for a kernel is then defined by

\begin{equation}
    \mathcal{N}(j)=\left\{i\in \mathcal{K}:\; \left\|(\boldsymbol{x_j^0}-\boldsymbol{x_i^0})\right\|_{\theta}^{r(1-\Delta\tilde{\kappa}_j)+\Delta\tilde{\kappa}_j,\;r^{-1}(1-\Delta\tilde{\kappa}_j)+\Delta\tilde{\kappa}_j)}<R^2_j\right\}
\end{equation}
such that for $\Delta\tilde{\kappa}_j\rightarrow 1$ the alignment neighbourhood approaches a fully isotropic neighbourhood. This correction is not guaranteed to overcome the problem completely, but it helps alleviate the most adverse effects. The direction of anisotropy should not be changed more than strictly necessary, as this may propagate into larger regions of the structure and cause unfavourable curvature there.

\subsection{Intermediate grid sampling and characteristics}\label{sec:sampling_and_filter}
Having performed the phase alignment to obtain the kernel phase shifts and neighbourhoods, sufficient information about the phasor kernels is known to perform sampling on an intermediate grid $\mathcal{T}_i$, defined by a set of points $\boldsymbol{x}\in \mathbb{R}$. The impact on the sampled phasor field by the kernel $j\in \mathcal{K}$ is determined by the phase-shifted complex wave of the kernel at the sample point, weighted by a Gaussian to ensure locality of the kernel's response on the domain. Formally, the contribution from kernel $j\in \mathcal{K}$ on sampling point $\boldsymbol{x}\in \mathbb{T}_i$ is given

\begin{equation}
\mathcal{G}_j(\boldsymbol{x})=\mathrm{e}^{-\beta_j\Delta_j(\boldsymbol{x})^2+2i\pi\omega\boldsymbol{d}_j\cdot (\boldsymbol{x}-\boldsymbol{x}_j^0)+i\varphi_j},\; \forall \boldsymbol{x}\in \mathcal{T}_i^j
\end{equation}

$\Delta_j(\boldsymbol{x})$ denotes a distance measure between the sampling point and the centre kernel, and $\mathcal{T}_i^j\subseteq \mathcal{T}_i$ is the domain of influence by kernel $j$. These measures determine, together with the bandwidth $\beta_j$, the magnitude of the signal registered at the sampling point. The choice of these two quantities have significant implications for both the smoothness of the phasor field, structural performance of the final dehomogenised design as well as the computational cost.

Firstly, in line with the anisotropy introduced for the phase alignment for the kernels, the radius of impact of a kernel in sampling space is now determined by an orientated anisotropic Gaussian kernel, replacing the isotropic version in previous works on phasors. The anisotropy follows the direction of the kernel neighbourhood to ensure that a kernel has limited influence on sampling points located in regions where its phase is not properly aligned. As such the chosen distance measure is $\Delta_j(\boldsymbol{x})=\left\|(\boldsymbol{x}-\boldsymbol{x_j^0})\right\|_{\theta_j}^{(r_1,\;r_2)}$, from \autoref{eq:align_neighbourhood}, for some $r_1,\;r_2\in \mathbb{R}^{+}$ where $r_1<r_2$ ensures the correct direction of anisotropy. This adaption is crucial for realising the reduction in local lamination curvature from the anisotropic phase alignment. Further details on how to choose degree of anisotropy is included in \autoref{app:parameter_select}.

The domain of influence is of great importance to maintain the benefits from the phase-alignment procedure and ensuring computational efficiency. The choice of domain of influence is influenced by the motivation to limit the unwanted impact on non-aligned regions by imposing a cutoff for the Gaussian weight of a kernels impact. This cutoff is imposed such that for a kernel $j\in \mathcal{K}$ it holds true that
\begin{equation}
    \mathrm{e}^{-\beta_j\Delta_j(x)}>10^{-6} \; \Leftrightarrow \; \Delta_j(x)<-\dfrac{\ln{(10^{-6})}}{\beta_j} ,\; \forall \mathcal{T}_i^j
\end{equation}
This truncation of the radius of impact further aids in reducing the computational time by limiting the number of sampling points for which a kernel-impact is to be computed.

Even with the smoothing effect of the phase-alignment there may be defects in the phasor field caused by abrupt variation in the phase field of the kernels. Therefore, as presented by \citealt{Tricard_orientable_2020}, a regularising filter is embedded in the evaluation of the phasor field response to reduce such locally distorting effects. By analytical integration the filtering operation is integrated directly in the kernels, reducing the filtering operation to a product of kernels to be summed when performing intermediate-grid sampling.

Let $\boldsymbol{\hat{d}_{\boldsymbol{x}}}$ denote the local lamination orientation at sampling point $\boldsymbol{x}\in \mathcal{T}_i$, such that
\begin{equation}
    \boldsymbol{\hat{d}_{\boldsymbol{x}}}{=}\left(\hat{d}^c_{\boldsymbol{x}},\hat{d}^s_{\boldsymbol{x}}\right)^T{=}\left(\cos(\hat{\theta}_{\boldsymbol{x}}),-\sin(\hat{\theta}_{\boldsymbol{x}})\right)^T
\end{equation}
where $\hat{\theta}_{\boldsymbol{x}}$ is the angle of the local orientation. $\hat{\theta}_{\boldsymbol{x}}$ is upsampled from the lamination directions on the coarse grid.
$\cos(\hat{\theta}_{\boldsymbol{x}})$ and $\sin(\hat{\theta}_{\boldsymbol{x}})$ are obtained by linear interpolation of $\cos(\theta_j)$ and $\sin(\theta_j)$ separately, from the coarse kernel mesh to the intermediate sampling grid. The upsampling is performed by firstly utilising linear interpolation of $\cos(\theta_j)$ and $\sin(\theta_j)$ to the intermediate grid to obtain candidate directions defined by ${\cos}(\tilde{\hat{\theta}})$ and ${\sin}(\tilde{\hat{\theta}})$. Where the field is subject to discontinued transitions due to the $2\pi$-periodic nature of the orientation field, the upsampled fields do not adhere to ${\cos}^2(\tilde{\hat{\theta}})+{\sin}^2(\tilde{\hat{\theta}})=1$ as expected for the trigonometric functions. Thus, a field $\textit{correct}_{\theta}\in[0,\;1]$ determining to what degree the interpolated values are inconsistent, with 0 meaning they are consistent, can be defined
\begin{equation}
    \textit{correct}_{\theta}=1-2\max\left\{{\cos}^2(\tilde{\hat{\theta}})+{\sin}^2(\tilde{\hat{\theta}})-0.5,\;0\right\}
\end{equation}
A second set of candidate directions is then obtained by nearest neighbour upsampling to achieve ${\cos}(\dot{\hat{\theta}})$ and ${\sin}(\dot{\hat{\theta}})$. The upsampled orientations can then be achieved by 
\begin{equation}
    \hat{\theta}_x=\text{atan2}\left({\sin}(\tilde{\hat{\theta}}),\;{\cos}(\tilde{\hat{\theta}})\right)(1-\textit{correct}_{\theta})+\text{atan2}\left({\sin}(\dot{\hat{\theta}}),\;{\cos}(\dot{\hat{\theta}})\right)\textit{correct}_{\theta}
\end{equation}
This adjustment ensures that the sharp transitions in the orientations on the coarse scale due to $2\pi$-periodic jumps are maintained on the upsampled scale.

Define $\Lambda_j(\boldsymbol{x})=\omega(\boldsymbol{d}_j-\boldsymbol{\hat{d}}_{\boldsymbol{x}})$ to be a weight of similarity between the orientation of the kernel $j\in \mathcal{K}$ and the local orientation at the sampling point $\boldsymbol{x}\in \mathcal{T}_i$. The filter-kernel integrated in the phasor field evaluation is then given as
\begin{equation}
\mathcal{K}_j(\boldsymbol{x})=\mathrm{e}^{\left(\dfrac{\beta_j^2\Delta_j(\boldsymbol{x})^2-\pi^2\|\Lambda_j(\boldsymbol{x})\|_2^2+2i\alpha_j\Lambda_j(\boldsymbol{x})\cdot(\boldsymbol{x}-\boldsymbol{x}_j^0)}{\alpha_j+\beta_j}\right)}
\end{equation}
where $\alpha_j\in \mathbb{R}^+$ defines the bandwidth of the filter kernel applied. The filtered response of kernel $j\in \mathcal{K}$ at sampling point $\boldsymbol{x}\in \mathcal{T}_i$ is as such given by the product $\mathcal{G}_j(\boldsymbol{x})\mathcal{K}_j(\boldsymbol{x})$. An example of the effect of the filter was illustrated for opposing angles in \autoref{fig:opposing_angle_alignment_exemplified_with_filter}. For details on how the filter integration is derived, the reader is referred to the supplemental material of \citealt{Tricard_orientable_2020}. The key observation is that when $\Lambda_j\rightarrow0$, kernel and sampling point have the same orientation and the filter simply results in a decreased bandwidth increasing the magnitude of the signal from the kernel measured at the sampling point.

\begin{algorithm}[h]\onehalfspacing
\caption{Sample aligned phase field on intermediate mesh}\label{alg:sample_intermediate_grid}
\begin{algorithmic}
\Require ${\mathcal{G}(\boldsymbol{x})}=0,\; \forall \boldsymbol{x}\in \mathcal{T}_i$
\For{$j \in \mathcal{K}$}
\State $\Delta_j(\boldsymbol{x})=\left\|(\boldsymbol{x}-\boldsymbol{x_j^0})\right\|_{\theta}^r$ \Comment{Anisotropic distances from kernel centre to sampling point locations}
\State $\mathcal{T}_i^j=\left\{\boldsymbol{x}\in\mathcal{T}_i\;:\;\mathrm{e}^{-\pi\beta_j\Delta_j(\boldsymbol{x})}>10^{-6}\right\}$ \Comment{Reduce sample space by anisotropic Gaussiam cut-off}
\State $\Lambda_j(\boldsymbol{x})=\omega(\boldsymbol{d}_j-\boldsymbol{\hat{d}}_{\boldsymbol{x}})$ \Comment{Similarity weight for kernel and sampling point directions}
\State $\mathcal{G}(\boldsymbol{x})+=\mathrm{e}^{-\frac{1}{\alpha_j+\beta_j}\left(\beta_j^2\Delta_j(\boldsymbol{x})^2+\pi^2\|\Lambda_j(\boldsymbol{x})\|_2^2-2i\alpha_j\Lambda_j(\boldsymbol{x})\cdot(\boldsymbol{x}-\boldsymbol{x}_j^0)\right)}\mathrm{e}^{2i\pi\omega\boldsymbol{d}_j\cdot(\boldsymbol{x}-\boldsymbol{x}_j^0)+i\varphi_j},\; \forall \boldsymbol{x}\in \mathcal{T}_i^j$ \Comment{Add kernel contribution with integrated filter}
\EndFor
\end{algorithmic}
\end{algorithm}

Combining these components the procedure of sampling the phasor field across the domain $\mathcal{T}_i$ from a set $\mathcal{K}$ of phasor kernels is described in \autoref{alg:sample_intermediate_grid}. When utilising phasors for dehomogenisation, sampling to an intermediate grid is only needed once per lamination direction. After sampling, $\mathcal{G}(\boldsymbol{x})$ has the form of a complex wave function on the domain $\boldsymbol{x}\in \mathcal{T}_i$. The phasor field is then obtained by the instantaneous phase of this field, $\phi(\boldsymbol{x})=\Arg(\mathcal{G}(\boldsymbol{x}))$. As such, the phasor field represents a periodic angular field, $\phi(\boldsymbol{x})\in [-\pi,\;\pi]$, with the prescribed periodicity and smooth contourlines along the lamination directions. The phasor sine-field is then $\hat{\phi}(\boldsymbol{x})=\sin(\phi(\boldsymbol{x}))$, and a triangularly varying density field can be obtained $\rho=\frac{1}{\pi}(\arcsin{\hat{\phi}}+1)$. This triangular translation allows for extraction of a body-fitted mesh from the level-set or by direct thresholding of $\rho\in[0,1]$ by local relative thicknesses to obtain the final solid-void design.

\subsection{Disconnections}\label{subseq:branches}
Branching occurs where there are point-singularities in the phasor field caused by the complex Gabor field vanishing due to destructive interference (\citealt{Tricard_orientable_2020}). Such branching artefacts are well known for periodic stripe patterns (\citealt{Lichtenbergetal2018,Knoppeletal2015,MaWalzeretal2020}). The branching is necessary for the resulting field to best adhere to the desired periodicity and the varying orientations simultaneously. A branch effectively adds or removes a stripe segment to best maintain congruent separation of the striped pattern. Branching is therefore expected for non-trivial vector fields and these point-singularities represent where the underlying vector field cannot be accurately represented by the phasor field.

The nature of these branches translates directly to the triangular density-field. Depending on the location of the singularity in relation to the phase of the sine-field the possible nature of the branch in the density field can be classified into three categories; a fully connected solid branch, a partially connected branch across intermediate densities, or a fully void branch leading to a complete disconnection. The sine-field for these three types are illustrated in \autoref{fig:branch_identify_degree}. When thresholded by the local relative lamination thickness a branch is only guaranteed to ensure structural connection when it is of the fully solid type, or the relative thickness is fully solid. Ensuring structural connectivity is crucial for mechanical performance, and thus a procedure for ensuring all branches are connected is necessary.

Methods for reducing or filtering away the branching point-singularities have been proposed, but these cannot be included in a procedural manner and do not allow for controlling the oscillation profile. \citealt{Tricard_phasornoise_2019} discussed these challenges in more detail and concluded that it is unlikely to achieve a procedural approach to tackling these disconnections. \citealt{MaWalzeretal2020} utilised periodic stripe patterns to design metal frame structures and tackled the branching points by encouraging orthogonal bars to connect the branching-points in the structure. However, due to the phasors not providing direct control of the branching point locations, this approach could require both a more significant violation of the prescribed periodicity as well as an increased area of violation around a branching point, and is thus not beneficial for phasor-based dehomogenisation. Instead, a connection procedure based on the nature of the phasor field and image morphology techniques is proposed. An outline of the procedure is given below, where each step will be explained in more detail in \autoref{subsubseq:identify_branch}-\SecNum{subsubsec:pinch_branch}.

\begin{figure}[!htb]
\noindent\fbox{%
    \parbox{\textwidth}{%
\begin{minipage}{0.54\textwidth}
\hspace{3mm}\textbf{Outline of branch-connection procedure}\\
\begin{flushleft}
\begin{myenumerate}
    \item Locate branching points in intermediate mesh
    \item For each branching point on the sine-field:
    \begin{myenumerate}
        \item Determine degree of disconnection 
        \item Determine direction in which to connect
        \item Determine centre of branch
        \item Perform local phase shift about centre to close branch
    \end{myenumerate}
    \item For each branching point on the triangular field:
    \begin{myenumerate}
        \item Incrementally pinch material towards branch centre 
    \end{myenumerate}
\end{myenumerate}
\end{flushleft}
\end{minipage}
\hfill
\begin{minipage}{0.445\textwidth}
    \centering
    \includegraphics[width=0.75\linewidth]{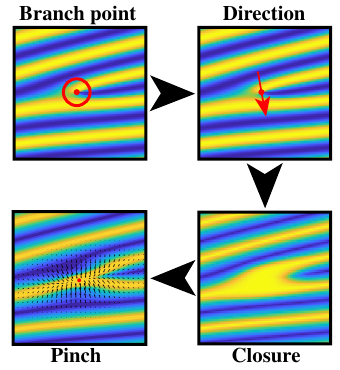}
\end{minipage}
}}
\end{figure}

\subsubsection{Identifying branching points}\label{subsubseq:identify_branch}
To connect branches it is first necessary to identify where in the density-field they are located. As the branching is known to occur at singular points in the phasor field, they can be directly obtained from the null-points in the underlying complex Gabor field $\mathcal{G}(\boldsymbol{x})$. The coordinates of the branching points can be determined by the local minima in regions where the complex wave-field vanishes. \autoref{fig:branch_identify_degree} illustrates examples of the three types of branches and how the local minima of $\mathcal{G}(\boldsymbol{x})$ are located at the centre of these branches. Let $\Gamma\subset \mathcal{T}_i$ denote the set of identified branching points in the intermediate density region, meaning that solid and void singularities are ignored. 

\subsubsection{Determining degree and direction of disconnection}\label{subsubseq:degree_disconnect}
As previously mentioned the nature of the branch may vary from a fully solid branch to a complete disconnection. Any attempt to ensure connection of branches will cause disturbances to the direction and periodicity of the phasor field, and thus it is beneficial to limit the extent of the manipulations performed. Therefore, to limit the distortion of the phasor field, a measure representing the degree of disconnection for a specific branch is introduced. This measure is to be used for controlling the remaining actions in the branch-connection procedure.

\begin{figure}[!htb]
    \centering
\includegraphics[width=0.95\linewidth]{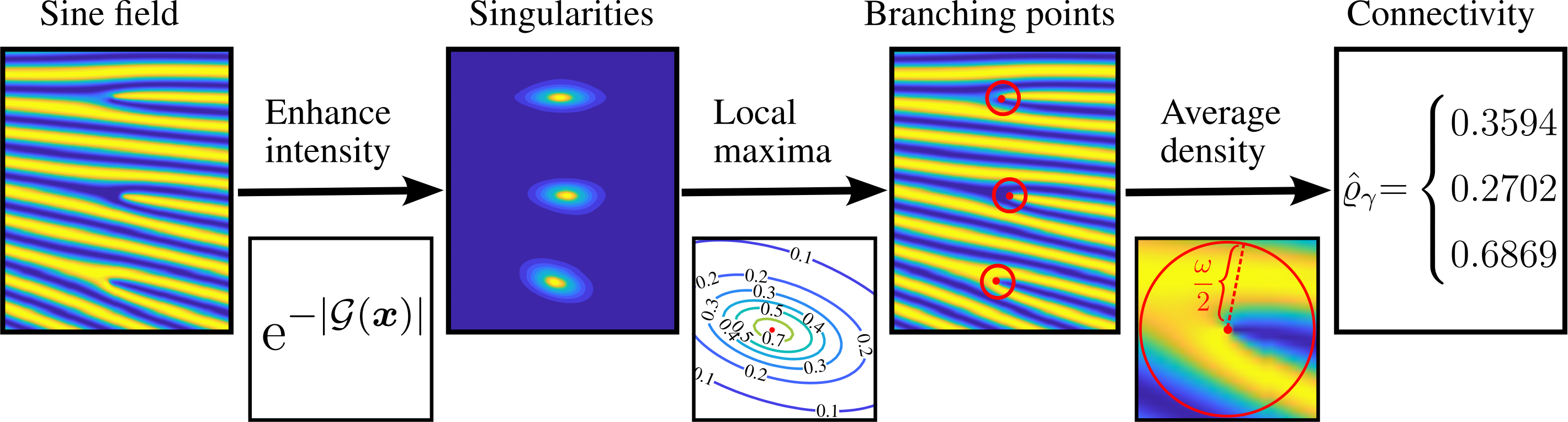}
\caption{Illustrating the pipeline for identifying branching points and determining degree of connection. Note that the degree of connectivity is measured as if the yellow is solid and blue is void, conversely, this measure would correspond to the degree of disconnection if blue was solid and yellow void.}
    \label{fig:branch_identify_degree}
\end{figure}

The degree of disconnection is directly related to the density of material around the disconnection. As the field is known to be approximately constant-periodic, the mean value of the density within a circle with diameter equal to one period is sufficient to extract a numerical value indicating this degree of disconnection. The mean value within this region is guaranteed to lie in the interval $[-1,1]$, with most values in the range $[-0.5,0.5]$. Let $\varrho_\gamma$ denote the average triangular-field value within the described circle of branching point $\gamma\in \Gamma$, the degree of connection is given by $\hat{\varrho}_\gamma=({\varrho_{\gamma}+1})/{2}$, and the degree of disconnection is conversely given by $1-\hat{\varrho}_\gamma$. This transformation ensures that the degree of connection lies within the interval $[0,1]$, and mostly within $[0.25,0.75]$, such that it can be used as a weight to control branch-connection actions according to the original branch separation. \autoref{fig:branch_identify_degree} illustrates how the degree of connection is computed for each identified branching point and provides the numerical value of the degree of connection for examples of each of the three main branching types.

The second important control-feature of a disconnection is the direction in which the branch should be connected. As the original singular point is located further away from the natural branch-centre for lower degrees of connection, the direction is needed to choose an appropriate branch centre. It is known that the branch should be closed orthogonally to the lamination direction and that a natural branch-centre is located near the singular point in the intermediate mesh. Further, most disconnections take some form of a partial disconnection, meaning that there are intermediate densities in the direction along which the branch should be closed. Based on this, the local density at points 1/3 period to each side of the singular point are compared to choose the direction in which the density is larger because this indicates a trend in connection favouring this direction.

Having determined the connection direction two new coordinates are determined for each branch. These points, $\boldsymbol{\gamma}^c$ and $\boldsymbol{\gamma}^o$, are based on moving in the determined connection direction with step-sizes determined by the degree of disconnection. The first, $\boldsymbol{\gamma}^c$ estimates the centre of the disconnection representing where the branch should be closed to ensure connection. The second, $\boldsymbol{\gamma}^o$, estimates the nearest point along a solid line, and represents towards where the branching point should be pushed to reduce the disturbance in lamination directions of the original disconnected field. 

\begin{algorithm}[h]\doublespacing
\caption{Identify $\boldsymbol{\gamma}^c$ and $\boldsymbol{\gamma}^o$ for branching point $\gamma \in \Gamma$.}\label{alg:branch_point_coordinates}
\begin{algorithmic}
\Require $\hat{\varrho}_{\gamma}$ \Comment{Degree of disconnection for current branching point}
\Require $\hat{\theta}_{\boldsymbol{\gamma}}$ \Comment{Local lamination angle at singular point}
\State $\boldsymbol{\gamma}^{+}=\boldsymbol{\gamma}+\begin{pmatrix}-\frac{\omega}{3}\sin(\hat{\theta}_{\boldsymbol{\gamma}})\\\frac{\omega}{3}\cos(\hat{\theta}_{\boldsymbol{\gamma}})\end{pmatrix}$ and $\boldsymbol{\gamma}^{-}=\boldsymbol{\gamma}-\begin{pmatrix}-\frac{\omega}{3}\sin(\hat{\theta}_{\boldsymbol{\gamma}})\\\frac{\omega}{3}\cos(\hat{\theta}_{\boldsymbol{\gamma}})\end{pmatrix}$ \Comment{Determine candidate closing-directions}
\State $\oplus=\text{sgn}(\sin(\phi(\boldsymbol{\gamma}^{+})){-}\sin(\phi(\boldsymbol{\gamma}^{-})))$ \Comment{Determine closure direction orthogonal to the lamination direction}
\State $\boldsymbol{\gamma}^{c}=\boldsymbol{\gamma}\oplus(1-\hat{\varrho}_\gamma)\begin{pmatrix}-\frac{\omega}{3}\sin(\hat{\theta}_{\boldsymbol{\gamma}})\\\frac{\omega}{3}\cos(\hat{\theta}_{\boldsymbol{\gamma}})\end{pmatrix}$ \Comment{Shift the disconnection centre}
\State $\boldsymbol{\gamma}^{o}=\boldsymbol{\gamma}\oplus(1-\hat{\varrho}_\gamma)\begin{pmatrix}-{\omega}\sin(\hat{\theta}_{\boldsymbol{\gamma}})\\{\omega}\cos(\hat{\theta}_{\boldsymbol{\gamma}})\end{pmatrix}$ \Comment{Determine branch orientation correction point}
\end{algorithmic}
\end{algorithm}
\autoref{alg:branch_point_coordinates} outlines the procedure for identifying the crucial coordinates for a given branching point $\gamma\in\Gamma$. Given the local lamination direction at the branching point given the angle $\hat{\theta}_\gamma$, $\gamma^+$ and $\gamma^-$ are computed as candidate move directions orthogonal to the local lamination direction. The move direction $\oplus{\in}\{-1,\;1\}$, representing whether the branch should be closed along the vector obtained by a $\pi/2$ clockwise ($\oplus{=}1$) or counter-clockwise ($\oplus{=}-1$) rotation of the local lamination direction-vector, is determined such that if $\sin(\phi(\boldsymbol{\gamma}^{+}))>\sin(\phi(\boldsymbol{\gamma}^{-}))$ the branch is to be closed in the direction of $\boldsymbol{\gamma}^{+}$ and in $\boldsymbol{\gamma}^{-}$ otherwise. This is because a higher sine-field value is directly related to a higher density when transformed to a triangular field and because the direction in which higher densities are found closer to the original branching point is more appropriate for connection. Note that for near-solid branches one may have $\boldsymbol{\gamma}^+\approx \boldsymbol{\gamma}^-$, but in these cases the choice of direction is less crucial as the degree of connection will be large and the magnitude of the branch closing measures will be significantly reduced by this weight. This effect is relevant already when computing the new set of branching point coordinates, $\boldsymbol{\gamma}^c$ and $\boldsymbol{\gamma}^o$, where the magnitude of the orthogonal step taken is weighted by the degree of disconnection.

\subsubsection{Perform local phase-shift}\label{subsubseq:branch_phase_shift}
The branching points occur where the complex wave function $\mathcal{G}(\boldsymbol{x})$ approaches zero. Shifting the phase of the entire phasor field allows for altering the profile of the phasor sine-field and thus the nature of a branch, but does not change the location of these branching points. This means that if a phase shift of $\pi$ is applied such that the sine-field considered is given by $\sin(\phi(\boldsymbol{x})+\pi)$, a redistribution of material is achieved where a previously fully disconnected branch becomes fully connected. This effect can be utilised to connect branches by applying a local phase shift in the region surrounding a branching point. An appropriate and smooth local phase shift allows for moving the material distribution around the branching point without significant disturbances to the periodicity and direction of the remainder of the field.

\begin{figure}[!htb]
    \centering
    \includegraphics[width=\linewidth]{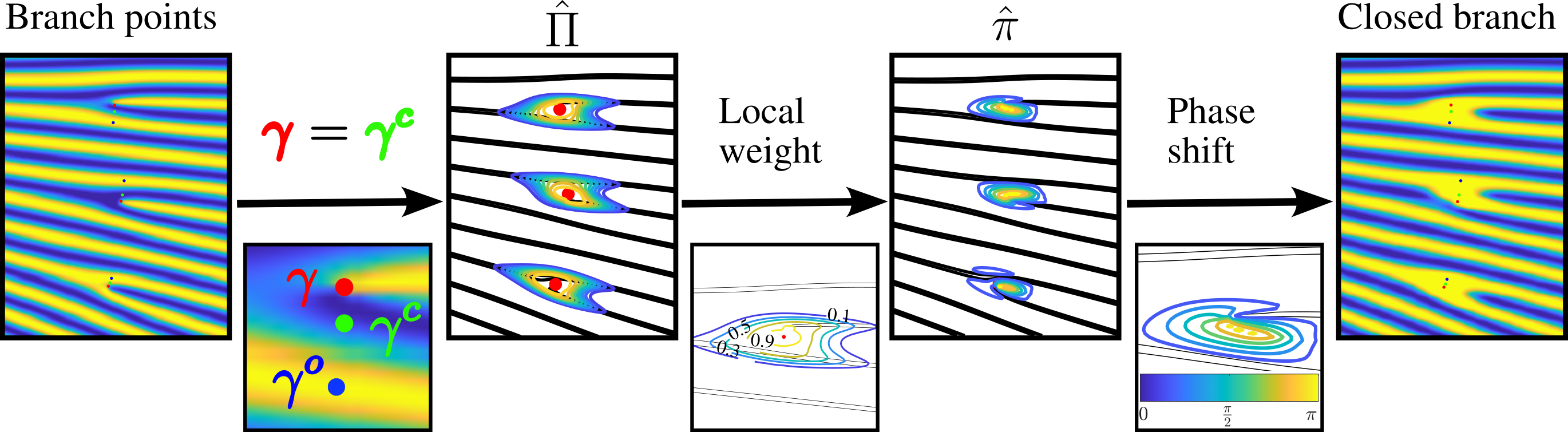}
    \caption{Pipeline illustrating the location of closing direction, new branching point coordinates $\boldsymbol{\gamma}^c$ and $\boldsymbol{\gamma}^o$, how the phase shift $\hat{\pi}$ is constructed and the nature of the closed branches on the sine field $\max\{\sin \phi^+,\;\sin \phi^-\}$.}
    \label{fig:new_coordinates_to_phase_shift}
\end{figure}

To best achieve such appropriate phase shifts several measures are needed to both ensure branch connection and to limit disturbances away from the disconnected region. To this end, a modified anisotropic Gaussian $\Pi(\boldsymbol{x})$ with $\boldsymbol{\gamma}^c$ is defined for each branching point $\boldsymbol{\gamma}\in\Gamma$.
\begin{equation}
    \Pi(\boldsymbol{x})=\mathrm{exp}\left(-{2\omega^2}\|\boldsymbol{x}-\boldsymbol{\gamma}^c\|^{(r_1,\;r_2)}_{\hat{\theta}(\boldsymbol{x})}
        {\left(1-0.5\sin\phi(\boldsymbol{x})\right)} \right),\;\|\boldsymbol{\cdot}\|_{\theta}^{(r_1,\;r_2)}=\left\|\begin{pmatrix}-r_1\sin\theta & -{r_1}\cos\theta\\ {r_2}\cos\theta & -{r_2}\sin\theta\end{pmatrix}(\boldsymbol{\cdot})\right\|_2
\end{equation}
The anisotropic distance measure $\|\boldsymbol{\cdot}\|_{\theta}^{(r_1,\;r_2)}$ is designed to follow the lamination direction as it varies with $\boldsymbol{x}$ when $r_1{<}{<}r_2$. The default radius of the Gaussian is half a period controlled by the weight $2\omega^2$ such that if $r_1=r_2=1$ the Gaussian forms a circle with main impact radius equal to one period. The modification of weighting by $1-0.5\sin\phi(\boldsymbol{x})$ is to decrease the impact on high density regions away from the kernel centre while increasing the impact on void regions. The motivation lies in wanting to allow for increasing densities in void regions about the branch but not cause disturbances of solid lines neighbouring the branching point. The presented weight is further modified by a smoothstep function (\autoref{eq:shift_mag_smoothstep}) to increase the separation between the weight and the remainder of the field as well as increase the the shift magnitude near the diconnection centre.
\begin{equation}\label{eq:shift_mag_smoothstep}
    \hat{\Pi}(\boldsymbol{x})=\text{smoothstep}\left(\Pi(\boldsymbol{x})\right)=3\Pi(\boldsymbol{x})^2-2\Pi(\boldsymbol{x})^3
\end{equation}
Based on this weight and the knowledge that a phase shift of $\pi$ transforms a void element to a solid element the local phase shift enclosing the considered branching point is given
\begin{equation}
    \hat{\pi}(\boldsymbol{x})=\hat{\Pi}(\boldsymbol{x})\pi\left(1-\dfrac{\frac{2}{\pi}\arcsin\phi(\boldsymbol{x})+1}{2}\right)
\end{equation}
Scaling the magnitude by the degree of void in the original disconnected field adjusts the phase shift as to encourage a transformation towards solid for any density element. To utilise this phase shift to connect a branch, the union of applying the positive and negative phase shifts forms a new intermediate density field $\hat{\rho}$.
\begin{align}
     &\phi^+(\boldsymbol{x})=\phi(\boldsymbol{x})+\hat{\pi}(\boldsymbol{x}) \text{ and } \phi^-(\boldsymbol{x})=\phi(\boldsymbol{x})-\hat{\pi}(\boldsymbol{x})\\
    &\hat{\rho}=\frac{1}{\pi}\arcsin(\max\{\sin \phi^+,\;\sin \phi^-\})+0.5
\end{align}
This new density field will have solidified branch connections, and the benefit of adding material in the described manner is that the shift around the branching point follows the shape of a branch. However, the now connected solidified branches do not follow periodicity and will be unnecessarily thick after thresholding. The steps of performing this local phase shift and the resulting solidified branches for the three main disconnection types are illustrated in \autoref{fig:new_coordinates_to_phase_shift}. The surplus material surrounding the branching points should be reduced to achieve better performance and consistency of the final de-homogenised solution. This motivates the pinch procedure described in \autoref{sec:pinch}.

\subsubsection{The pinch operator}\label{sec:pinch}
The pinch procedure is designed to reduce the material around a branching point by locally ``compressing'' density values towards the branch centre. The purpose is to maintain the connectivity achieved by $\hat{\rho}$ at the centre of a branch, but reduce the overall material added by extending nearby void-regions towards this centre. The procedure for performing this material reduction constitutes an iterative application of what is termed the pinch operator. This operator is defined as a set of pixel-wise move directions and magnitudes locally near a branching point.

\begin{figure}[!htb]
    \centering
    \def\svgwidth{\textwidth}
    \includegraphics[width=\linewidth]{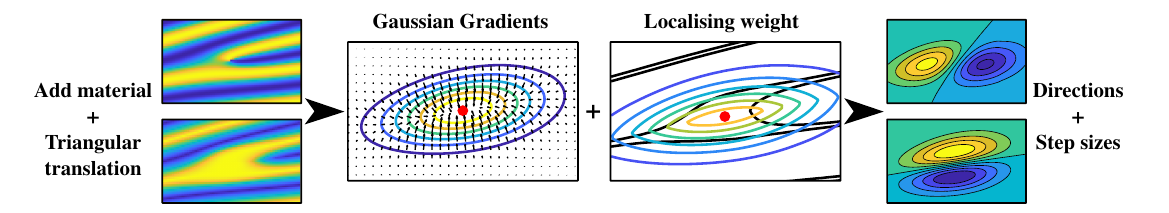}
    \caption{The outline of how the pinch-operator about a branching point is defined.}
    \label{fig:pinch_procedure}
\end{figure}

\autoref{fig:pinch_procedure} illustrates how a pinch operator is defined. Given a chosen pinch-centre point on the triangular translation of a solidified branch an anisotropic Gaussian is defined. The gradients of this Gaussian are directed towards the defined pinch centre, which is the direction in which material is to be squeezed to reduce material and maintain branch closure, and the gradients are of increasing relative magnitude towards this centre. The pinch operator requires element-wise information about direction and distance to the point from which the element should adopt its new pinched density-value. The negative normalised Gaussian gradients provide relative magnitudes and directions for this operation as adopting the density from a point away from the branch centre results in a morphological pinch effect towards the branch centre. A localising weight is utilised to keep the pinch effect local near the branch. Additionally, the normalisation of the gradients allows for prescribing a pinch-magnitude corresponding to the maximal distance from any point to that of its new density value by multiplying by a scalar.

\begin{figure}[!htb]
\includegraphics[width=\linewidth]{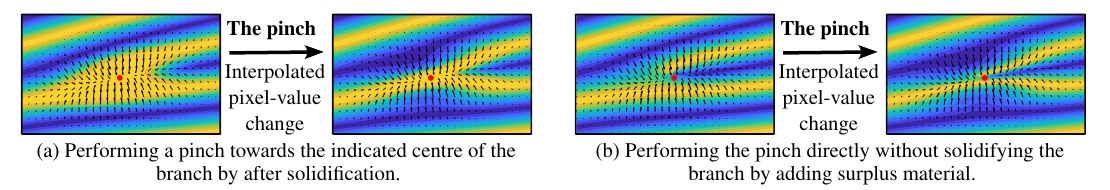}
\caption{Illustrating how the continuous weighted interpolation-based pinch operator is applied in a discretised manner to close the branch as described in \autoref{fig:pinch_procedure} both with (a) and without (b) added material. }
\label{fig:performing_pinch_example_wwo_material}
\end{figure}

The scaled and weighted gradients are described by a continuous operator, but as the pinch is to be performed on a sampled fixed grid discretisation and interpolation are required to obtain a smooth pixel-wise density change. The continuity of the operator is thus maintained for each element by linear interpolation between the discretised pixel-values surrounding the point from which an element is to adopt its new density-value. \autoref{fig:performing_pinch_example_wwo_material}(a) illustrates how the pinch operator defined in \autoref{fig:pinch_procedure} changes the material distribution about the branching point to reduce the material while ensuring that branch-shape and connectivity is maintained. If one were to not ensure branch solidification before pinching one would encounter the issue of the branch centre taking on intermediate densities. This would mean that connectivity is not guaranteed when thresholding to a solid-void structure, depending on the local relative thickness of the lamination layer at this branch. This effect is illustrated in \autoref{fig:performing_pinch_example_wwo_material}(b).

\subsubsection{Incremental pinch to close branch}\label{subsubsec:pinch_branch}
There are many options for how to apply the pinch operator to close a given branch. The impact radius and degree of anisotropy of the Gaussian determines the relative difference in magnitude of the pinch as a function of distance to the pinch centre as well as the difference in magnitude of the pinch parallel and orthogonal to the lamination direction. The Gaussian thus controls the underlying shape of the pinch. The localising weight has the power to modify this shape slightly, and is especially important in restricting the pinch to only affect the intended branch. Also the scale applied to the locally weighted gradients determines the step-size in terms of the distance allowed between an element and the point from which it adopts its new density. Generally it is found that better coherence with the periodic wave-field is ensured if pinches of smaller magnitude are applied in sequence, as larger magnitudes may remove the smooth intermediate density transition between solid and void. This motivates the use of incrementally applying pinch operators of smaller magnitudes to close the branch.

It can also be observed that pinches of larger magnitudes lead to pressing the material towards a thin middle-line in the direction of anisotropy of the Gaussian and the localised weight. The intermediate density transition between fully solid and void contourlines is also reduced compared to the periodic behaviour of the remainder of the field. This means that the magnitude to a great extent determines the relative thickness of the branch when the periodic field is later thresholded in a global manner. Thus, to avoid thresholded branches from being too thin compared to the target relative thickness, but also to ensure that low density regions do not result in too thick branches, it is beneficial to also control the magnitude of the pinch by the local relative thickness.

These considerations have lead to the general guidelines of ensuring that the Gaussian is strongly anisotropic along the lamination direction, the localising weight ensuring a cutoff near the void contourlines on each side of the branch and the magnitude of the pinch is kept within $\omega/2$ and is negatively proportional to the local relative thickness. To get smoother and more coherent results from larger magnitude pinches it is recommended to incrementally perform smaller magnitude pinches to build to an overall larger magnitude pinch. Based on this, it is chosen to incrementally apply the pinch-procedure in a step-wise procedure with moving centre between $\boldsymbol{\gamma}^c$ and $\boldsymbol{\gamma}^o$ and reduction in the area of the localising weight. This strategy means that the first steps are detrimental to the branch shape, whereas the later pinch steps reduce material only near the branch connection point.

\begin{algorithm}[H]
\caption{Pinch branch about disconnection $\boldsymbol{\gamma}\in \Gamma$}\label{alg:branch_pinch_procedure}
\begin{algorithmic}
\Require $\mathcal{I}=\{(i,j)\in \Z^+\}$: pixel index set encapsulating area about $\boldsymbol{\gamma}\in \Gamma$ to be pinched
\For{$k=1\text{ to } k_{max}$}
\State $\Delta_{k}=(1-\hat{\varrho}_\gamma)\dfrac{k-1}{k_{max}-1}$ \Comment{Step-size to move pinch centre-point}
\State $\boldsymbol{\gamma}_k=\Delta_{k}\boldsymbol{\gamma}^{o}+(1-\Delta_{k})\boldsymbol{\gamma}^{c}$ \Comment{Current pinch centre}
\State \scalebox{1.}{$\Pi_{pinch}(\boldsymbol{x})=\mathrm{exp}\left({-\omega^2\|\boldsymbol{x}-\boldsymbol{\gamma}_k\|_{\hat{\theta}(\boldsymbol{\gamma}_k)}^{(r_1,\;r_2)}}\right)$} \Comment{Anisotropic Gaussian to define pinch directions}
\State $\boldsymbol{\bar{\gamma}}_k=\Delta_{k}^2\boldsymbol{\gamma}^{o}+(1-\Delta_{k}^2)\boldsymbol{\gamma}^{c}$ \Comment{Shifted centre of weight localiser}
\State $\bar{r}_1^k=\left(1-\dfrac{k-1}{k_{max}-1}\right)r_1$ \Comment{Anisotropy adjustment for reducing localising weight}
\State \scalebox{1.}{$\Pi_{local}(\boldsymbol{x})=\mathrm{exp}\left({-2\omega^2\|\boldsymbol{x}-\bar{\boldsymbol{\gamma}_k}\|_{\hat{\theta}(\boldsymbol{x})}^{(\bar{r}_1^k,\;r_2)}-\dfrac{2\omega}{(2-\mu_{\boldsymbol{x}})}|(\cos{\hat{\theta}_{\boldsymbol{x}}},\;-\sin{\hat{\theta}_{\boldsymbol{x}}})\cdot(\boldsymbol{x}-\boldsymbol{\gamma}_k)|}\right)$} \Comment{Pinch magnitude localising weights}
\State $\begin{pmatrix}p_x\\p_y\end{pmatrix}=\begin{pmatrix}\nabla_x \Pi_{pinch} \\ \nabla_y \Pi_{pinch}\end{pmatrix}\dfrac{1}{\max(|\nabla \Pi_{pinch}|)}\Pi_{local}(1-\mu_{\boldsymbol{x}})\dfrac{\hat{\omega}}{k_{max}}$ \Comment{Continuous pixel-based step sizes and directions}
\State $\hat{\rho}(i,j)=\begin{pmatrix}\lceil p_y\rceil{-}p_y\\p_y{-}\lfloor p_y\rfloor\end{pmatrix}^T
\begin{pmatrix}\hat{\rho}(i{+}\lfloor p_x\rfloor,\;j{+}\lfloor p_x\rfloor) &\hat{\rho}(i{+}\lceil p_x\rceil,\;j{+}\lfloor p_y\rfloor)\\\hat{\rho}(i{+}\lfloor p_x\rfloor,\;j{+}\lceil p_y\rceil) &\hat{\rho}(i{+}\lceil p_x\rceil,\;j{+}\lceil p_y\rceil) \end{pmatrix}
\begin{pmatrix}\lceil p_x\rceil{-}p_x\\p_x{-}\lfloor p_x\rfloor\end{pmatrix},\; \forall (i,j)\in \mathcal{I}$ \Comment{Linear interpolation}
\EndFor
\end{algorithmic}
\end{algorithm}

\autoref{alg:branch_pinch_procedure} outlines the procedure defining the pinch steps taken to close a branch. First, a step-length $\Delta_{k}$ is defined to determine where the pinch centre $\boldsymbol{\gamma}_k$ of the current iteration is to be located on the line between $\boldsymbol{\gamma}^c$ and $\boldsymbol{\gamma}^o$. The step size is set to be zero in the first iteration as to ensure sufficient closure near where the material was added. In the following iterations the step size is incrementally increased, proportional to the degree of disconnection. If the degree of disconnection of the branch approaches zero, it has an appropriate centre and shape and outline inherited from the original disconnected field, and thus the centre of this branch should not move during the pinch-procedure. 

The normalised gradients of an anisotropic Gaussian $\Pi_{pinch}$, with centre at pinch centre $\boldsymbol{\gamma}_k$, define the directions and maximal potential magnitudes of the pinch. To localise the pinch, the magnitude of the gradients are scaled according to a modified Gaussian $\Pi_{local}$ centred at a point $\bar{\boldsymbol{\gamma}}_k$ between $\boldsymbol{\gamma}^c$ and the current pinch centre. The adjusted weight in the direction anisotropy $\bar{r}_1^k$ utilised in this weight-function is what reduces the impact of the pinch in each step. The modifications made to $\Pi_{local}$ reduces the pinch magnitudes away from the main part of the branch in the orthogonal direction to the laminations, which in turn reduces the risk of affecting neighbouring waves and causing unwanted curvature when pinching.

The pinch magnitudes are further weighted by how thin the relative thickness is to avoid removing too much material in thicker branches. Lastly, the maximal pixel-change is defined relative to the number of pixels corresponding to one wave-length divided by the maximum number of iterations. This ensures that the overall effect of the pinch procedure is contained to only affect the branch itself, avoiding distortion of neighbouring stripes in the periodic pattern. What this also means, however, is that the maximal number of steps should be chosen such that $k_{max}\geq 2$. The final continuous pinch magnitudes and directions $(p_x,\;p_y)^T$ are then utilised in a linear interpolation step to transfer the pinch effect to the discretised image.

\section{Implementation} \label{sec:Implementation_main}

Combining the different subprocedures introduced in \autoref{sec:phasor_main} an overall procedure overview for the phasor-based dehomogenisation procedure is given in \autoref{alg:procedure_outline}. Selected aspects of the presented methodology can also be combined to create a varying thickness structural boundary, reducing the effects of artefacts adopted from the underlying coarse mesh. The details of this proposed procedure is detailed in \autoref{sec:boundary_method}.  

The specific parameter choices not defined in the previous section are summarised in \autoref{tab:Parameter_choices}. In theory the only parameter specification needed from the user is the desired periodicity of the realised structure, but the parameters presented in \autoref{tab:Parameter_choices} can also be varied to obtain different behaviour depending on the considered problem instance. The definitions presented here are used throughout all tests in \autoref{sec:reuslts_main} proving the stable performance across varying problem characteristics. To best balance computational accuracy and efficiency three different resolutions, in addition to that of the coarse-scale optimised solution, are considered throughout the dehomogenisation procedure: the first intermediate mesh $\mathcal{T}_i^1$, the second intermediate mesh $\mathcal{T}_i^2$ and the final fine resolution mesh $\mathcal{T}_f$. For more elaborate discussions of parameter choices, specific mesh resolutions, what to consider if changing these base case settings, and other implementational details, the reader is referred to \autoref{app:parameter_select}.


\begin{algorithm}[htb!]
\setstretch{1.1}
\caption{Phasor-based dehomogenisation procedure overview}\label{alg:procedure_outline}
\begin{multicols}{2}
\raggedbottom
\begin{algorithmic}[1]
\Require A chosen wavelength specifying the realised periodicity of the finalised structure
\State Translate homogenised optimised solution to a set of phasor kernels (\autoref{sec:phasor_intro_sampling})
\State Construct phasor-based boundary and smoothed structure indicator field on fine scale (\autoref{sec:boundary_method})
\State Use phasor-boundary method to smooth fully solid regions
\State Phase alignment on coarse mesh (\autoref{sec:phase_align})
    \begin{enumerate}[(i)]
        \item Construct anisotropic kernel neighbourhoods
        \item Align according to \autoref{alg:phase_alignment}
    \end{enumerate}
\State Sample on first intermediate mesh (\autoref{sec:sampling_and_filter})
    \begin{enumerate}[(i)]
        \item Determine upscaling factor to first intermediate mesh (\autoref{eq:iup1})
        \item Construct grid of element centres 
        \item Upscale optimised orientation field to intermediate mesh
        \item Perform filtered sampling at each intermediate mesh element
        \item Locate branching points on intermediate grid (\autoref{subsubseq:identify_branch}-\SecNum{subsubseq:degree_disconnect})
    \end{enumerate}
\State Upscale to second intermediate grid
 \begin{enumerate}[(i)]
        \item Determine second upscaling factor (\autoref{eq:iup2})
        \item Construct corresponding intermediate mesh
        \item Upscale the sampled complex wave field to second intermediate mesh using cubic interpolation
        \item Translate branching point locations to second intermediate grid by spatial locations
    \end{enumerate}
\State Connect branches on second intermediate grid (\autoref{subsubseq:degree_disconnect}-\SecNum{subsubsec:pinch_branch})
    \begin{enumerate}[(i)]
        \item Translate upscaled complex wave field to a real sine-wave
        \item Connect branches by local phase-shift (\autoref{subsubseq:branch_phase_shift}) and pinch-procedure (\autoref{sec:pinch})
        \item Translate connected field to triangular wave
    \end{enumerate}
\State Combine final structure
    \begin{enumerate}[(i)]
        \item Upscale triangular wave to final fine resolution by linear interpolation
        \item Threshold each layer and combine union of thresholded fields
        \item Multiply with smoothed indicator field to cut structural boundary
        \item Add the structural boundary and the smoothed solid regions to obtain the final structure (\autoref{fig:Final_struct_different_boundary_corrections}(c))
    \end{enumerate}
\end{algorithmic}
\end{multicols}
\end{algorithm}


\begin{table}[H]
\setstretch{1.25}
    \centering
        \caption{The specific parameter choices considered unless otherwise specified for each sub-procedure. The resolution column indicates at what upscaling level the procedure is applied. To ensure that the minimal features in the structure are resolved by at least three pixels, $h_{min}=3$, when the triangular field is upsampled from intermediate mesh $\mathcal{T}_i^2$ to fine mesh $\mathcal{T}_f$ the upscaling factor between the meshes $f_{up}$ should satisfy $f_{up}\geq \frac{h_{min}}{\hat{\omega}\mu_{min}}$. $F_{up}=f_{up}i_{up}^2$ denotes the full upscaling factor from the coarse to final fine resolution.}
    \label{tab:Parameter_choices}
    \begin{tabularx}{\textwidth}{l|cl|l}
    \toprule
        \textbf{Procedure} & \textbf{Parameter} & \textbf{Value} & \textbf{Resolution}\\[0.1ex]\midrule
        \multirowcell{4}[0pt][l]{Phase\\alignment} & $R$& $2h_c$ ($h_c$ coarse mesh element size) & \multirowcell{4}[0pt][l]{Coarse mesh from homogenised\\solution with $n_x\times n_y$ elements. }\\
         &  $(r_1,\;r_2)_j$ & $(\frac{1}{\pi},\; \pi)(1-\Delta\tilde{\kappa}_j)+\Delta\tilde{\kappa}_j$ \autoref{eq:phasealignrad}\\
                        & Iterations  & 20 \\
                        & Order & \autoref{eq:alignment_order_in_article}\\\midrule
        \multirowcell{2}[0pt][l]{Sampling} & $\beta=\alpha$ & ${\omega/h_c}$ ($\omega$ the user-specified periodicity) & \multirowcell{2}[0pt][l]{First intermediate resolution  with \\ $(n_xi_{up}^1)\times(n_yi_{up}^1)$ elements (\autoref{eq:iup1}). } \\
                 & $(r_1,\;r_2)$ & $(1/\pi,\;\pi)$ \\\midrule
        \multirowcell{2}[0pt][l]{Branch\\connection} & $(r_1,\;r_2)$ & $(1/\pi,\;1)$  &\multirowcell{2}[0pt][l]{Second intermediate resolution  with \\ $(n_xi_{up}^2)\times(n_yi_{up}^2)$ elements (\autoref{eq:iup2}) }\\
         & $k_{max}$ & 3   \\\midrule
        \multirowcell{3}[0pt][l]{Boundary\\construction} & $\beta=\alpha$ & $\bar{\omega}/h_c$ ($\bar{\omega}{=}\min\left(1/(4h_c),\;\omega/2\right)$ ) &\multirowcell{3}[0pt][l]{Depending on subprocedure.}\\
                 & $r_1$ & \autoref{eq:boundary_r1_in_article}
                  \\
                & $r_2$ & $\min\{1/r_1,\;1\}$     \\     \bottomrule
    \end{tabularx}
\end{table}
\begin{center}
\noindent\fbox{%
    \parbox{0.9\textwidth}{%
    \begin{align}
      \begin{split}&\bar{\Delta}x^l_j=\min_{b\in \bar{\mathcal{B}}}\left\{||(x_j^0)^l-x_b^0||\right\} \\ &\bar{\mathcal{B}}=\left\{b\in \mathcal{B}: \left|\cos \tilde{\tau}_b\cos\theta_j^l+\sin\tilde{\tau}_b\sin\theta_j^l\right|\geq 0.95 \;\land\; \min_{k\in \mathcal{N}^{3\times3}}\left\{\cos \tilde{\tau}_b\cos\tilde{\tau}_k+\sin\tilde{\tau}_b\sin\tilde{\tau}_k\right\}\geq0  \right\} \label{eq:alignment_order_in_article}\end{split}\\
   & r_1= \left(0.5+\left(\min_{k\in \mathcal{N}^{3\times3}}\left\{\cos \tilde{\tau}_b\cos\tilde{\tau}_k+\sin\tilde{\tau}_b\sin\tilde{\tau}_k\right\}\geq 0.95\right)\right)^{-1}\label{eq:boundary_r1_in_article}\\
   &i_{up}^1\geq \frac{h_c\omega}{0.1} \label{eq:iup1}\\
  &i_{up}^2\geq {2h_c\omega}/{\mu_{min}} \; \land \; i_{up}^2\geq i_{up}^1\label{eq:iup2}
\end{align}
}}
\end{center}

\subsection{Adding structural boundary}\label{sec:boundary_method}
\begin{figure}[!htb]
    \centering
    \includegraphics[width=\linewidth]{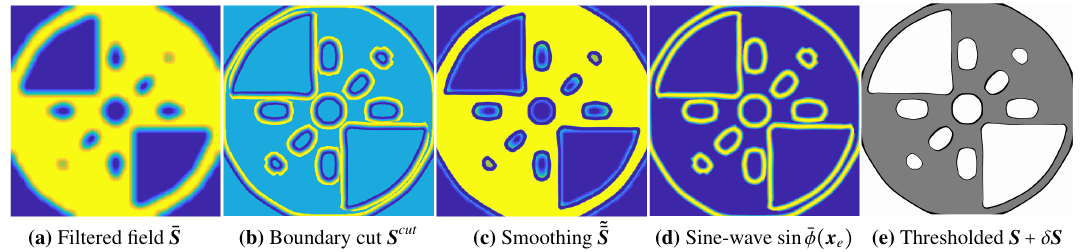}
    \caption{Illustrating how the upscaled filtered indicator field (a) is combined with the boundary cut-field (b) to obtain a smoothing of the filtered indicator field (c). Combined with the boundary sine-wave (d) thresholding is utilised to obtain the boundary-smoothed indicator field $\boldsymbol{S}$ and the varying thickness  boundary $\delta\boldsymbol{S}$ are obtained (e).}
    \label{fig:cutfield_threshold_fine_mesh}
\end{figure}

The phasor methodology can also be utilised to smooth the structural domain to reduce staircase artefacts and structural appendices introduced by the underlying coarse mesh, and in the same process add a varying thickness boundary along the outer edges of the domain. The main idea is built on constructing an additional lamination layer with active kernels along the boundary of the structural domain on the coarse mesh, ensuring smoothly transitioning orientations along the boundary (\autoref{fig:boundary_coarse_mesh}). The corresponding sampled phasor field can then be utilised both to introduce a smoothing cut along the staircase of the coarse indicator field and to add a varying thickness boundary along the smoothed indicator field (\autoref{fig:cutfield_threshold_fine_mesh}). This methodology is also directly applicable for smoothing fully solid regions in the dehomogenised design.

\subsubsection{Constructing coarse mesh set of phasor kernels}
Firstly, let $\boldsymbol{s}^l$ denote the field indicating what coarse mesh elements contain material in lamination layer $l\in \mathcal{L}$, such that the union of these fields, $\boldsymbol{s}$, serves as an indicator field for the combined structure.
\begin{align}
   {s_e}=\bigcup_{l\in \mathcal{L}}s_e^l \;:\; {s_e^l}=\begin{cases}1& \text{if }\mu_e^l\geq \mu_{min}\\ 0 &\text{otherwise} \end{cases},\;\; e\in \mathcal{T}_c,\; l\in \mathcal{L}
\end{align}
Apply a 3x3 mean box filter with zero-padding to obtain the filtered global indicator field $\boldsymbol{\tilde{s}}$ and and a 3x3 mean box filter with replication padding to each lamination layer indicator field to obtain the filtered fields $\boldsymbol{\tilde{s}^l}$. The union $\boldsymbol{\bar{s}}=\cup_{l\in \mathcal{L}}\boldsymbol{\tilde{s}^l}$ will serve as a the filtered global indicator field with replication padding for ensuring structural coherence along the domain boundary. This approach is chosen because it is beneficial to obtain individual layer indicator fields at the different resolutions to restrict the other processes in the de-homogenisation to not consider void regions, as well as the individual indicator fields being useful in adjusting the structural boundary where only one layer is active.

To obtain the orientational field along the structural boundary the Sobel operator is applied to $\boldsymbol{\hat{\tilde{s}}}$,  ${\hat{\tilde{s}}}_e=s_e\tilde{s}_e,\; \forall e\in \mathcal{T}_c$, i.e. the product between the global indicator field and its filtered field. This filter provides the gradients orthogonal to the boundary of the filtered field. The gradients of the modified filtered field are considered, rather than the 0-1 indicator field $\boldsymbol{s}$ or the filtered field $\boldsymbol{\tilde{s}}$, to increase the region where the gradient information is reliable inside the material domain and encourage larger gradient magnitudes along the outer boundary of $\boldsymbol{s}$. This will be useful for defining the locations of the boundary phasor kernels as well as improving the quality of the phasor field along the boundary by allowing for a larger area of filtering. The orientation along the boundary $\tau_e$ of kernel $e\in \mathcal{T}_c$ is obtained by rotating the direction of the gradients by $\pi/2$, such that it is coherent with the phasor definitions.

Define $\mathcal{B}\subseteq \mathcal{T}_c$ to be the set of elements where the magnitude of the boundary gradients $|\nabla \boldsymbol{\hat{\tilde{s}}}|$ is strictly positive, and thus indicates the set of elements with reliable boundary orientations. Let $\tau_e$ denote the orientation of kernel $e\in \mathcal{T}_c$ obtained from the indicator gradients. To improve the coherence between the structural boundary and the lamination layers these orientations are then subjected to corrections based on the degree of alignment with the nearest lamination orientation. 

 \begin{align}
&\tilde{\theta}_e=\text{argmax}\{|\cos{\tau}_e\cos\theta^{l}_e+\sin{\tau}_e\sin\theta^{l}_e|\;:\; l\in \mathcal{L}\}\\
&w_e=\cos{\tau}_e\cos\tilde{\theta}_e+\sin{\tau}_e\sin\tilde{\theta}_e\\
&\hat{w}_e=\mathcal{H}\left(w_e,\;0\right)\quad \land \quad \tilde{w}_e=\dfrac{\max\{|{w}_e|-0.75,0\}}{0.75}\\
&{\tau}_e{:=}\Arg\left((1-\tilde{w}_e)\mathrm{e}^{i{\tau}_e}+\tilde{w}_e\mathrm{e}^{i(\tilde{\theta}_e+\hat{w}_e\pi)}\right)
\end{align}

As such, if the orientation of $\tau_e$ is close to the nearest lamination direction $\tilde{\theta}_e$, it is updated to better align with the nearest lamination orientation, but maintain its original direction. Secondly, to ensure a smoothly transitioning directional field along the boundary after this update, the orientations are updated in an average smoothing procedure considering the orientations of a kernel's nearest neighbours in $\mathcal{B}$. The smoothing procedure starts from $\bar{\tau}_e=\tau_e,\; \forall e\in \mathcal{B}$ and is applied to each boundary region element in an iterative procedure in order of increasing $|w_e|$-value.
\begin{align}
&\bar{\tau}_e:=\Arg\left(\displaystyle\sum_{m\in\mathcal{N}^{3\times 3}_e}\dfrac{\tilde{w}_m+1}{2}\mathrm{exp}\left(i(\bar{\tau}_e+(\boldsymbol{d}^{\bar{\tau}}_e\cdot\boldsymbol{d}^{\bar{\tau}}_m<0)\pi)\right) \right)\\
    &\boldsymbol{d}^{\tau}_e=(\cos{\tau_e},-\sin{\tau_e})^T\quad \land \quad \mathcal{N}^{(n+2)\times (n+2)}_e=\{m\in \mathcal{B}\;:\; \max\{|\boldsymbol{x}_e-\boldsymbol{x}_m|\}\leq n\},\; n\in \mathbb{N}
\end{align}
\begin{figure}[!htb]
    \centering
    \includegraphics[width=\linewidth]{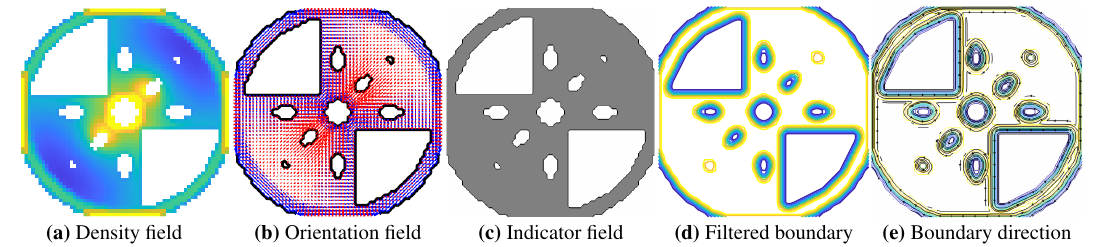}
    \caption{Illustrating the main steps of obtaining the boundary region phasor kernel directions on the coarse mesh. Given the density field (a) the indicator field (c) can be obtained. Filtering of the indicator field gives a smoothed structural boundary (d) on the coarse mesh, where yellow indicates where the filtered region extends fully into the solid structural domain given buy the indicator field. Gradients of the smoothed boundary combined with the orientation field of the lamination directions (b) are combined to obtain the boundary region kernel directions (e).}
    \label{fig:boundary_coarse_mesh}
\end{figure}

\autoref{fig:boundary_coarse_mesh} illustrates the main components of obtaining a smoothly oriented set of boundary region phasor kernels. A subset of these kernels are extracted to define the active boundary phasor kernels, while the remaining allow for improving the quality of the sampled boundary-wave away from the active kernel centres. The set of active kernels defining the structural boundary is then $\mathcal{\hat{B}}=\left\{e\in \mathcal{B}\;:\;s_e{=}1\; \land \; \tilde{s}_j{<}1\right\}$, ensuring that only elements with reliable lamination directions and boundary orientation touching the boundary of the indicator field are included. As such, the boundary is now defined by a set $\mathcal{\hat{B}}$ of phasor kernels with smoothed orientations $\bar{\tau}_e,\; e\in \mathcal{\hat{B}}$ along the structural boundary. To ensure appropriate cut-off of the indicator field the phase-shifts $\hat{\varphi}_b$ of the boundary phasor kernels $b\in \mathcal{\hat{B}}$ are predetermined such that the 1-contourline lies at the intended boundary of the final structural domain. 

\begin{minipage}{0.495\textwidth}
    \begin{align}
    & \delta\tau_e^l=|\cos{\tilde{\tau}}_e\cos\theta^{l}_e+\sin{\tilde{\tau}}_e\sin\theta^{l}_e|,\; e\in \mathcal{\hat{B}},\; l\in \mathcal{L}\\
    &\delta\tau_e=\begin{cases}1 & \text{if } \max_{l\in\mathcal{L}}\{\delta\tau_e^l\}>0.95\\ 0 & \text{otherwise} \end{cases}\\
    & c_e^1=\min_{b\in \mathcal{N}^{3\times 3}}\left\{\dfrac{\boldsymbol{d}^{\tilde{\tau}}_e\cdot \boldsymbol{d}^{\tilde{\tau}}_b+1}{2}\right\}\\
    & c_e^2=\max\left\{\displaystyle\sum_{l\in \mathcal{L}}\mu_e^l,\; 0.5(1-c_e^1)\right\}
\end{align}
\end{minipage}
\hfill
\begin{minipage}{0.495\textwidth}
    \begin{align}
& c_e^3=\mathcal{H}\left(1-\dfrac{\min_{l\in\mathcal{L}}\{\tilde{s}_e^l\}}{{\bar{s}_e}},\;0\right)\mathcal{H}\left(\displaystyle\sum_{l\in\mathcal{L}}\tilde{s}_e^l-\bar{s}_e,\;0\right)\\
    &c_e^4=\delta\tau_e\displaystyle\sum_{l\in \mathcal{L}}\mu_e^lc_e^3\\
    & \delta s_e=c_e^2\left(1-c_e^4\right)-0.5\\
    & \hat{\varphi}_e=\begin{cases}\dfrac{\pi}{8}  & \text{if } \displaystyle\sum_{l\in \mathcal{L}}\mu^l\geq 0.99 \; \lor\; e\in \partial\mathcal{T}_c\\ \dfrac{\pi}{2}(1-\delta s_e)&\text{otherwise}\end{cases}
\end{align}
\vfill
\end{minipage}

The premise of this definition builds on the observation that a phase shift of $\pi/2$ in isolation results in the 1-contour of the resulting phasor sine-wave cutting through the origin of the considered kernel. Therefore, this phase-shift is the baseline of the individual boundary kernel phase shifts. Multiple corrections are introduced to locally adjust the baseline phase shift combined in $\delta s_e$. Firstly, $c_e^1$ measures the smallest degree of orientation-alignment between kernel $e\in \mathcal{{B}}$ and its nearest neighbours, such that $1-c_e^1$ can be interpreted as how much the boundary orientation varies within a 3x3 neighbourhood of the considered kernel. $c_e^2$ is the main phase-shift correction control parameter and is determined by the amount of material in a kernel element and its degree of orientation variation. $c_e^3$ is an indicator field for areas along the boundary where only one layer is active, but is near the boundary of another layer.

Whenever there is material in all layers along the boundary $c_e^4=0$ and thus $\delta s_e=c_e^2-0.5$. Here the phase shift is larger for kernels containing a smaller volume fraction of material and becomes smaller as the amount of material increases. This ensures that the 1-contour line if shifted further through the elements that contain less material and reduces the cut-off of elements that contain more material. The inclusion of $c_e^1$ in $c_e^2$ of how much the boundary orientation varies within a 3x3 neighbourhood of the boundary kernel ensures that if the variation is large the phase shift becomes smaller to ensure a smoothly connected boundary and avoid cutting across narrow structural members.

\begin{figure}[htb!]
    \centering
    \includegraphics[width=\linewidth]{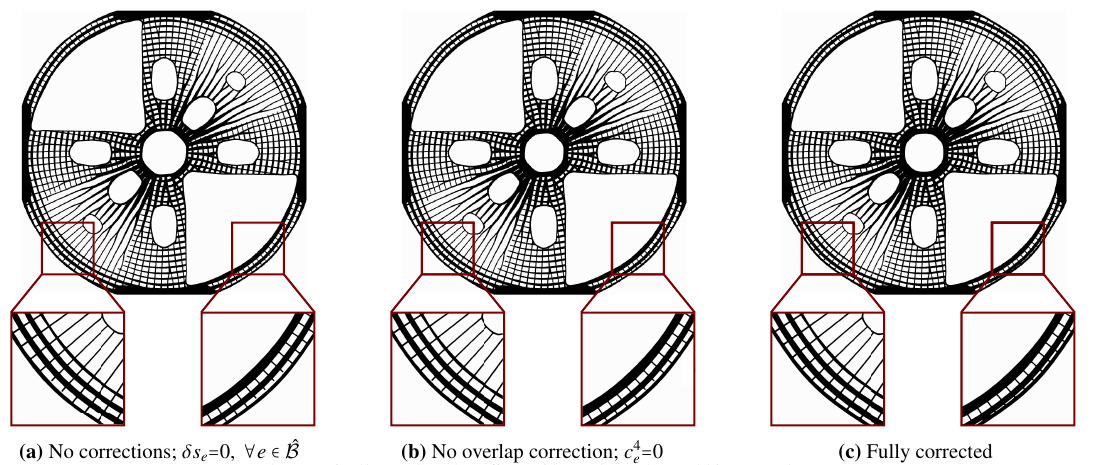}
    \caption{Illustrating the effect of boundary phase shift corrections.}
    \label{fig:Final_struct_different_boundary_corrections}
\end{figure}

The correction $c_e^3$ is only active if the orientation of a boundary kernel follows the nearest lamination direction closely and not all lamination layers contain material in this kernel. If there is only material in one layer at the boundary parallel to the lamination direction of this layer there is an issue of non-supported bars along this part of the boundary (\citealt{JensenGroen2022}). Therefore, this correction is designed to shift of the boundary further into the structural domain when this occurs. The effect of this correction can be observed in \autoref{fig:Final_struct_different_boundary_corrections}(c).

Lastly, the phase-shift definition is subjected to a correction to avoid cut-off of fully solid parts of the domain as well as a separation from the domain outer boundary $\partial\mathcal{T}_c$ by enforcing a phase shift of $\pi/8$ along these parts of the structural boundary, such that the underlying coarse element is inside the structure. Note that fully solid domains here are defined as regions with total relative thickness $\geq 0.99$ to ensure robustness against potential numerical deviations in the representation of solid caused by the optimiser or employed filter and to maintain length-scale control for the void-regions as well (\citealt{JensenGroen2022}, \citealt{Groen18}). The second correction pertains to the elements along the domain boundary $\partial \mathcal{T}_c$, where the phase shift is reduced based on how close the kernel origin is to the structural domain and the maximal thickness at the centre of this kernel.

\subsubsection{Obtaining the boundary on a finer mesh}
The phasor field along the boundary is to be used to smooth the staircase artefact of upscaling the indicator field from the coarse to the fine mesh. Therefore, the  periodicity of the boundary kernels are set to $\bar{\omega}=\min(1/(4h_c),\omega/2)$ to ensure that the positive part of one wave-length in the resulting phasor sine field is wide enough to cover two coarse elements and at least twice the wave-length of the lamination fields. Phasor field sampling with filtering is then used directly without phase alignment to obtain the complex boundary wave field $\bar{\mathcal{G}}(\boldsymbol{x})$ on an intermediate mesh, $\boldsymbol{x}\in \mathcal{T}_i$.

Let $\boldsymbol{\bar{S}}$ denote the upsampled filtered indicator field obtained by linear interpolation of $\boldsymbol{\bar{s}}$ on some mesh resolution $\mathcal{T}$ and the boundary phasor wave $\bar{\phi}(\boldsymbol{x})=\text{Arg}(\bar{\mathcal{G}}(\boldsymbol{x}))$. The structural boundary smoothing given the sampled phasor field is then achieved by $\boldsymbol{\tilde{\bar{S}}}=\boldsymbol{\bar{S}}+\boldsymbol{S}^{cut}$ where $\boldsymbol{S}^{cut}$ denotes the boundary cut-field obtained for each sampled element $e\in \mathcal{T}_i$ by
\begin{equation}
S_e^{cut}=\begin{cases}\dfrac{1}{2}+\dfrac{1}{2}\tanh{\left(\sin{\left(\bar{\phi}(\boldsymbol{x}_e)+\dfrac{\pi}{2}\right)}\right)} & \text{if } \sin{\left(\bar{\phi}(\boldsymbol{x}_e)\right)}\geq 0 \\ 0 & \text{otherwise}\end{cases},\; e\in \mathcal{T}_i
\end{equation}

As such, the sampled boundary field is utilised to increase the value of the filtered indicator field on the inside of the 1-contour line of the boundary sine-wave $\bar{\phi}$ and decrease the value outside this line. The hyperbolic tangent ensures that the imposed transition increases smoothly but rapidly near the 1-contour line. The boundary-smoothed indicator field $\boldsymbol{S}$ at the current resolution $\mathcal{T}$ is obtained by thresholding $\boldsymbol{\tilde{\bar{S}}}$ directly. The thresholded field is then utilised to cut the boundary-wave along the new structural boundary, such that it is only positive within the structural domain. The restricted boundary sine-wave is then transformed to a triangular wave and thresholded according to the desired boundary thicknesses $\hat{\mu}$, adjusted to adhere to the modified boundary periodicity and that only the positive part of the boundary sine-wave is considered, to obtain the structural boundary $\partial\boldsymbol{S}$.
\begin{align}
    &\boldsymbol{S}=\mathcal{H}(\boldsymbol{\tilde{\bar{S}}},\;0.5)\quad \land \quad\partial\boldsymbol{S}=\mathcal{H}\left(\boldsymbol{S}\left(\frac{1}{\pi}\arcsin(\sin(\bar{\phi}))+0.5\right),\;1-\dfrac{2\bar{\omega}\hat{\mu}}{\omega}\right)\mathcal{H}\left(\sin(\bar{\phi}),\;0\right)
\end{align}

In this way, the phasor methodology allows for sampling a smooth structural boundary at any suitable finer resolution from a set of phasor kernels on the coarse grid. Keeping the computations for boundary defining parameters on the coarse grid makes the procedure computationally efficient. Further, the fact that the sampled complex wave field $\bar{\mathcal{G}}(\boldsymbol{x})$ can be upscaled to any resolution or body-fitted mesh before translation and thresholding allows for good flexibility and scalability. Lastly, the use of the nearest lamination orientations to adapt the direction of the boundary phasor kernels only serves as an improvement to the structural boundary, but a sufficiently smooth boundary following the structural domain as defined by the indicator field on the coarse mesh can be obtained without this information. Thus, this boundary procedure can be implemented for any structure where a coarse-scale indicator field is available, and is therefore only to a limited degree restricted by the solution method or structural representation used. 

The existing method proposed in \citealt{Christensenetal2023} relies on applying a PDE-filter to the indicator field on the fine resolution after applying boundary refinement as described in \citealt{JensenGroen2022}. The smoothness of the obtained boundary is in this case dependent upon the degree of staircase artefacts caused by the underlying coarse density fields as the refinement method applied in most staircase occurrences cuts the boundary elements from the coarse mesh in half. Additionally the construction of a PDE-filter on the fine resolution to only be applied once is computationally expensive relative to the overall running time of the phasor-based methodology. The benefit of the proposed phasor-based framework is that it inherits the procedural benefit in terms of coarse-mesh parameterisation and control while allowing for greater smoothing of the mentioned staircase artefacts.

\section{Numerical examples} \label{sec:reuslts_main}
This section quantifies the added value from the proposed dehomogenisation method, by comparing its performance to existing methods as well as testing its reliability and robustness.

To allow for direct comparisons to existing frameworks, the compliance-volume fraction measure $\mathcal{S}=VC$ (\citealt{Elingaardetal2022}) will be utilised as well as the corresponding ratio to the homogenised solution $\mathcal{R}={\mathcal{S}_{dehom}}/{\mathcal{S}_{hom}}$. Here $\mathcal{S}_{dehom}$ and $\mathcal{S}_{hom}$ are the compliance-volume fraction measures for the dehomogenised and homogenised solutions respectively. It is here noted that the most accurate comparison to the homogenised solution is obtained by upscaling the coarse mesh optimised solution to the same fine-scale resolution used for the dehomogenised, but as some existing methods have compared to the coarse mesh homogenised solution, the benchmark for comparison will vary throughout the tests. This will be specified accordingly.

Existing methods such as the CNN-based (\citealt{Elingaardetal2022}) and the growth-based (\citealt{Garnieretal2022}) have specified the periodicity in terms of $\varepsilon_i$ on the intermediate mesh and $\varepsilon_{f}$ on the final mesh, i.e. the wavelengths as functions of the element size at the corresponding resolution. Instead, $\omega$, i.e. the maximal number of periods that can fit within the structure, will be referred to as the periodicity measure in most tests throughout this section. This measure is resolution independent, and thus a more general option. Unless otherwise specified, the intermediate upscaling factors $i_{up}^1$ and $i_{up}^2$, as well as the fine resolution upscaling factor $f_{up}$, are chosen such that \autoref{eq:iup1} and \autoref{eq:iup2} are satisfied with equality. Additionally, if not otherwise specified, the parameters given in \autoref{tab:Parameter_choices} are utilised.

All results reported for the phasor-based method are obtained by a single-core Matlab implementation on an Apple M1 chip. The tests performed consist firstly of comparing to existing methods considering problem instances from \citealt{Elingaardetal2022} in \autoref{seq:compare_existing}, with additional focus on the effect of the added structural boundary, the impact of varying the number of phase alignment iterations and the stability benefits of local methods (CNN-based, growth-based, phasor-based etc.) compared to the global projection-based methods. Secondly, the phasor method will be tested for periodicity convergence in \autoref{sec:periodicity_convergence} with special focus on the limits of reliability for this approach compared to the chosen mesh resolutions. Furthermore, the proposed framework is tested for dehomogenisation of intermediate design solutions in \autoref{sec:iteration_history} during the optimisation process. The performance for selected multiload problem instances for Rank-3 and Rank-4 optimised structures is presented in \autoref{sec:additional_tests} and the benefit of geometry extraction to a body-fitted mesh is exemplified in \autoref{sec:bodyfitted}. Lastly, the time complexity of the method is addressed in \autoref{sec:time_test}.

\subsection{Comparing to existing methods} \label{seq:compare_existing}
One of the main motivations for the suggested phasor-based approach was whether one could obtain better and more stable performance in terms of solution quality, without much increase in computational time, compared to the CNN-approach presented in \citet{Elingaardetal2022}. To this end a selection of test cases used to verify the validity of the CNN will be considered. For fair comparisons of results the same fine resolution upscaling factor $f_{up}$ is ensured in each case, with $i_{up}^1$ and $i_{up}^2$ adjusted according to \autoref{eq:intermediate_resolution_limits} with equality. The fine scale resolution of some of the high-periodicity small minimal relative thickness examples from \citet{Elingaardetal2022} pushes the fine-scale resolution of the phasor-dehomogenised structures to their lower limit. This means that the proposed procedure is expected to perform better if finer resolutions were allowed in these cases, which will be illustrated by the periodicity test presented in \autoref{sec:periodicity_convergence}. In addition to serving as a foundation for comparison, the problem instances considered in this subsection will also serve as basis for tests determining fundamental properties of the phasor-based method in terms of the effect of number of phase-alignment iterations, the gain by adding the phasor-boundary layer and the overall benefit of the local nature of the method.


\subsubsection{Symmetric cantilever} \label{sec:elin_cantilever_results}
Different versions of the minimal compliance symmetric cantilever beam are the most prominent test cases presented in \citealt{Elingaardetal2022} and a subset of these examples will therefore also be the main focus of comparing the phasor-based method to the CNN-based method. The selected cantilever problem instances are given in \autoref{tab:cantilever_elingaard_instances}, each uniquely identified by the homogenised solution element size $n_c$, minimal relative thickness $\mu_{min}$, volume fraction constraint $V_{frac}$, dehomogenised design element size $h_f$ and the wavelength in pixels/period on the fine mesh $\varepsilon_f$. Each instance is assigned an identifying ID reference to be used throughout this test section. In addition to comparing to the results from \citealt{Elingaardetal2022} the more conventional dehomogenisation approach also allowing for branching from \citealt{GroenFork2019} as well as the reaction-diffusion pattern-generation approach from \citealt{Garnieretal2022} are also considered. The latter idea of treating dehomogenisation as a
pattern-generating procedure is the one closest related to the phasor-based methodology.

\begin{table}[H]
\caption{Summarising the cantilever problem instances from \citealt{Elingaardetal2022} to be considered for method comparisons. The instance ID is unique to each case and will be used as identifier in the following result-reporting.}
\label{tab:cantilever_elingaard_instances}
\resizebox{\textwidth}{!}{
\begin{tabular}{cccccc|cccccc|cccccc}
\toprule
ID & $h_c$ & $\mu_{min}$ & $V_{frac}$ & $h_f$     & $\varepsilon_f$ & ID & $h_c$ & $\mu_{min}$ & $V_{frac}$ & $h_f$     & $\varepsilon_f$ & ID & $h_c$ & $\mu_{min}$ & $V_{frac}$ & $h_f$     & $\varepsilon_f$ \\ \midrule
A1 & 1/30  & 0.05        & 0.25       & $1/24h_c$ & $60h_f$         & B1 & 1/30  & 0.05        & 0.25       & $1/40h_c$ & $50h_f$         & C1 & 1/120 & 0.05        & 0.25       & $1/24h_c$ & $60h_f$         \\
A2 & 1/30  & 0.05        & 0.40       & $1/24h_c$ & $60h_f$         & B2 & 1/30  & 0.05        & 0.40       & $1/40h_c$ & $50h_f$         & C2 & 1/120 & 0.05        & 0.40       & $1/24h_c$ & $60h_f$         \\
A3 & 1/30  & 0.10        & 0.25       & $1/24h_c$ & $60h_f$         & B3 & 1/30  & 0.10        & 0.25       & $1/40h_c$ & $50h_f$         & C3 & 1/120 & 0.10        & 0.25       & $1/24h_c$ & $60h_f$         \\
A4 & 1/30  & 0.10        & 0.40       & $1/24h_c$ & $60h_f$         & B4 & 1/30  & 0.10        & 0.40       & $1/40h_c$ & $50h_f$         & C4 & 1/120 & 0.10        & 0.40       & $1/24h_c$ & $60h_f$         \\
A5 & 1/30  & 0.20        & 0.25       & $1/24h_c$ & $60h_f$         & B5 & 1/30  & 0.20        & 0.25       & $1/40h_c$ & $50h_f$         & C5 & 1/120 & 0.20        & 0.25       & $1/24h_c$ & $60h_f$         \\
A6 & 1/30  & 0.20        & 0.40       & $1/24h_c$ & $60h_f$         & B6 & 1/30  & 0.20        & 0.40       & $1/40h_c$ & $50h_f$         & C6 & 1/120 & 0.20        & 0.40       & $1/24h_c$ & $60h_f$  \\\bottomrule      
\end{tabular}}
\end{table}

\begin{figure}[!htb]
    \centering
 \includegraphics[width=0.95\linewidth]{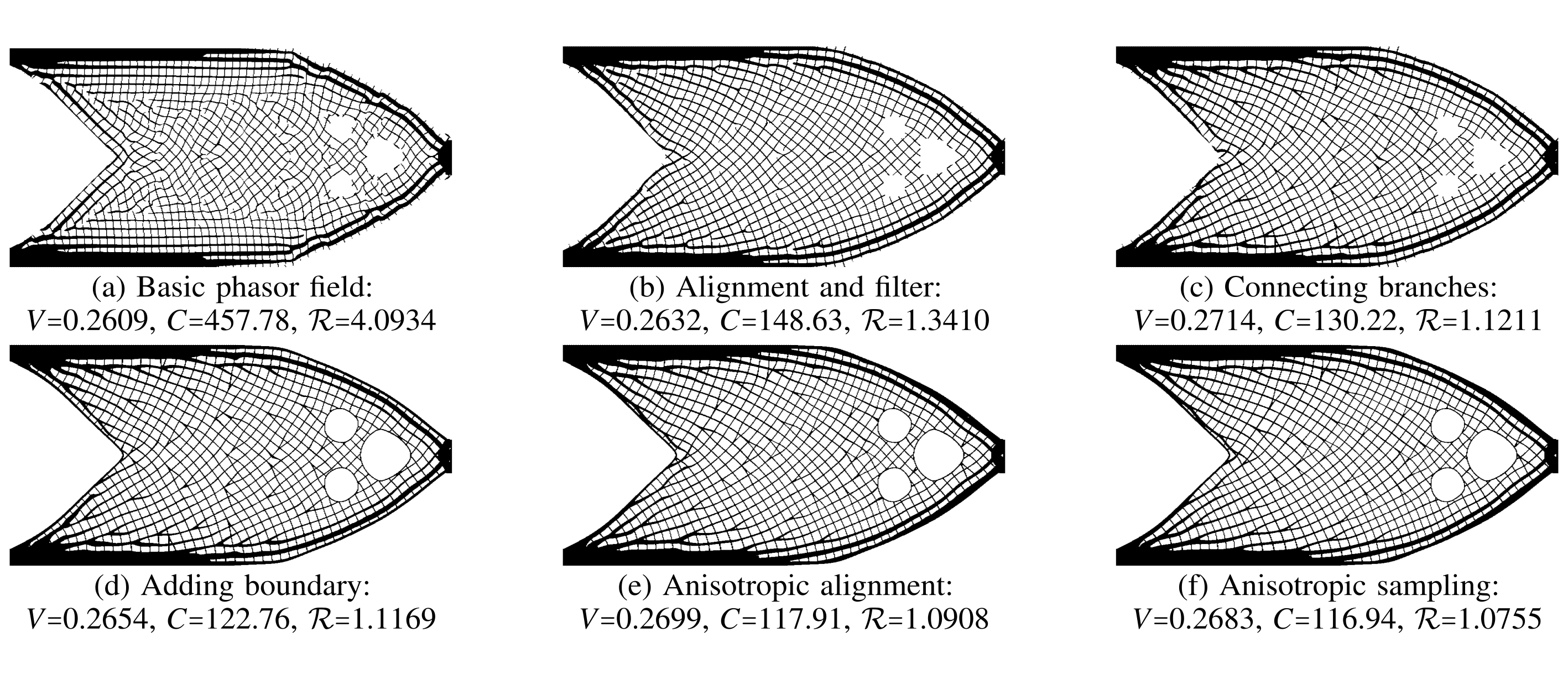}
    \caption{Illustrating the effect of the added measures to elevate the basic phasor noise definition to one suitable for dehomogenisation by considering the cantilever instance B3 in \autoref{tab:cantilever_elingaard_instances}. The presented results reflect the results obtained by utilising the fixed parameters presented in \autoref{tab:Parameter_choices} for each additional feature. The sequence starts  in (a) by presenting the results for direct isotropic phasor kernel sampling without any phase alignment or filtering. The isotropic alignment and sampling filter is added in (b), branch-connection in (c) and boundary in (d). The effect of the anisotropic alignment neighbourhood is presented in (e), and finally the completed procedure with anisotropic sampling result in (f).}
    \label{fig:cantilever_improvement_effect}
\end{figure}

First, to indicate the importance of the procedural modifications introduced to the original phasor noise formulation, \autoref{fig:cantilever_improvement_effect} illustrates the dehomogenised designs for cantilever instance B3 where the crucial procedure additions are added in sequence. If the pure phasor definition is used directly (a), the dehomogenised design exhibits a lacking ability to follow lamination directions and a large degree of local curvature around the disconnections which results in poor stiffness. Including 20 alignment iterations and the sampling filter (b) improves the smoothly varying orientation of the lamination directions and results in a significant improvement in structural performance, already sufficient to obtain a similar solution quality to the corresponding solution presented in \citealt{Elingaardetal2022} (\autoref{tab:results_cantilever_Elin_Groen}). The addition of connecting the branches (c) further improves the design quality approaching the quality that was obtained by the ``conventional'' method introduced by \citealt{GroenFork2019}. Adding the structural boundary (d) ensures the design is improved beyond this benchmark. Adding anisotropy to the phase-alignment (e) procedure and sampling processes (f) reduces the local curvature around the branches, improves the branch distribution as well as the branch shapes and local thicknesses.

\begin{table}[H]
\centering
\caption{Comparing dehomogenisation time and solution quality of the proposed phasor-based method with added structural boundary using 20 phase alignment iterations for the Cantilever with the alternative dehomogenisation methods from \citealt{Elingaardetal2022} and \citealt{GroenFork2019}. $V_m$, $C_m$, $\mathcal{R_m}$ and $T_m$ denote the volume fraction, compliance, volume-compliance ratio to the homogenised solution and time for method $m$, respectively. }
\label{tab:results_cantilever_Elin_Groen}
\resizebox{\textwidth}{!}{
\begin{tabular}{c|cc|cccc|cccc|cccc}
\toprule
ID &  $V_{hom}$ & $C_{hom}$ & $V_{Elin}$ & $C_{Elin}$ & $\mathcal{R}_{Elin}$ & $T_{Elin}$ & $V_{Groen}$ & $C_{Groen}$ & $\mathcal{R}_{Groen}$ & $T_{Groen}$ & { $V_{phasor}$} & { $C_{phasor}$} & { $\mathcal{R}_{phasor}$} & { $T_{phasor}$} \\ \midrule
A1  & 0.2535    & 106.21    & 0.2695     & 140.16     & {\color[HTML]{CC313D} 1.4030} & \textbf{{1.28}}  & 0.2514      & 130.79      & {\color[HTML]{F8A102} 1.2212}          & 17.53       & 0.2711            & 115.72            & {\color[HTML]{1AAA1A} \textbf{1.1652}} & 1.56                                      \\
A2 & 0.4024    & 68.58     & 0.4329     & 75.25      & {\color[HTML]{CC313D} 1.1804} & \textbf{{1.27}}  & 0.3984      & 73.31       & {\color[HTML]{1AAA1A}\textbf{1.0584}} & 17.31       & 0.4214            & 74.80             & {\color[HTML]{F8A102} 1.1421}          & 1.41                                      \\
A3  & 0.2568    & 113.61    & 0.2661     & 167.32     & {\color[HTML]{CC313D} 1.5260} & 1.26                                     & 0.2497      & 139.27      & {\color[HTML]{1AAA1A}\textbf{1.1920}} & 17.33       & 0.2794            & 124.90            & {\color[HTML]{F8A102} 1.1960}          & \textbf{{ 1.10}}  \\
A4   & 0.4080    & 69.00     & 0.4393     & 73.68      & {\color[HTML]{CC313D} 1.1499} & 1.33                                     & 0.4052      & 72.82       & {\color[HTML]{1AAA1A}\textbf{1.0482}} & 17.34       & 0.4289            & 70.86             & {\color[HTML]{F8A102} 1.0795}          & \textbf{{ 1.05}}  \\
A5  & 0.2614    & 122.86    & 0.2665     & 141.57     & {\color[HTML]{CB0000} 1.1747} & 1.28                                     & 0.2455      & 143.30      & {\color[HTML]{F8A102} 1.0955}          & 17.82       & 0.2852            & 115.38            & {\color[HTML]{1AAA1A}\textbf{1.0247}} & \textbf{{ 0.65}}  \\
A6  & 0.4165    & 73.37     & 0.4389     & 77.07      & {\color[HTML]{CB0000} 1.1070} & 1.28                                     & 0.4018      & 83.35       & {\color[HTML]{F8A102} 1.0960}          & 18.29       & 0.4455            & 72.52             & {\color[HTML]{1AAA1A}\textbf{1.0573}} & \textbf{{ 0.95}}  \\ \midrule
B1 & 0.2535    & 106.21    & 0.2581     & 143.65     & {\color[HTML]{CB0000} 1.3770} & \textbf{{2.23}}  & 0.2538      & 114.94      & {\color[HTML]{F8A102} 1.0836}          & 36.51       & 0.2615            & 110.67            & {\color[HTML]{1AAA1A}\textbf{1.0748}} & 3.30                                      \\
B2   & 0.4024    & 68.58     & 0.4201     & 78.79      & {\color[HTML]{CB0000} 1.1993} & \textbf{{2.18}}  & 0.4017      & 72.86       & {\color[HTML]{1AAA1A}\textbf{1.0606}} & 30.70       & 0.4101            & 71.75             & {\color[HTML]{F8A102} 1.0663}          & 2.97                                      \\
B3   & 0.2568    & 113.61    & 0.2569     & 149.53     & {\color[HTML]{CB0000} 1.3167} & \textbf{{2.10}}  & 0.2549      & 128.36      & {\color[HTML]{F8A102} 1.1213}          & 31.14       & 0.2683            & 116.94            & {\color[HTML]{1AAA1A}\textbf{1.0755}} & 2.16                                      \\
B4   & 0.4080    & 69.00     & 0.4224     & 75.16      & {\color[HTML]{CB0000} 1.1279} & 2.19                                     & 0.4068      & 72.20       & {\color[HTML]{1AAA1A}\textbf{1.0434}} & 31.01       & 0.4171            & 70.90             & {\color[HTML]{F8A102} 1.0506}          & \textbf{{ 2.07}}  \\
B5  & 0.2614    & 122.86    & 0.2566     & 152.86     & {\color[HTML]{CB0000} 1.2214} & 2.21                                     & 0.2404      & 143.34      & {\color[HTML]{F8A102} 1.0729}          & 29.82       & 0.2644            & 120.785           & {\color[HTML]{1AAA1A}\textbf{0.9946}} & \textbf{{ 1.31}}  \\
B6  & 0.4165    & 73.37     & 0.4323     & 77.39      & {\color[HTML]{CB0000} 1.0949} & 2.16                                     & 0.4083      & 79.16       & {\color[HTML]{F8A102} 1.0576}          & 33.80       & 0.4262            & 74.16             & {\color[HTML]{1AAA1A}\textbf{1.0344}} & \textbf{{ 1.75}}  \\ \midrule
C1  & 0.2532    & 105.98    & 0.2587     & 130.40     & {\color[HTML]{CB0000} 1.2575} & \textbf{{23.30}} & 0.2529      & 112.02      & {\color[HTML]{1AAA1A}\textbf{1.0559}} & 297.49      & 0.2632            & 110.53            & {\color[HTML]{F8A102} 1.0842}          & 26.48                                     \\
C2  & 0.4023    & 68.61     & 0.4216     & 73.97      & {\color[HTML]{CB0000} 1.1297} & \textbf{{23.47}} & 0.4016      & 71.48       & {\color[HTML]{1AAA1A}\textbf{1.0399}} & 309.59      & 0.4114            & 71.33             & {\color[HTML]{F8A102} 1.0631}          & 25.16                                     \\
C3  & 0.2566    & 111.55    & 0.2586     & 128.85     & {\color[HTML]{CB0000} 1.1642} & 23.41                                    & 0.2579      & 117.42      & {\color[HTML]{1AAA1A}\textbf{1.0579}} & 345.38      & 0.2692            & 114.32            & {\color[HTML]{F8A102} 1.0752}          & \textbf{{ 20.81}} \\
C4  & 0.4078    & 68.94     & 0.4237     & 72.25      & {\color[HTML]{CB0000} 1.0885} & 23.71                                    & 0.4090      & 70.87       & {\color[HTML]{1AAA1A}\textbf{1.0310}} & 312.30      & 0.4194            & 70.24             & {\color[HTML]{F8A102} 1.0478}          & \textbf{{ 20.90}} \\
C5   & 0.2572    & 113.19    & 0.2607     & 120.00     & {\color[HTML]{CB0000} 1.0745} & 24.02                                    & 0.2548      & 118.53      & {\color[HTML]{F8A102} 1.0375}          & 260.15      & 0.2775            & 105.66            & {\color[HTML]{1AAA1A}\textbf{1.0070}}  & \textbf{{ 9.16}}  \\
C6   & 0.4154    & 71.86     & 0.4323     & 73.72      & {\color[HTML]{CB0000} 1.0677} & 23.55                                    & 0.4159      & 73.91       & {\color[HTML]{F8A102} 1.0298}          & 344.80      & 0.4309            & 71.16             & {\color[HTML]{1AAA1A}\textbf{1.0273}} & \textbf{{ 17.29}} \\ \bottomrule
\end{tabular}}
\end{table}
\autoref{tab:results_cantilever_Elin_Groen} compares the performance of the phasor-based dehomogenisation method using 20 phase-alignment iterations to that of the CNN-based method from \citealt{Elingaardetal2022} and the projection based method from \citealt{GroenFork2019}. The performances for each problem instance from \autoref{tab:cantilever_elingaard_instances} are compared in terms of the compliance-volume ratio to the homogenised solution, where the best performing method is highlighted in {\color[HTML]{1AAA1A}\textbf{green}}, the second best in {\color[HTML]{F8A102}{orange}} and the worst performing in {\color[HTML]{CB0000}{red}}. The most time-efficient method is highlighted in \textbf{bold}. It becomes evident that the phasor-based method significantly outperforms the CNN-based method in terms of structural performance requiring similar running times. In terms of solution quality, the phasor based method produces results similar to those of the projection-based method, but is about ten to thirty times more time-efficient.

It is expected that the structural boundary added by an additional phasor-based layer, as introduced in \autoref{sec:boundary_method}, will have a significant improving effect on the structural performance, compared to the design without this boundary. An indication of this effect has already been observed in \autoref{fig:cantilever_improvement_effect} when extending from (c) to (d). This improving effect is especially expected to be of importance for structures with low periodicity, many holes or thin structural members (\autoref{app:supp_res}). The dehomogenised structures are more sensitive to the overall phase shift of the lamination field in these cases, as void cut-off of supporting members may occur. As the benchmark methods from \citealt{Elingaardetal2022} and \citealt{GroenFork2019} do not include adding a structural boundary, it is relevant to also consider the structural performance without the added boundary for a more complete comparison to these approaches. It is, however, noted that the boundary-generating procedure is an integrated part in the phasor-based approach and information obtained is utilised throughout other subprocesses.

As observed in the phase alignment test in \autoref{fig:circle_phase_align_perfect_from_zero}, the iterative procedure, after reaching a desired inter-kernel phase alignment, continues iterating by gradually imposing a global phase-shift of this alignment. This development causes phase-shifts of the lamination layers which impacts the location of supporting bars relative to the void cut-off. The effect of the number of alignment iterations used will, after a stable alignment is achieved, be closely related to how beneficial the added boundary is. Therefore, the effect of adding the structural boundary and the number of fixed alignment iterations will be tested in combination.

\begin{figure}[!htb]
    \centering  
    \includegraphics[width=0.9\linewidth]{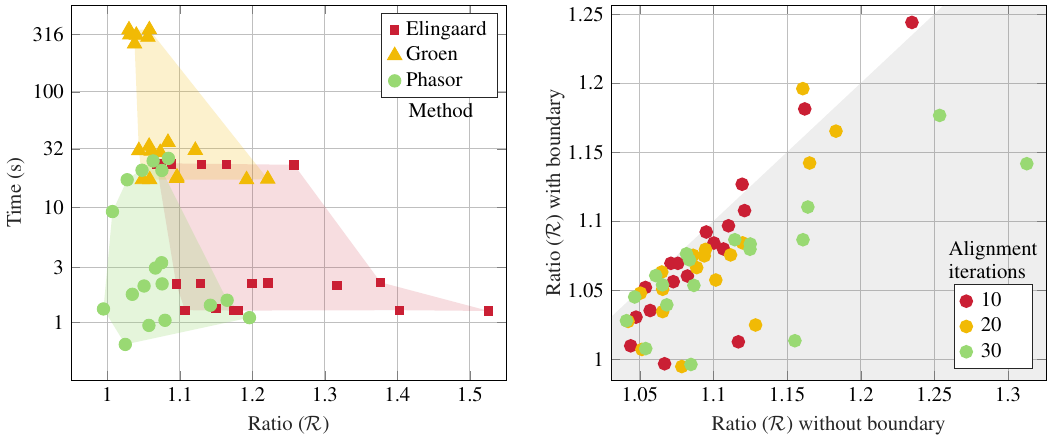}
    \caption{\textbf{Left:} Summarising the results for comparing the performance of the dehomogenisation methods in \autoref{tab:results_cantilever_Elin_Groen} in terms of the compliance-volume ratio to the homogenised solution against the dehomogenisation time, grouped by method. \textbf{Right:} Comparing the dehomogenisation results of the phasor-based method with and without added structural boundary grouped by the number of alignment iterations used, in terms of the compliance-volume ratio to the homogenised solution. The shaded area indicates where the added boundary benefits the solution quality.}
\label{fig:cantilever_method_alignment_boundary_compare}
\end{figure}

\autoref{fig:cantilever_method_alignment_boundary_compare} provides both a comparison of the solution time and quality of the dehomogenisation methods in \autoref{tab:results_cantilever_Elin_Groen} (left) and a comparison of the effect on the solution quality of adding the structural boundary and varying the numbers of alignment iterations for the phasor-based method (right). The method comparison indicates that the phasor-based method is more consistent than the CNN-based method in term of solution quality with limited compromise in terms of running time. Further, the phasor-based method is significantly more efficient than the projection-based method (10-30 times speed-up) without notable loss in solution quality. From the investigations of the effect of adding the structural boundary and varying the number of alignment iterations it can be observed that the structural boundary in almost all cases improves the structural performance as well as making the phasor-based method less sensitive to the number of alignment iterations. Overall, it is found that increasing the number of alignment iterations from 10 to 20 reduces the variance in volume and compliance of the tested instances, while the further increase to 30 iterations does not necessarily improve the solution quality. This is an indication of how performing some alignment iterations is necessary to obtain a sufficiently smooth transition between the kernels, but further iteration corresponds roughly to a global phase shift of the wave-field. It is found that 20 iterations are suitable for the test cases presented in this paper. The instances most sensitive to the number of iterations are those with lower periodicity and without an added boundary (\autoref{app:supp_res}). Adding the boundary and increasing the periodicity reduces the variance in solution quality as a function of the number of iterations, which is tied to the similar effect of applying global phase-shifts, which will be further investigated in \autoref{sec:periodicity_convergence}.

\begin{figure}[!htb]
\begin{minipage}{0.52\textwidth}
\begin{table}[H]
\centering
\caption{Comparing performance of the phasor based method to the Cantilever dehomogenisation examples from \citealt{Garnieretal2022}.}
\label{tab:compare_garnier}
\resizebox{\textwidth}{!}{
\begin{tabular}{c|ccccc|ccc}
\toprule
ID & $h_i$     & $V_{Garn}$ & $C_{Garn}$  & $\mathcal{R}_{Garn}$ & $T_{Garn}$  & $h_i$            & $\mathcal{R}_{phasor}^{20}$ & $T_{phasor}^{20}$ \\ \midrule
B1 & 1/10$h_c$ & 0.2570     & 135.46      & 1.2930               & \textit{NA} & (1/8, 1/32)$h_c$ & \textbf{1.0748}                      & 3.30              \\
B2 & 1/10$h_c$ & 0.4281     & 78.17       & 1.2126               & \textit{NA} & (1/8, 1/32)$h_c$ & \textbf{1.0663}                      & 2.97              \\
B3 & 1/10$h_c$ & 0.2667     & 156.12      & 1.4271               & \textit{NA} & (1/8, 1/16)$h_c$ & \textbf{1.0755}                      & 2.16              \\
B4 & 1/10$h_c$ & 0.4332     & 81.16       & 1.2489               & \textit{NA} & (1/8, 1/16)$h_c$ & \textbf{1.0506}                      & 2.07              \\
B5 & 1/10$h_c$ & 0.2808     & 134.78      & 1.1784               & \textit{NA} & (1/8, 1/8)$h_c$  &\textbf{ 0.9946}                      & 1.31              \\
B6 & 1/10$h_c$ & 0.4432     & 75.22       & 1.0909               & \textit{NA} & (1/8, 1/8)$h_c$  & \textbf{1.0344}                      & 1.76              \\ \midrule
C1 & 1/3$h_c$  & 0.2415     & \textit{NA} & \textit{NA}          & \textbf{20.06}       & (1/4, 1/16)$h_c$ & 1.0842                      & 26.48             \\
   & 1/4$h_c$  & 0.2473     & \textit{}   & \textit{}            & 23.07       &                  &                             &                   \\
   & 1/8$h_c$  & 0.2449     &             &                      & 61.8        &                  &                             &                   \\ \midrule
C3 & 1/3$h_c$  & 0.2524     & \textit{NA} & \textit{NA}          & 22.06       & (1/4, 1/8)$h_c$  & 1.0752                      & \textbf{20.81}             \\
   & 1/4$h_c$  & 0.2541     & \textit{}   & \textit{}            & 22.07       &                  &                             &                   \\
   & 1/8$h_c$  & 0.2442     &             &                      & 63.4        &                  &                             &                   \\ \midrule
C5 & 1/3$h_c$  & 0.2603     & \textit{NA} & \textit{NA}          & 22.04       & (1/4, 1/4)$h_c$  & 1.007                       & \textbf{9.16}              \\
   & 1/4$h_c$  & 0.2639     & \textit{}   & \textit{}            & 23.03       &                  &                             &                   \\
   & 1/8$h_c$  & 0.2609     &             &                      & 71.3        &                  &                             &                   \\ \bottomrule
\end{tabular}}
\end{table}
\end{minipage}
\hfill
\begin{minipage}{0.48\textwidth}
\vspace{12mm}
\includegraphics[width=\linewidth]{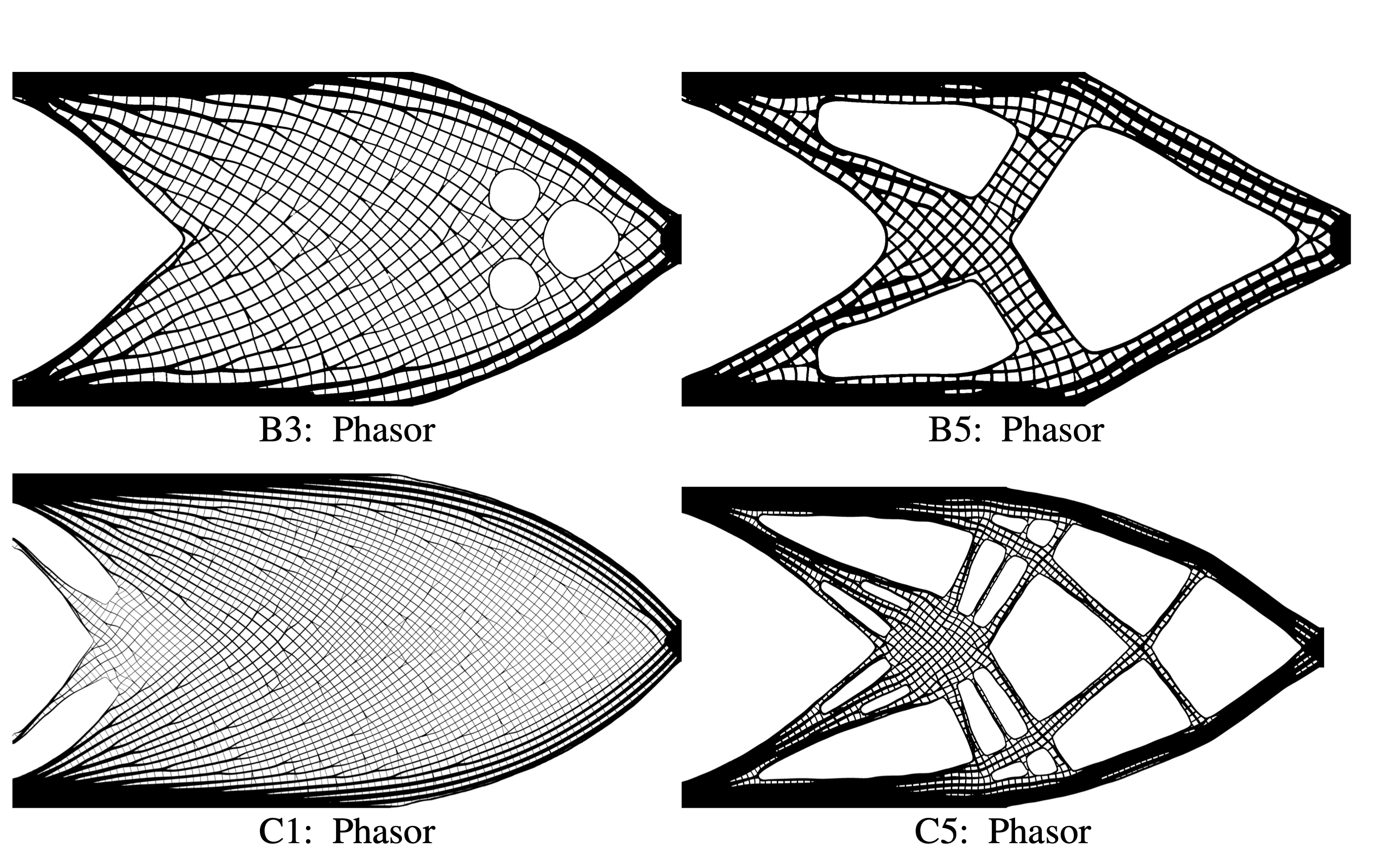}
    \caption{Selected dehomogenised B- and C-instance designs obtained by the phasor-based method.}
    \label{fig:selected_struct_compare_Garnier}
\end{minipage}
\end{figure}

\autoref{tab:compare_garnier} compares the performance of the phasor based method to the reaction-diffusion pattern generating method from \citealt{Garnieretal2022}. Crucial measures for an appropriate comparison are missing for the pattern generating approach, but as the authors state in their article, and as can be observed in the presented comparison, the performance appears to be similar to that of the CNN-based method in terms of solution quality. Solution times are only reported for the larger instances considered, for different intermediate mesh resolutions, but without any measure of solution quality. The efficiency of the phasor-based method approaches the pattern-generating times using coarser intermediate resolutions. As such, the phasor-based method is presumed to be equally beneficial compared to the pattern-generating method as compared to the CNN-based method, where the phasor-based obtains better designs. \autoref{fig:selected_struct_compare_Garnier} illustrates the dehomognenised designs obtained by the phasor-based method for selected B- and C-instances of the cantilever beam considered for throughout these tests.

\subsubsection{Double clamped beam}
For further comparison to existing methods and to highlight an additional major benefit of the localised nature of the phasor-based method, compared to the globality of the more conventional projection-based methods, the double-clamped beam instances from \citealt{Elingaardetal2022} are also considered for testing.

\begin{figure}[!htb]
\begin{minipage}[t]{0.495\textwidth}
 \begin{center}
\includegraphics[width=1\textwidth]{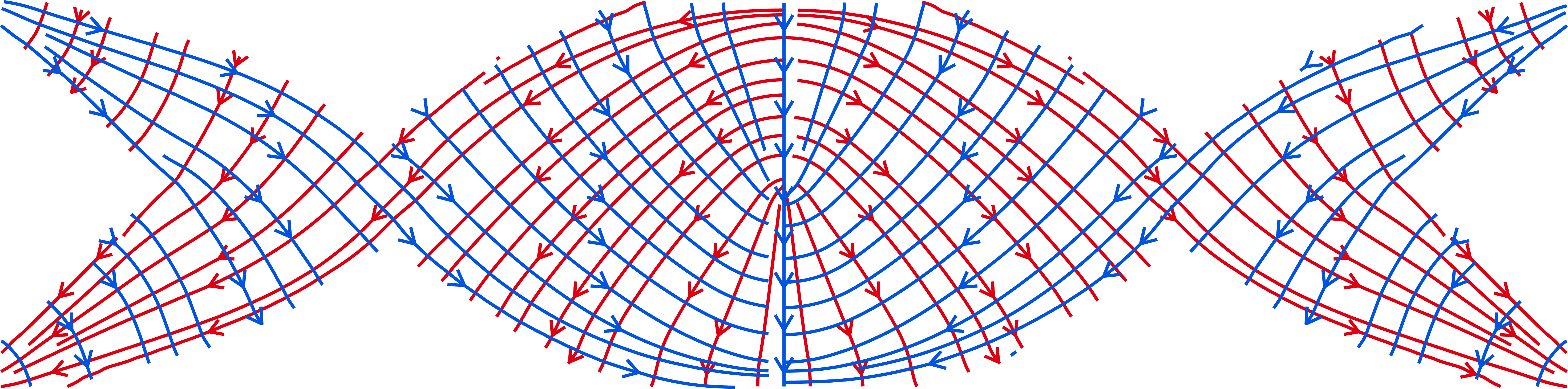}
{(a) The presence of singularities and neighbouring elements with opposite directions in the homogenised optimised orientation field.}
 \end{center}
\end{minipage}
\hfill
\begin{minipage}[t]{0.495\textwidth}
 \begin{center}
\includegraphics[width=1\textwidth]{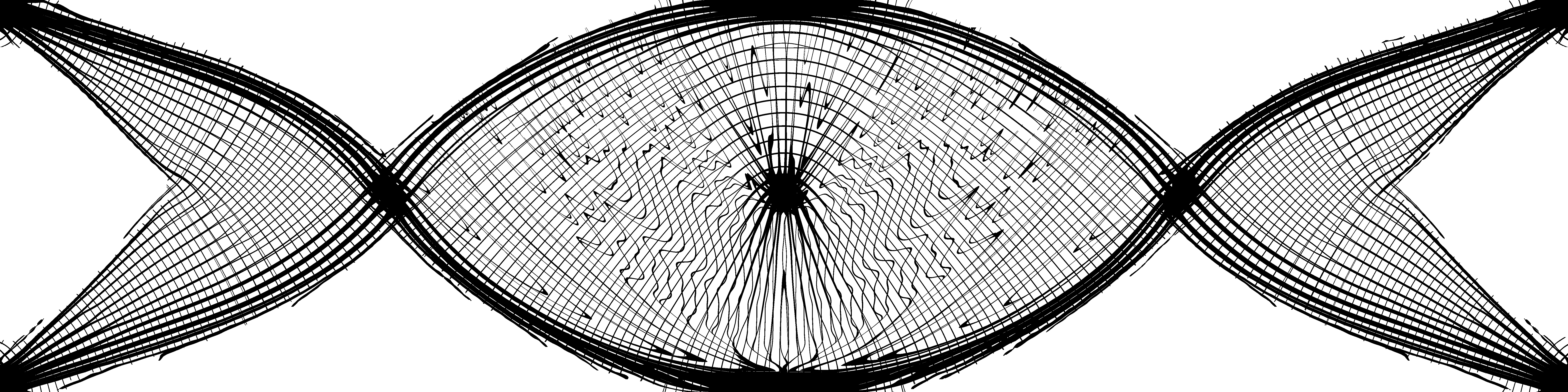}
{(b) The challenge of singularities for conventional global methods, generated by the projection-based approach from \citealt{GroenFork2019}.}
 \end{center}
\end{minipage}
\vfill
 \begin{center}
\includegraphics[width=\linewidth]{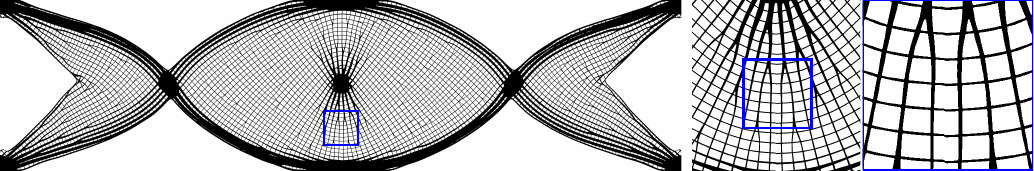}
{(c) The benefits of the locality of the phasor-based approaches, including comparison of branch closure quality.}
 \end{center}
    \caption{Illustrating how the double-clamped beam promotes additional benefits associated with locally based dehomogenisation methods like the phasor-based method compared to the conventional global methods.}
    \label{fig:double_clamped_singularity}
\end{figure}
\autoref{fig:double_clamped_singularity} presents the double clamped beam instance with minimum relative thickness $\mu_{min}=0.1$ and volume fraction constraint $V=0.25$ presented in \citealt{Elingaardetal2022}, where the streamlines corresponding to the orientations of the homogenised solution is illustrated in (a). The streamline-plot indicates the singularity located at the fully solid region in the centre of the structure, where the load is applied, and how this causes neighbouring elements to have orientations with 180 degree rotation between them. This instance will therefore exemplify the effect of the phase-alignment measures introduced to align kernels with opposing directions. The singularity in the middle poses a challenge for projection-based methods like that presented in \citealt{GroenFork2019} due to the least squares problem being solved attempting to obtain a globally smooth solution to construct the intermediate field required for such methods. The structural consequences without additional measures taken can be observed in (b), which was the motivation for the mixed-integer heuristic proposed in \citealt{Stutzetal2020}. The corresponding dehomogenised solution obtained by the phasor-based method is presented in (c), where it is evident that the effects of the singularity is reduced to a small local shift preventing perfect alignment along a central vertical line of 180 degree rotation, but lamination orientations and structural connectivity is maintained.

\begin{table}[H]
\caption{Comparing the results obtained by the phasor-based dehomogenisation method to those obtained by the CNN for the double-clamped beam from \citealt{Elingaardetal2022}. The best compliance-volume ratio and the shortest dehomogenisation time is highlighted for each case.}
\label{tab:elin_double_clamped}
\resizebox{\textwidth}{!}{
\begin{tabular}{ccc|cc|ccccc|ccccccc}
\toprule
\multicolumn{3}{c|}{Homogenisation} & \multicolumn{2}{c|}{Fine scale} & \multicolumn{5}{c|}{\citealt{Elingaardetal2022}}                                         & \multicolumn{7}{c}{Phasor}\\ \midrule
$\mu_{min}$ & $V_{hom}$ & $C_{hom}$ & $h_f$       & $\varepsilon_f$   & $\varepsilon_i$ & $V_{Elin}$ & $C_{Elin}$ & $\mathcal{R}_{Elin}$ & $T_{Elin}$ & $(i_{up}^1,i_{up}^2,f_{up})$ & $\omega$ & $V_{phasor}$ & $C_{phasor}$ & $\mathcal{R}_{phasor}$ & $T_{phasor}$ & $\bar{\mathcal{R}}_{phasor}$ \\ \midrule
0.05        & 0.2510    & 25.89     & 1/24$h_c$   & 60$h_f$           & 20$h_i$         & 0.2590     & 31.35      & 1.2492               & \textbf{7.69}       & (4, 16, 24)                  & 80       & 0.2690       & 26.65        & \textbf{1.1033}                 & 9.33         & 0.8830                       \\ 
0.10        & 0.2538    & 27.11     & 1/24$h_c$   & 60$h_f$           & 20$h_i$         & 0.2619     & 32.51      & 1.2369               & 7.12       & (4, 8, 24)                   & 80       & 0.2767       & 27.12        & \textbf{1.0908 }                & \textbf{6.38}         & 0.8814                       \\ \midrule
0.05        & 0.2510    & 25.89     & 1/40$h_c$   & 50$h_f$           & 10$h_i$         & 0.2568     & 31.50      & 1.2446               & \textbf{15.80}      & (8, 32, 40)                  & 160      & 0.2599       & 27.73        & \textbf{1.1091 }                & 19.24        & 0.8910                       \\ 
0.10        & 0.2538    & 27.11     & 1/40$h_c$   & 50$h_f$           & 10$h_i$         & 0.2574     & 31.12      & 1.1657               & 14.70      & (8, 16, 40)                  & 160      & 0.2619       & 28.05        & \textbf{1.0678}                 & \textbf{11.97}        & 0.9172 \\\bottomrule
\end{tabular}}
\end{table}
The CNN-based method is also localised in nature, and thus circumvents the negative impact of the mentioned singularities. \autoref{tab:elin_double_clamped} provides a comparison of the dehomogenised designs obtained by this CNN to the phasor-based method, for different instances of the double-clamped beam distinguished by the minimal relative thickness and the dehomogenised periodicity. The best solution in terms of compliance-volume ration and the shortest dehomogenisation time is highlighted in bold for each instance. The phasor-based approach provides significantly better dehomogenised designs in each cases, even with the potential of a small shift causing partial disconnections along the 180 degree orientation change. It is further observed that the CNN-based method is more computationally efficient for the two instances with minimum relative thickness $\mu_{min}=0.05$, while the phasor-based approach is faster for the two instances with the larger minimum relative thickness of $\mu_{min}=0.10$. The dehomogenisation time of the CNN-based method is mostly dependant upon the periodicity and thus, due to the network being trained to produce fixed-periodicity intermediate fields, the required resolution of the intermediate and fine meshes. For the phasor-based method the resolution of the intermediate meshes are also the main factors in determining the running time, but, as the periodicity on the intermediate field is controllable, the minimal thickness in combination with the periodicity determines the intermediate upscaling factors $i_{up}^1$ and $i_{up}^2$. The required resolutions of the intermediate meshes, which largely determines the dehomogenisation time required for the phasor-based method, are therefore smaller for larger minimal thicknesses at the same periodicity, explaining why the phasor-based method is faster for the instances with $\mu_{min}=0.10$.


\subsection{Periodicity Convergence} \label{sec:periodicity_convergence}
The homogenisation formulation used for optimisation assumes infinite periodicity. Therefore it is expected that the dehomogenised solution will be better for decreasing wavelengths. To investigate this effect for the phasor-based method the same homogenised optimisation solution will be considered for different wavelengths but equal $f_{up}$ to ensure fair comparison of compliance. 

\subsubsection{Two-load wheel and dehomogenisation limitations}\label{sec:periodicity_convergence_wheel}
\begin{figure}[!htb]
\begin{center}
\includegraphics[width=0.95\linewidth]{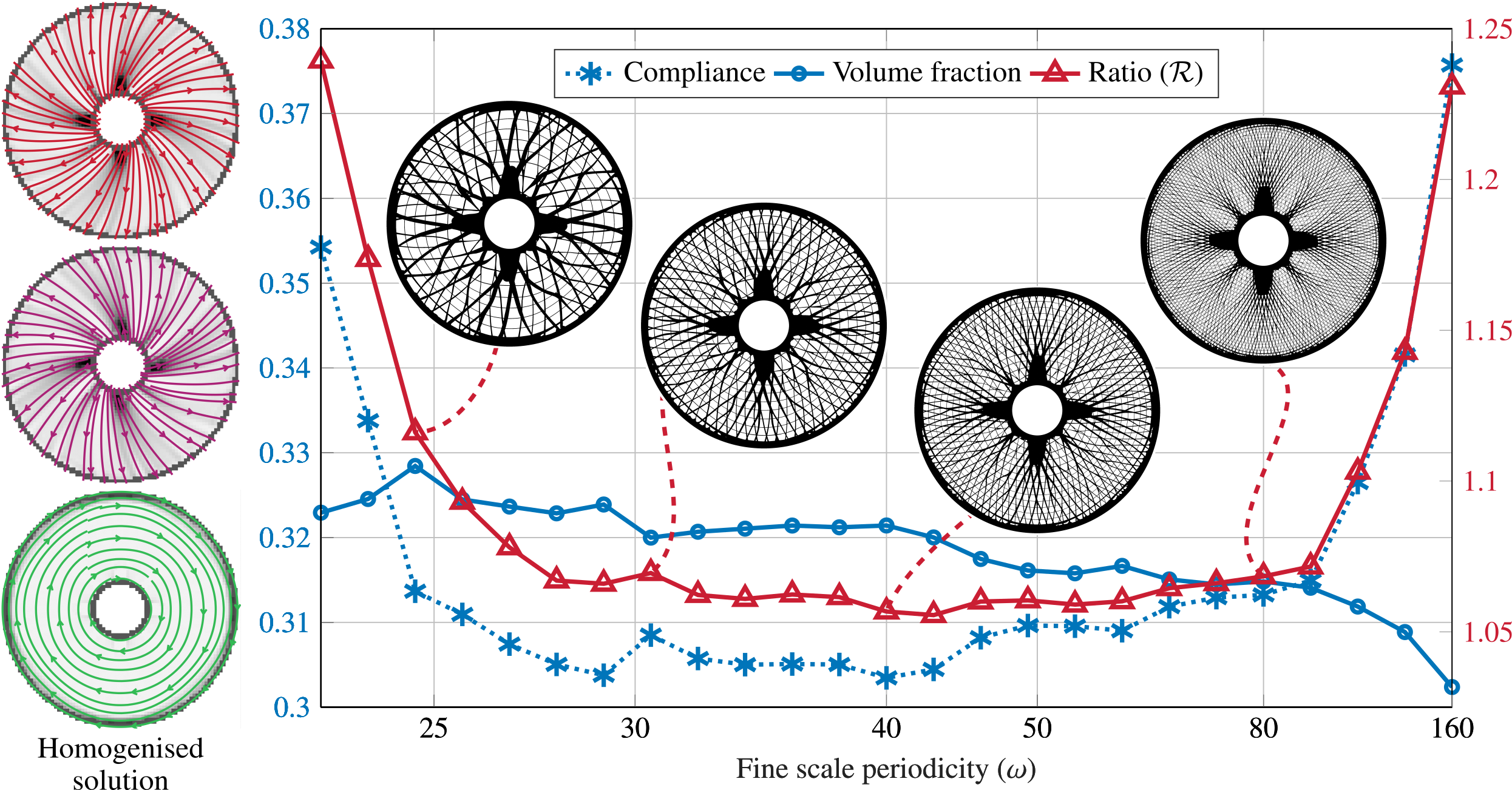}
\caption{\textbf{Left:} The homogenised solution from \citealt{JensenGroen2022} using the $\textit{SG}_A$ starting guess is illustrated by separately considering the relative thicknesses and orientations for each of the three lamination layers. The coarse scale (80x80 elements) homogenised solution has a compliance of 0.308 and volume fraction of 0.300. \textbf{Right:} Illustrating how the compliance value and volume fraction  (left axis) as well as the volume-compliance ratio compared to the fine-scale homogenised solution (right axis) depends on increasing periodicity, measured by the phasor field frequency $\omega$, for the two-load wheel example.}
\label{fig:periodicity_convergence_wheel_lines}
\end{center}
\end{figure}

The first test case considered for the periodicity convergence is the two load wheel presented in \citealt{JensenGroen2022}, where the boundary of the inner circle is prescribed clamped boundary conditions while the outer circular boundary is loaded by distributed pressure and torsion loads. Considering the rank-3 homogenised solution obtained by starting from the approximated principal orientation field starting guess denoted $\textit{SG}_A$, the three independent layers are represented in the left \autoref{fig:periodicity_convergence_wheel_lines}. For testing the periodicity convergence of the phasor-based dehomogenisation procedure for the two-load wheel a fixed fine resolution upscaling factor $F_{up}{=}40$ is considered, while $i_{up}^1$ and $i_{up}^2$ are updated according to \autoref{eq:iup1} and \autoref{eq:iup2} with strict equality. Upscaling according to $F_{up}$ provides a fine-resolution homogenised solution with compliance 0.308 and volume fraction 0.300, which will be used to benchmark the homogenised designs.

\autoref{fig:periodicity_convergence_wheel_lines} illustrates the results from dehomogenising the two-load wheel with decreasing period, where the performance deteriorates at high values of $\omega$, due to meshing and disconnection problems. It can be observed that for very short or very long wavelengths the performance diminishes significantly, but for $\omega\in[24,80]$ periodicity convergence is observed. Within this interval, the volume fraction of the dehomogenised design approaches the desired volume fraction of the homogenised solution while maintaining a low compliance, ensuring improvements in terms of the volume-compliance ratio.

\begin{figure}[!htb]
    \centering
\includegraphics[width=0.95\linewidth]{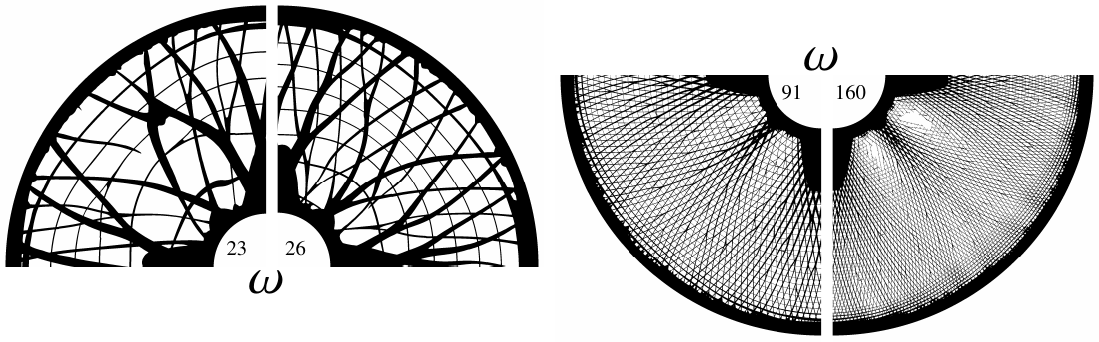}
\caption{Illustrating a selection of dehomogenised wheel designs from the range of periodicities tested in \autoref{fig:periodicity_convergence_wheel_lines}. The extraction of longest and shortest wave-lengths show the limitations of the phasor-based method firstly in terms of very low periodicity being unstable in terms of periodic consistency, and secondly how a very high periodicity relative to the minimal thickness and projected mesh resolutions results in structural disconnections due to insufficient resolution.}
\label{fig:periodicity_wheel_selected_structs}
\end{figure}

\autoref{fig:periodicity_wheel_selected_structs} focuses on a part of the dehomogenised solution for some selected periodicities outside of the reliable periodicity interval. For very long wavelenghts, or equivalently very low periodicities (i.e. $\omega{=}23$), the dehomogenised design is not able to adhere to the prescribed periodicity and exhibits undesirable curvature and structural disconnections. This is related to the relation between the underlying phasor kernel bandwidth and frequency. If the wavelength becomes relatively long compared to the size of the Gaussian kernel controlling the kernel's signal magnitude, the frequency control of the sampled response is known to decline (\citealt{Tricard_orientable_2020}). For the increased periodicity $\omega{=}26$ the frequency control is increased, but still not strong enough to ensure sufficiently consistent periodicity in all layers, which is also the reason for the branch connection procedure failing to connect one of the branches. The structural performance is however still significantly improved compared to the lower periodicity. When the periodicity increases beyond the recommended limits, and becomes very high, the thinnest structural members will start to disconnect and disappear due to the inability to capture such fine detail in the phasor field at these resolutions for a fixed mesh. As long as the considered periodicity is within the reliable range, dehomogenised designs with approximately constant local periodicity are achieved.

\subsection{Cantilever and phase shift sensitivity}
As an extension to the periodicity convergence test, the cantilever A4 in table \autoref{tab:cantilever_elingaard_instances} is considered for both increasing periodicity and intermediate wave-field phase shift. Adding a constant $\bar{\varphi}$ such that $\phi(\boldsymbol{x}):=\phi(\boldsymbol{x})+\bar{\varphi}$ allows for shifting the wave-field profile in a $2\pi$-periodic manner. Due to the nature of the branching points being stationary regardless of a phase shift this allows for extracting alternative solutions from the same sampled complex wave-field and thus the same underlying phase-aligned set of phasor kernels. As the default phase shift of the resulting structure is different depending on the number of alignment iterations, this test also provides a baseline idea of how sensitive the solution is to the fixed number of iterations used in this sub-procedure.

For the purpose of this test, a fixed upscaling factor $F_{up}=40$ will be considered, and the correspondingly upscaled homogenised solution as the benchmark for the compliance-volume ratio measure. For each tested periodicity, within the reliability limits of the phasor method, the dehomogenisation is performed considering $\pi/4$ evenly spaced phase-shifts $\bar{\varphi}\in[0,\;2\pi]$.

\begin{figure}[!htb]
\begin{center}
\includegraphics[width=0.9\linewidth]{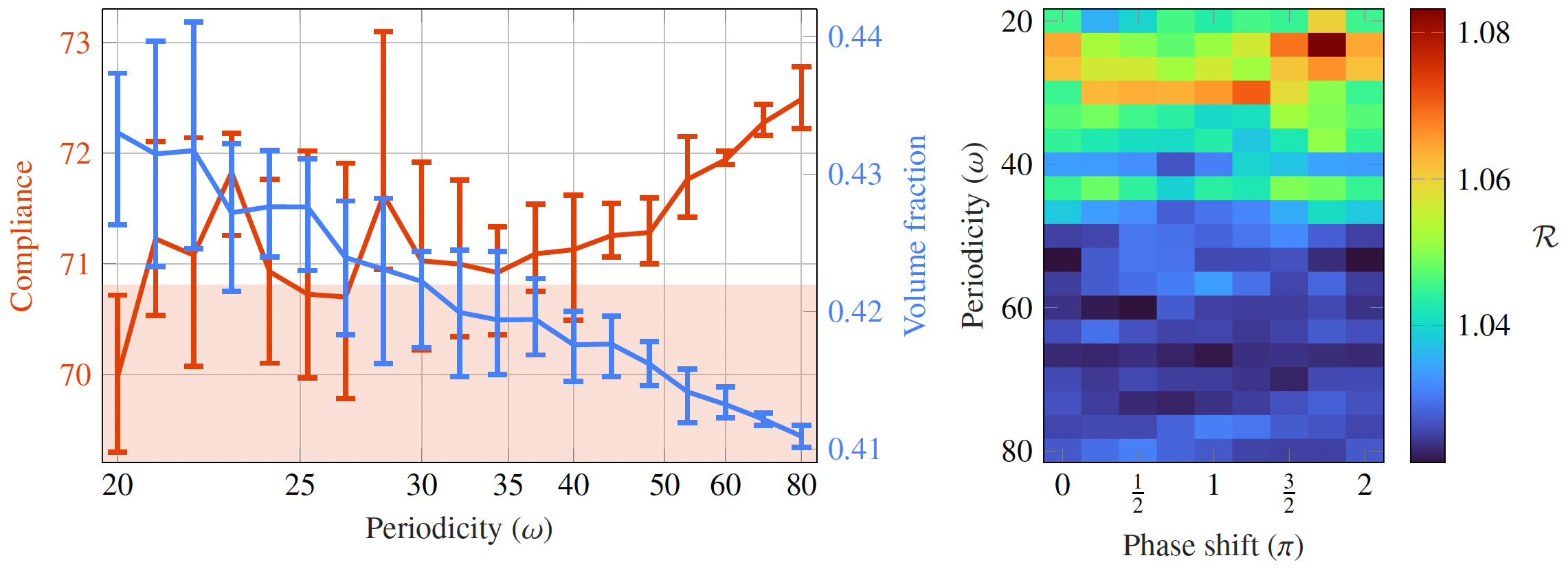}
\caption{Left plot shows how the mean compliance and volume fraction across $\pi/4$ evenly spaced phase-shifts $\bar{\varphi}\in[0,\;2\pi[$, with phase-shift dependent variation given by the errorbars, increases with increasing wavelength. The compliance of the homogenised solution is indicated by the shaded area. The right heatmap shows the individual compliance-volume ratios as a function of periodicity and phase shift relative to the upscaled homogenised solution ($C_{fine}{=}70.81$,  $V_{fine}{=}0.4088$).}
\label{fig:cant_periodicity_pshift_convergence_w_heatmap}
\end{center}
\end{figure}

\autoref{fig:cant_periodicity_pshift_convergence_w_heatmap} summarises the results from the described series of phase shift tests. The left plot illustrates the compliance and volume-fraction convergence for increasing periodicity. The error bars indicate the span across all phase shifts at each periodicity, while the curve outlines the average trend. The right heatmap presents the compliance-volume ratio for each dehomogenised structure, identified by its phase shift and periodicity. Firstly, it can be observed that the variance in compliance and volume fraction is significantly reduced as the periodicity increases. As the volume fraction approaches the desired value from above, while the compliance has a tendency to increase as the periodicity increases, the ratio-heatmap is included to clarify that the overall structural performance is indeed improving for increased periodicity, while the variance in performance is decreased. 

\begin{figure}[!htb]
\begin{center}
\includegraphics[width=0.8\linewidth]{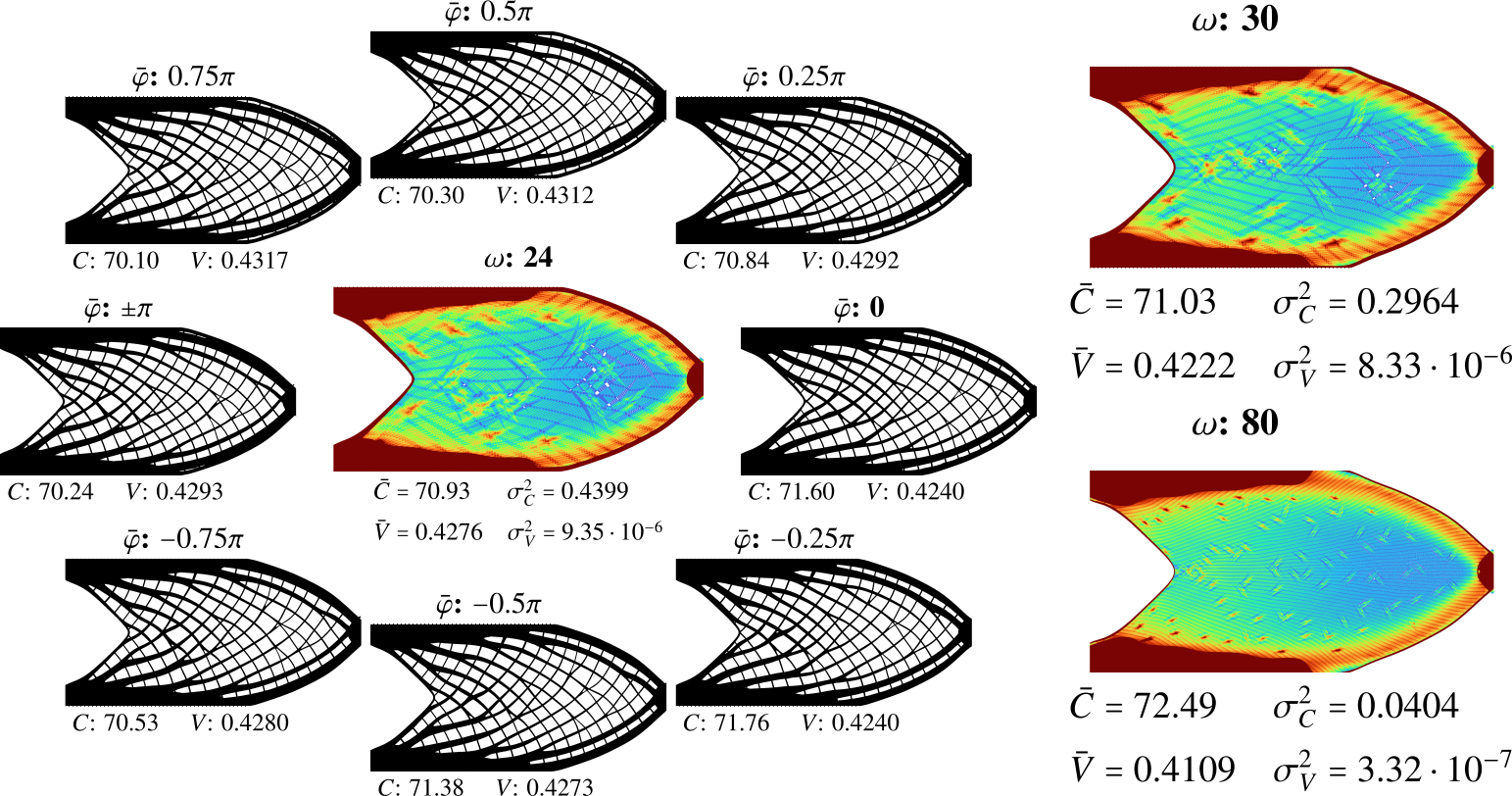}
\caption{Illustrating how the constant phase-shift affects the final dehomogenised structure for increasing periodicity. The phase-shift is $2\pi$-periodic and the central plot in each case illustrates the union of the phase-shifted structures indicating the unaffected solid-regions and structural boundary as well as how the adherence to periodicity vanishes locally near a branching point.
Higher periodicity reduces the size of the branching regions as well as the variation in structural design and performance.}
\label{fig:eps130_phase_shift_circle}
\end{center}
\end{figure}

\autoref{fig:eps130_phase_shift_circle} illustrates (left) the dehomogenised structures including compliance and volume fraction for each considered global phase shift, $\bar{\varphi}$, for the periodicity $\omega=24$. The central illustration represents the union of all the phase-shifted structures, with the mean and variance of their compliance and volume fraction values. Dark red represents areas where all structures in the circle are solid, whereas light green represents where few-to-none of the structures are solid. Similar illustrations for the periodicities $\omega\in\{30,80\}$ are given to the right. It is evident that the largest overlaps occur in the fully solid regions of the structural domain while there is little overlap in regions where the relative thickness is low or a branching occurs. These branching regions are also found where the structure locally does not cohere to the constant periodicity, thus it is expected that they will cause discontinuities in the otherwise smoothly varying overlap-field. The smoothness of this field improves while the size of the branching regions decreases with increased periodicity, which is closely tied to the decrease in variance of the solution quality of the phase-shifted structures. The structural performance is more sensitive to global phase-shifts if the periodicity is low. These results further underline the observations made for the effect of varying number of alignment iterations in \autoref{sec:elin_cantilever_results}, which is also expected to be reduced as the periodicity increases due to the global phase-shift stagnation of the phase-alignment procedure. 
\subsection{On-the-fly dehomogenisation} \label{sec:iteration_history}
The computational efficiency of the phasor-based method promotes the option of obtaining on-the-fly dehomogenisation during the homogenisation based optimisation. A potential gain from this would be to perform real-time multi-scale TO in a manner similar to the popular interactive single-scale TopOpt Apps (\citealt{AageNobel2013}). 

To facilitate this, it is necessary to investigate the ability of the phasor-based method to appropriately dehomogenise the intermediate solution at different stages throughout the iteration process. The aim of the dehomogenisation process is not in itself to optimise the design, but translate the homogenised solution in a coherent manner. This means that for any iteration during the optimisations process the ideal dehomogenisation procedure would produce a design that has approximately the same structural performance as the intermediate homogenised design. Therefore, a Rank-2 cantilever beam is considered for minimum compliance optimisation. The coarse homogenised mesh considered has $60\times30$ elements, the volume fraction constraint is set to $V{=}0.35$, the minimum relative thickness to $\mu_{min}=0.1$, and a density filter radius of $R{=}1.2$ is deployed. The intermediate design in each iteration of the optimisation process is dehomogenised for a periodicity $\omega{=}48$ and evaluated on a fine resolution $1440\times720$, corresponding to an upscaling factor of $F_{up}{=}24$, within the reliable bounds of the phasor-based method for the current minimal thickness and periodicity. 

\begin{figure}[!htb]
\begin{center}
\includegraphics[width=0.9\linewidth]{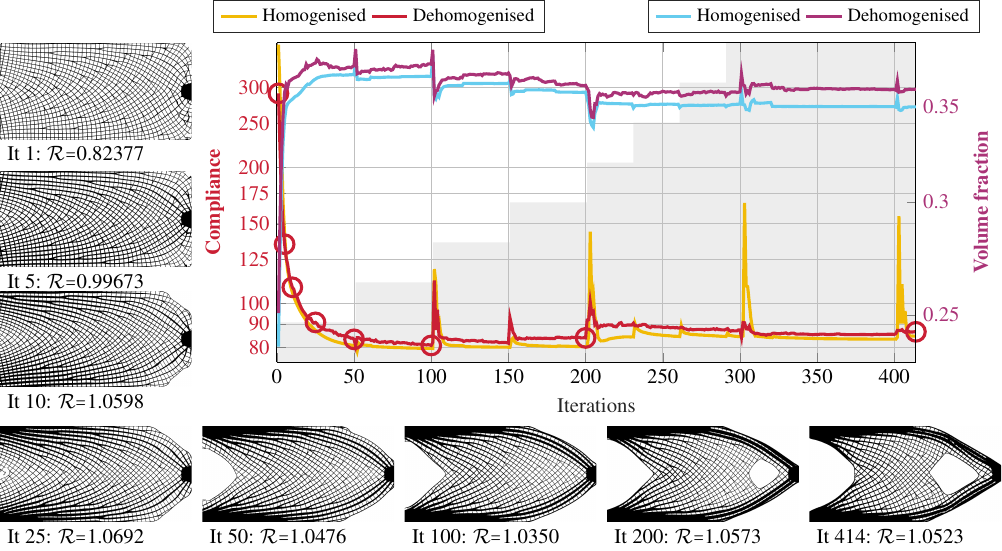}
\caption{Illustrating iteration-wise compliance (left axis) and volume fraction (right axis) convergence for the dehomogenised design compared to the coarse-scale homogenised structure being optimised. The filled area indicates the beta continuation during the optimisation process, where the value is doubled at each step, which causes a local disturbance in the convergence sequence. The markers indicate from when in the iteration process the dehomogenised structures illustrated are extracted for further investigation. These show how the structure changes during iterations and the ratio to the homogenised solution is specified in each case. Note that log-scale is utilised to magnify the differences in the converging designs.}
\label{fig:opt_conv_cantilever}
\end{center}
\end{figure}

\autoref{fig:opt_conv_cantilever} shows the iteration history with respect to the compliance (left axis) and volume fraction (right axis) for both the homogenised and dehomogenised design representations. The circled compliance values highlight the iterations from which the included structures are extracted. The convergence of the homogenised and dehomogenised design formulations follow the same overall trend during optimisation, with the dehomogenised design typically having a slightly higher volume fraction and compliance, which is to be expected when it is attempting to translate the current homogenised design. There is a clear relation between the beta-continuation and the largest compliance deviations, while the volume deviation is more stable throughout. It is interesting to note that the later beta-continuation updates having a major impact on the quality of the homogenised design, have little effect on the dehomogenised ones. The iterations just after these continuation steps are also those where the performance of the dehomogenised design is the furthest away from the homogenised. This is however not a concern, as it is known that the beta-continuation causes disturbances in the optimisation process, and thus these intermediate solutions can be discarded as not being reasonable homogenised designs.


The changes in the homogenised solution diminishes quickly during the iteration process and the updates after 100 iterations incur small changes in especially compliance but also volume fraction. It is observed that the compliance increases slightly towards the end of the optimisation process, but this is to be expected as the increase in $\beta$ removes thicknesses in $]0,\;\mu_{min}[$, ensuring that the design adheres to the minimal relative thickness requirement, and as a result that the volume fraction is sufficiently reduced to adhere to the volume fraction constraint. For the dehomogenised design the trend during these last iterations is different, in that the same reduction in volume fraction is not observed, while the relative increase in compliance is less significant. The overall performance in terms of compliance-volume ratio is however still improved. Overall, the dehomogenisation procedure translates the homogenised solution with only a few percentage loss in structural performance throughout the iteration process. 

\subsection{Multiload cases} \label{sec:additional_tests}
\begin{figure}[H]
    \centering
    \includegraphics[width=0.9\linewidth]{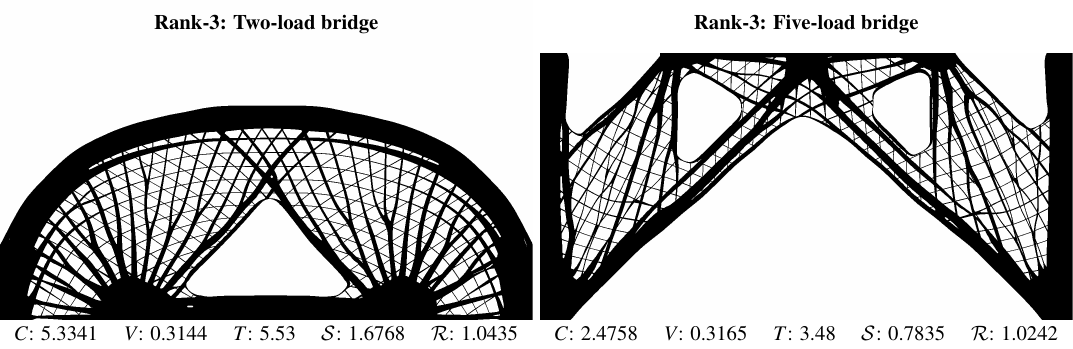}
    \caption{Presenting the results from the phasor-based dehomogenisation procedure when applied to the homogenised solutions obtained for the two-load (left) and five-load (right) bridge from \citealt{JensenGroen2022} using starting guess $SG_{A}$ in both cases. The results in terms of compliance ($C$), volume fraction ($V$), dehomogenisation time ($T$), the compliance-volume product ($\mathcal{S}$) and compliance-volume ratio to the upscaled homogenised solution ($\mathcal{R}$) are reported for each case.}
   \label{fig:multiload_short}
\end{figure}

Considering the multi-load optimisation instances presented, optimised for different starting guesses and dehomogenised using projection-based methods in \citealt{JensenGroen2022}, the phasor-based method has already proven valuable for the two-load wheel example, as seen in \autoref{sec:periodicity_convergence_wheel}. For the remaining two cases, the two-load and five-load bridge, the phasor-based dehomogenised designs of the best-performing optimised results using the $SG_{A}$ starting guess are presented in \autoref{fig:multiload_short} to further validate the flexibility and reliability of the framework.

\subsection{Body-fitted mesh}\label{sec:bodyfitted}
Throughout the presented numerical tests, the dehomogenised designs have all been extracted to a fine resolution structured grid before analysis. The pixel-based representation is, however, not well suited for the purpose of geometry extraction needed for manufacturing preparations and does not model local stress distributions well. A more flexible alternative is to extract the dehomogenised design, by exploiting the implicit nature of the underlying wave-field representations, in terms of its contour-lines. This representation defines the structure in terms of piecewise smooth curves along its solid-void boundary. Doing so, allows for importing the structure to a CAD-tool like COMSOL and construct a body-fitted mesh.

\noindent\begin{minipage}{0.485\textwidth}
    \begin{figure}[H]
    \centering
    \includegraphics[width=\linewidth]{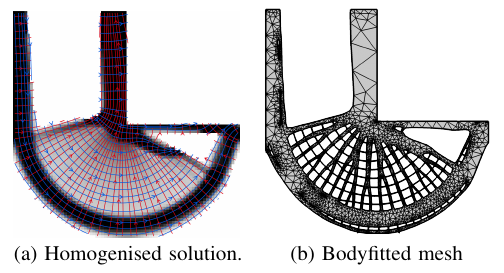}
    \caption{Illustrating the homogenised solution of an Rank-2 optimised L-beam instance on a 80x80-element mesh in terms of the combined density field with lamination orientations represented by streamlines (a). The extracted geometry of the corresponding dehomogenised design obtained by the phasor-based method with periodicity $\omega=26$ is imported to COMSOL to illustrate a body-fitted mesh with triangular elements (b) containing $6.2\cdot 10^4$ degrees of freedom.}
    \label{fig:body_fitted_example}
\end{figure}
\end{minipage}
\hfill
\begin{minipage}{0.485\textwidth}
\begin{table}[H]
\centering
\caption{Comparing the analysis results obtained when sampling the dehomogenised solution on structured fixed grids and extracting the geometry for body-fitted meshes for different mesh-resolutions specified in terms of degrees-of-freedom in the FE-analysis. The comparison represents both how a better performing structure is obtained when the smooth boundaries are obtained as well as how significantly fewer degrees-of-freedom are needed for a body-fitted mesh to converge.}
\label{tab:bodyfitted_comsol}
\resizebox{\textwidth}{!}{
\begin{tabular}{cc|ccc}
\toprule
Mesh & DOFs & Compliance & Volume & $\mathcal{S}$\\ \midrule
\multirowcell{2}[0pt][l]{Structured Q4} & 6.78e6 & 77.69 & 0.3237 & 25.15\\ 
 & 2.71e7 & 73.64 & 0.3237 & 23.84\\ 
& 6.10e7 & 70.92 & 0.3237  & 22.96\\\midrule 
\multirowcell{6}[0pt][l]{Body-fitted\\triangular} & 6.23e4 & 70.84 & 0.3193 & 22.62\\
&8.53e4 & 70.98 & 0.3192 & 22.66\\
&3.00e5 & 71.05 & 0.3190 & 22.66\\
&4.61e5 & 71.07 & 0.3190 & 22.67\\
&5.92e6 & 71.09& 0.3190& 22.68\\\bottomrule
\end{tabular}}
\end{table}
\end{minipage}
\vspace{2mm}

The body-fitted mesh allows for a smoother representation of structural boundaries requiring fewer degrees-of-freedom. Typically, as a result from this, a more accurate analysis result can be found using fewer computational resources. \autoref{fig:body_fitted_example} and \autoref{tab:bodyfitted_comsol} provide examples of these effects. Note that due to small structural holes in the structure there are regions of concentrated smaller elements within larger structural members in the presented mesh. The geometry extraction and unstructured mesh utilisation reduces the resolution dependency of the analysis accuracy as well as the structural quality, compared to the density-based representation. An additional potential benefit of this implicit extraction is the reduction in volume fraction deviation compared to the homogenised solution. Although some deviation is expected, due to the choice of a finite periodicity and having to add material for closing branches, the restrictions of sampling to a finite structured mesh is also of great influence. If infinitely fine resolution was allowed, the deviation is expected to be reduced significantly, approaching a negligible level, but this is computationally intractable. Extracting the implicit geometry offers a more realistically obtainable alternative to this concept. This motivates moving away from sampling to a fixed grid when implicit geometry representations are available, like for dehomogenised solutions obtained by the phasor-based method.

\subsection{Computational time} \label{sec:time_test}
From the tests performed for the cantilever problem instances in \autoref{tab:cantilever_elingaard_instances} it is evident that the computational time increases with the size of the coarse mesh and the upscaling factor. However, the associated increase is not consistent across instances of the same resolutions and periodicities, as can be seen by the large variation between the running times within each of the A, B and C parent categories in \autoref{tab:results_cantilever_Elin_Groen}. It is also found that the number of alignment iterations has little impact on the overall dehomogenisation time. Within each parent category there is a trend of the time required decreasing with an increasing minimum relative thickness. This could be related to several key aspects of the phasor-based method. Firstly, a smaller relative minimal thickness requires finer resolutions for the intermediate meshes, both for sampling ($i_{up}^1$) and for connecting the branches ($i_{up}^2$). These subprocedures present the most demanding subprocedures of the method, where especially the sampling (unless parallelised) is detrimental to the overall time. Secondly, for the cantilever examples, the instances with a greater relative minimal thickness combined with lower volume fractions have fewer intermediate density elements on the coarse mesh to translate. One phasor kernel is constructed for each of these intermediate elements. As they are treated in sequence in both the construction and alignment of kernels, as well as the intermediate sampling, the number of phasor kernels has a large influence on the required computational time.

As such, factors likely to contribute to longer solution times have been identified. For the cantilever example it can be verified that there is a strong linear relation between the time required and the upscaling factor used for the intermediate sampling grid multiplied by the number of active kernels, i.e. the number of coarse mesh intermediate-density elements. The number of phase alignment iterations has a negligible effect on this linear relation. 


To quantify the running time, a synthetic dehomogenisation test case is considered where all coarse-mesh elements are of intermediate density. This removes the effect of the number of active kernels from any comparison, which is preferable as this size is very problem instance and optimisation settings dependent. Instead, the aim is here to establish an idea of the computational complexity of the phasor-based method based on more general problem characteristics. The tests will be focused on the case of one lamination layer, such that the indicated complexity sizes correspond to the theoretical limits from \autoref{app:computational_complexity}. The implication is then that the computational complexity is to be multiplied by the number of layers for more complex cases. It should however be noted that, at least in cases where the meshes utilised in each layer are consistent, there are potential time-savings available, but to limit the degree of specialised assumptions and allow for a representative time-estimate in cases where the dehomogenisation of the separate layers is parallelised this effect is not considered.

\begin{figure}[!htb]
        \centering
    \includegraphics[width=0.75\linewidth]{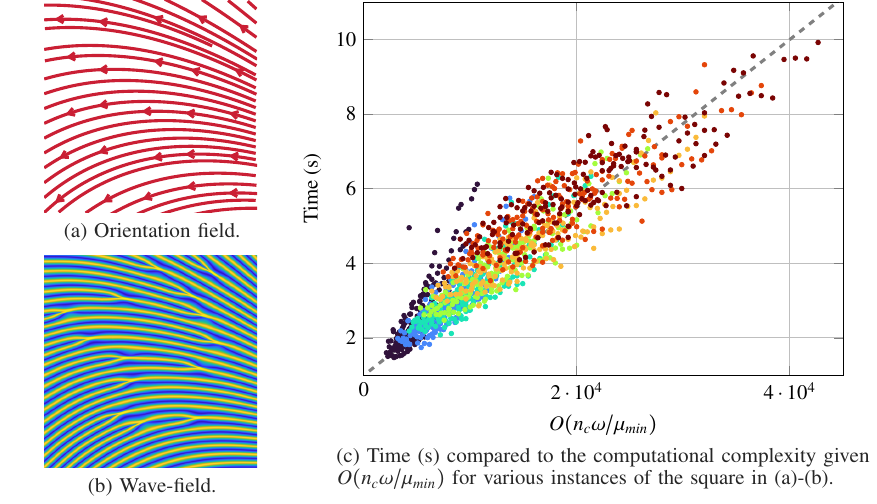}
   \caption{An illustration of the dehomogenisation problem considered for testing the computational complexity of the phasor-based dehomogenisation method. The problem instance is described by a Rank-1 laminate on a square domain with uniform relative thickness equal to the minimal thickness set and orientations as illustrated in (a). The synthetic definition allows for varying the size of the coarse mesh, the minimum relative thickness and the periodicity. The dehomogenised triangular wave field with connected branches of an example instance is shown in (b), and the time-complexity of the dehomogenisation process for different instances of the square problem is given in (c).}
    \label{fig:time_square_test}
\end{figure}

\autoref{fig:time_square_test} illustrates the nature of the problem being dehomogenised in the considered test cases in terms of the orientation field (a) and an example dehomogenised patch (b). As an initial indication of the computational complexity, the dehomogenisation time considering a varying coarse mesh discretisation of $n_c\times n_c$ elements, minimum thickness $\mu_{min}$, periodicity $\omega$ and the minimal feature resolution $h_{min}$ which controls the $f_{up}$ upscaling factor. Note that in relation to the theoretical bounds determined in \autoref{app:computational_complexity}, $N_c=n_c^2$ for the considered test cases. The time as a function of the complexity $O(n_c\omega/\mu_{min})$ (right) verifies a strong linear relation. The minimal feature resolution has little impact on the overall computation time when varied from $h_{min}\in\{3,...,12\}$. This is because the more computationally expensive procedures are conducted at intermediate resolution before the final upscaling to ensure the desired final resolution, such that only the final interpolations and thresholding operators are performed on the finest scale. If the minimal feature resolution was to be increased to much higher values this conclusion is likely to change, but as bodyfitted meshes are better suited for design-extraction for future manufacturing needs, a very large resolution pixelised design is nonsensical.

A more detailed test of the dehomogenisation time, considering also the individual computation times of the different sub-procedures is covered in \autoref{app:time_test}. Overall the phasor field sampling constitutes the most expensive subprocedure even when performed locally at the coarser of the two intermediate resolutions. This procedure is therefore also the main contributor to the overall complexity of the procedure $O(n_c^2(i_{up}^1+i_{up}^2))$. Thus, the computational time is determined by the number of coarse mesh elements, the chosen periodicity and the minimal relative thickness, and grows sublinearly with the number of fine-scale elements.

\section{Conclusion} \label{sec:conclusion_main}
This work has proposed a methodology utilising phasor noise for efficient 2D dehomogenisation. The main novelties introduced are: the anisotropic kernel definition, an efficient and mechanically sound branch connection procedure and the varying thickness boundary generation and smoothing method. Extensive numerical tests have established that the phasor-based approach achieves speed-ups comparable to the CNN-based method from \citealt{Elingaard2021} without any significant compromise in solution quality, compared to the more conventional projection-based methods. The method is localised by definition, which aids in overcoming the most severe consequences of singularities in the orientation field observed for the global projection methods. The underlying continuous nature of the intermediate field and the branch connection operator allow for more consistent branch connections than the post-processing procedure proposed for the CNN. Scalability has been established for increases in problem size and complexity, and the localised structure of the different subprocedures makes the procedure highly parallelisable. The main computational steps in the procedure are based on floating-point operations, making also GPU-implementations an accelerating option. This combined with the capabilities of the method in terms of translating intermediate homogenised designs during the optimisation process, allows for efficient on-the-fly dehomogenisation for visual investigations and interactive design processes.

The phasor-based dehomogenisation has been implemented and tested for fixed grid problems in 2D, exclusively. It is, however, worth noting that the methodology in itself is inherently mesh independent as all operators are discretisations of underlying continuous definitions. This means that the concept should be readily translatable to unstructured meshes or field-based representations. Due to the local control and blending properties of the phasor noise, the concepts can also be translated to work for applications requiring spatial grading of different periodicities as well as orientations. 

Associated with the proposed heuristic are some challenges related to the nature of the orientation fields provided. Firstly, connecting disconnections or fully disjointed wave-fields occurring for large and abrupt directional changes will require additional mechanisms dealing with the associated types of phasor field singularities, but such occurrences are unlikely for well-posed and optimised problems. Secondly, the adaptions made to the phase-alignment procedure significantly improves the structural connectivity in cases of 180 degree orientation changes, but does not currently guarantee perfectly smooth transitions. Especially prudent for these occurrences is the limitation imposed by the sampling filter which relies on a smoothly interpolated orientation field, which may be challenging to obtain due to the periodic behaviour. Lastly, ordered orientation fields have been considered a prerequisite for the proposed method. Such smooth orientation fields will not always be easily achievable, and thus mechanisms to circumvent the disturbances associated with non-ordered orientation fields are beneficial extensions. 

The phasor framework in itself is directly extendable to 3D, but the nature of branches and alignment may introduce new challenges for this transition. As it is highly beneficial to transition from fixed grid to unstructured meshes when considering 3D cases, it is also recommended to consider how the underlying continuous nature of the 2D procedure can be exploited. It is also in 3D where handling of non-ordered orientation fields will become especially important.  







\section*{Declarations}
\noindent \textbf{Acknowledgements:} Homogenised optimised results for comparing to the CNN in \citealt{Elingaardetal2022} are obtained from the associated Git-repository. All other optimisation results are provided by Peter D.L. Jensen\orcidauthorD or obtained by optimisation codes from \citealt{JensenGroen2022}. 

\noindent \textbf{Conflict of interest:} The authors state that there is no conflict of interest.

\noindent \textbf{Funding:} The authors acknowledge funding from the Villum Fonden through the Villum Investigator Project ``InnoTop''. 

\noindent\textbf{Replication of results:} The results presented in this review can be reproduced by the provided descriptions of the experiments. For homogenised optimised results and a description of the optimisation implementations the reader is referred to \citealt{Elingaardetal2022} and \citealt{JensenGroen2022}. 

\normalsize
\begingroup
\raggedright

\bibliography{referencesm.bib}
\endgroup

\newpage
\appendix
\renewcommand{\thesection}{\Alph{section}}
\section{Parameter selection}\label{app:parameter_select}
\numberwithin{equation}{section}
\setcounter{equation}{0}
\renewcommand{\theequation}{(\Alph{section}.\arabic{equation})}
\numberwithin{figure}{section}
\setcounter{figure}{0}
\renewcommand{\thefigure}{\Alph{section}.\arabic{figure}}
\numberwithin{table}{section}
\setcounter{table}{0}
\renewcommand{\thetable}{\Alph{section}.\arabic{table}}

This section elaborated on the parameter settings chosen in \autoref{tab:Parameter_choices}, as well as the upscaling factor requirements for the intermediate and fine-resolution meshes. The implications of the resolutions on the computational expenses of the subprocedures are addressed in a theoretical manner, highlighting how localisation of operators allows for better scalability of the method in \autoref{app:computational_complexity}.

\subsection{Phase alignment}
A default isotropic alignment neighbourhood is defined by the radius $R=2h_c$ for all kernels $j\in \mathcal{K}$. This setting ensures that the neighbourhood is kept small for efficiency, but allows fir increasing the degree of anisotropy of the neighbourhood while still ensuring alignment with at least one neighbour in the direction orthogonal to the direction of anisotropy, i.e. orthogonal to the lamination direction. Ensuring this is important as to not risk neighbouring kernels with isolated phase shifts. As the alignment propagates through the entire set of kernels, there is no particular gain in increasing the neighbourhood size.

The degree of anisotropy is chosen based on the number of active kernels within the neighbourhood radius as described in \autoref{sec:phase_align}, such that the weights defining the anisotropy are given $(r_1,\;r_2)_j=(\frac{1}{\pi},\; \pi)(1-\Delta\tilde{\kappa}_j)+\Delta\tilde{\kappa}_j$ for each kernel $j\in \mathcal{K}$. When $\Delta\tilde{\kappa}_j\rightarrow0$ the degree of anisotropy is such that the alignment neighbourhood spans $\pm 1$ kernels in the direction orthogonal to the direction of anisotropy.

In most cases a fixed number of 20 alignment iterations are sufficient, but some tests with differing numbers will be conducted to investigate the effect of alignment, especially in terms of size of the coarse-scale homogenised solution. For future studies it would be relevant to determine a suitable convergence criteria for this method for better choosing problem instance dependant settings. The phases of the coarse scale kernels are aligned in increasing order of $\bar{\Delta}x^l_j$, which is defined in \autoref{eq:alignment_order}. This definition corresponds to the distance of the kernels centre to the nearest boundary kernel that is smoothly aligned with the orientation of the current lamination layer. Currently, the breakeven measure in case of equal distances is simply the element number in an ordered mesh.

\begin{equation}\label{eq:alignment_order}
    \bar{\Delta}x^l_j=\min_{b\in \bar{\mathcal{B}}}\left\{||(x_j^0)^l-x_b^0||\right\},\;\;\bar{\mathcal{B}}=\left\{b\in \mathcal{B}: \left|\cos \tilde{\tau}_b\cos\theta_j^l+\sin\tilde{\tau}_b\sin\theta_j^l\right|\geq 0.95 \;\land\; \min_{k\in \mathcal{N}^{3\times3}}\left\{\cos \tilde{\tau}_b\cos\tilde{\tau}_k+\sin\tilde{\tau}_b\sin\tilde{\tau}_k\right\}\geq0  \right\}
\end{equation}

\subsection{Phasor field sampling}
For the phasor field sampling the bandwidths of the signal-localising Gaussians are set to $\beta=\omega/h_c$ for all kernels. The kernel size, in terms of the radius of impact of its signal, should be large enough to capture the structural feature size determined by the desired periodicity, but also small enough for the sampling to capture local changes in the underlying orientation field. Note that imposing this value will increase the lower limit for periodicities, in terms of $\omega$, where the periodic control of the phasor method diminishes. Tests, presented in \autoref{sec:periodicity_convergence_wheel}, do however indicate a reasonable interval of reliability for the choice of periodicity.\\

Considering the bandwidth of the integrated sampling filter, preliminary tests show that $\alpha=\beta$ is a consistent and reliable choice, while it is beneficial to choose a degree of anisotropy corresponding to the phase-alignment base case given $(r_1,\;r_2)=(1/\pi,\;\pi)$ for all sampled kernels.

\subsection{Branch connection}
The shape of the branch closure should span the maximal width of the branching region in the orthogonal direction ($r_2$) while the structural performance benefits from an elongated effect along the lamination direction ($r_1$). Therefore, the anisotropy in the branch connection procedure is given $(r_1,\;r_2)=(1/\pi,\;1)$. Smaller values of $r_1$ could be beneficial in cases of perfectly straight lines, while a more isotropic shape could perform better for cases with more rapid variation in the orientation field near the branching point. Tests have, however, shown that the chosen value is reasonably robust to directional changes, as it keeps the branch closure region sufficiently local to the branching points, while improving the branch shape by elongation along the lamination direction.\\

For the pinch procedure a fixed number of $k_{max}=3$ pinch-steps is chosen in each case. One could here consider adapting the number of steps to the local relative thickness, but in the proposed procedure the choice was made to include this defining metric as a weight to control the degree of pinching instead.

\subsection{Structural boundary}
The bandwidths of the kernels and sampling filter in the phasor-field sampling for structural boundary generation are chosen based on the same principals as for the general sampling procedure. Due to the change in periodicity to adapt the wavelength to the boundary region these bandwidths are here defined by $\beta=\alpha=\bar{\omega}/h_c$.\\

\begin{equation}\label{eq:boundary_r1}
r_1= \left(0.5+\left(\min_{k\in \mathcal{N}^{3\times3}}\left\{\cos \tilde{\tau}_b\cos\tilde{\tau}_k+\sin\tilde{\tau}_b\sin\tilde{\tau}_k\right\}\geq 0.95\right)\right)^{-1}
\end{equation}
The degree of anisotropy of a kernel signal is adjusted depending on the variation in neighbouring kernel orientations. For neighbouring boundary kernels with highly aligned directions the degree of anisotropy is increased by defining $r_1$ as in \autoref{eq:boundary_r1}. In cases of larger local direction variations near a kernel, a the impact radius is reduced and a small degree of anisotropy in the direction orthogonal to the kernel's direction is introduced by defining $r_2=\min\{1/r_1,\;1\}$.

\subsection{Intermediate mesh resolution}
The choice of intermediate mesh $\mathcal{T}_i$ depends on the minimum thickness $\mu_{min}$, and the desired wavelength on the fine mesh $\varepsilon_f$ or the periodicity on the fine mesh $\omega{=}\frac{1}{\varepsilon_f}$. The triangular density field sampled on the intermediate mesh, $\rho=\frac{1}{\pi}\arcsin(\sin(\phi(\boldsymbol{x}))),\; \boldsymbol{x}\in \mathcal{T}_i$, represents a discretisation of a smooth triangular wave-field. The resolution of this discretisation determines the tolerances for how well the sampled field represents the underlying continuous field. To ensure structural connectivity after thresholding it is crucial that the local maximal values of the density field ensure that $\rho\geq1-\mu_{min}$ results in a connected structure where the thinnest structural members are constituted by at least a few elements.\\

Let $h_i$ denote the element size on the intermediate mesh, then $\hat{\omega}=\frac{1}{h_i\omega}$ is the number of pixels covering a wavelength on the intermediate field. Half of a wavelength on the triangular field covers the full interval [0,1] once. Thus, the best expected representation of the continuous wave field given the intermediate mesh discretisation is given by a step-size $\frac{2}{\hat{\omega}}$. This means that the local maxima of a wave takes on a value in the interval $]1-\frac{2}{\hat{\omega}},\;1]$. For structural connectivity given the minimal relative thickness $\mu_{min}$ it is required that the local maximal values on the intermediate field lies in the interval $[1-\mu_{min},\;1]$. To ensure this the requirement that $\frac{2}{\hat{\omega}}\geq \mu_min$ is imposed, which implies that $h_i\leq\frac{\mu_{min}}{2\omega}$. Therefore, the intermediate mesh $\mathcal{T}_i$ should ensure a upscaling factor $i_{up}$ from the coarse mesh, with element size $h_c$, satisfying
\begin{equation}
    i_{up}\geq \frac{2h_c\omega}{\mu_{min}}
\end{equation}
As such, the size of the intermediate mesh increases with higher periodicity and smaller minimal relative thickness. However, as the most extensive operations on the intermediate mesh are based on local image morphology operations about the branching points, the scaling of the computational cost associated with the size of $i_{up}$ remains limited.\\

However, even if the local morphological operations associated with closing the branches offer good scalability, the sampling of the phasor-field on the intermediate mesh is less scalable, and thus the computational cost of sampling will increase greatly with $i_{up}$. To overcome this, it is recommended to utilise two different intermediate scales, one for sampling the phasor-field and one for connecting the branches. Relative to the coarse mesh these scales are defined by the upscaling factors $i_{up}^1$ and $i_{up}^2$ respectively, satisfying

\begin{equation}\label{eq:intermediate_resolution_limits}
    i_{up}^1\geq \frac{h_c\omega}{0.1} \; \land\;i_{up}^2\geq \frac{2h_c\omega}{\mu_{min}}
\end{equation}

It is found that $i_{up}^1$ is a sufficient upscaling factor to sample the phasor field given that the further upsamling to $i_{up}^2$ is conducted on the complex wave field $\mathcal{G}(\boldsymbol{x})$ which is then transformed to the phasor field $\phi(\boldsymbol{x})$ after upsampling.

\subsection{Fine mesh resolution (fixed grid)}
The relation between the intermediate mesh resolution given by the upscaling factor $i^2_{up}$ and the final fine mesh resolution is determined in a similar way as from \citealt{Elingaardetal2022}. To ensure that the minimal features in the structure are resolved by at least three pixels, $h_{min}=3$, when the triangular field is upsampled from intermediate mesh $\mathcal{T}_i^2$ to fine mesh $\mathcal{T}_f$ the upscaling factor between the meshes $f_{up}$ should satisfy $f_{up}\geq \frac{h_{min}}{\hat{\omega}\mu_{min}}$ and based on the recommendations for the intermediate mesh resolutions one has that $f_{up}\geq \dfrac{h_{min}h_{i^2}\omega}{\mu_{min}} \; \land \; \left(h_{i^2}= \dfrac{h_c}{i^2_{up}}\; \Leftrightarrow\; h_{i^2}\leq \dfrac{\mu_{min}}{2\omega}\right)$ such that if $i_{up}^2$ is chosen exactly, $f_{up}\geq \dfrac{h_{min}}{2}$ is sufficient. For future convenience, let $F_{up}=f_{up}i_{up}^2$ denote the full upscaling factor from the coarse to final fine resolution.

\section{Computational efficiency} \label{app:computational_complexity}
The theoretical computational complexity of the different subprocedures in the dehomogenisation procedure will here be addressed in an per lamination layer approach. Numerical tests to show how these derivations are realised and contribute to the overall dehomogenisation time will be further elaborated on in \autoref{sec:time_test}.

\subsection{Phase alignment}
The computational cost of the phase alignment procedure is inherently low, and becomes negligible compared to the other subprocedures involved in the phasor-based dehomogenisation method. Only the phase shifts of the kernels, $\varphi_j,\;j\in \mathcal{K}$, changes during one iteration of the phase-alignment procedure (\autoref{alg:phase_alignment}). The other factors involved in computing these updates are based on fixed distances and directions, meaning they can be precomputed and re-used in each iteration. One iteration of phase-alignment therefore corresponds to computing, for each active kernel $j\in \mathcal{K}$, its new phase shift as the weighted average of its neighbours, with weight, neighbours and alignment order determined in advance. The most computationally expensive part of the phase-alignment is the mentioned precomputation of necessary measures, where each kernel is traversed to determine its neighbourhood, the weight of the impact from these neighbours as well as the distance required for determining the phase-alignment order. As the neighbourhood radius $R$ in \autoref{tab:Parameter_choices} is defined as a function of the coarse resolution element size $h_c$ the location of neighbours can be limited to investigating a fixed number of nearby kernels regardless of coarse resolution. The cost of the precomputations are as such constant per active phasor kernel. As the number of kernels is directly tied to the coarse resolution elements, the computational complexity of the phase-alignment precomputations is $O(N_c)$, with $N_c$ denoting the number of coarse resolution elements. Given a fixed number of alignment iterations, the phase-alignment procedure can also be considered to be of constant complexity per phasor kernel, effectively meaning the computational complexity of the alignment procedure as a whole, with precomputations, can be described by $O(N_c)$. This means the alignment can be executed at a relatively low computational cost only increasing with the size of the coarse scale homogenised solution, such that the exact number of alignment iterations has a low impact on the overall running time. 

\subsection{Sampling}
Sampling is the most computationally expensive component of the phasor-based dehomogenisation procedure, but is in itself considered efficient due to its procedural nature. The response of the set of phasor kernels at any point in space can be computed independently in constant time. The signal at this given point is based on the average contribution from the coarse mesh phasor kernels. For each sampling point at the intermediate resolution given $i_{up}^1$ the contribution from each phasor kernel is summed, resulting in a worst case computational complexity of $O((N_ci_{up}^1)^2)$. Considering the case where \autoref{eq:intermediate_resolution_limits} is satisfied with equality, this is equivalent to $O(N_c\omega^2)$.\\

As the kernels emits signals with a localised response, the procedure can be made more efficient by only considering the response from kernels with impact radius covering the sampling point. This is exemplified by the impact set $\mathcal{T}_i^j$ defined for each kernel $j\in \mathcal{K}$ in the proposed procedure. The kernel bandwidth is defined as a function of the coarse-resolution element size and periodicity (\autoref{tab:Parameter_choices}) such that for lower periodicity, or equivalently longer wave-lengths, and finer coarse scale resolution the spatial impact radius of each kernel is reduced. Considering the prescribed bandwidth and $i_{up}^1=\frac{h_c\omega}{0.1}$ the size of the sampling space $\mathcal{T}_i$ for a given kernel can be described by complexity $O(\omega h_c)$, meaning the computational complexity of the sampling procedure is reduced to $O(N_c\omega)$.


\subsection{Closing branches}
The current branch closure procedure is defined based on the sampled field at the finer intermediate resolution ($i_{up}^2$). Due to the location of the branching points being directly accessible by localising the singular points where $|\mathcal{G}|\rightarrow 0$ on the sampled field, and the branch-closure being executed by only considering the field locally near these points, the associated computational cost is kept limited. The cost of this procedure will still depend upon the intermediate resolution and the number of branching points. The number of branching points, considering the same underlying orientation field, depends upon the periodicity, where higher periodicity induces more disconnections in the intermediate sampled field. A higher periodicity also requires a finer intermediate resolution to capture sufficient detail in the intermediate field.\\

Locating the branching points can be achieved by searching for the sampling points where $|\mathcal{G}|\rightarrow0$. Performing this search at the sampled intermediate scale given $i_{up}^1$ implies a computational complexity of $O(N_c(i_{up}^1)^2)$ for searching all sampling points. The determination of the closure direction, phase shift to branch closure and following pinch procedure can all be performed by considering only a small area around each branching point separately. The size of the localised area considered, in terms of sampling points covered, is dependant upon the periodicity $\omega$ and the second intermediate resolution given the coarse resolution $N_c$ and the upscaling factor $i_{up}^2$. For a sufficiently smooth orientation field $\omega^2$ may serve as a conservative upper bound on the number of possible branching points, while the necessary size of the localised area considered near each branching point is of complexity $O(\hat{\omega}^2)$ where $\hat{\omega}$ is the number of pixels covering a wavelength on the $i_{up}^2$ intermediate field. Utilising equality in \autoref{eq:intermediate_resolution_limits} to decide $i_{up}^2$, the combined computational complexity of determining direction, solidifying and pinching the branches becomes $O((\omega/\mu_{min})^2)$.

\subsection{Adding structural boundary}
The structural boundary is constructed by considering the coarse resolution homogenised solution directly, and the set of phasor kernels used for later sampling the boundary is fully determined by the coarse elements along the structural boundary. Thus, the boundary is defined by a relatively small set of phasor kernels from which the sampled phasor field can be directly transformed and thresholded to obtain the desired variable thickness boundary. Constructing and sampling the boundary corresponds to a more restricted version of the combined phase alignment and sampling procedure such that the computational complexity of the sampling, $O(N_c\omega)$, is a conservative upper bound for this procedure. The most costly components of the boundary construction becomes the upscaling to the fine resolution and subsequent thresholding operations. For the finest scale, and the leading in terms of computational complexity, these operations can be bounded by $O(N_cf_{up}^2)$ which for equality in the choice of upscaling factors is equivalent to $O((\omega/\mu_{min})^2)$.

\subsection{Additional subprocesses}
In addition to the main procedural components of the proposed phasor-based dehomogenisation method grid-related operations such as linear interpolation, translation and thresholding for the different upscaling steps are crucial to extract the information needed from one subprocess to the next. The computational complexity of these processes is determined directly from the fine scale resolution as $O(N_cf_{up}^2)$, which with exact choice of $i_{up}^2$ is equivalent to $O((\omega/\mu_{min})^2)$.

\section{Supplementary results} \label{app:supp_res}
\begin{figure}[!htb]
    \centering
    \includegraphics[width=0.85\linewidth]{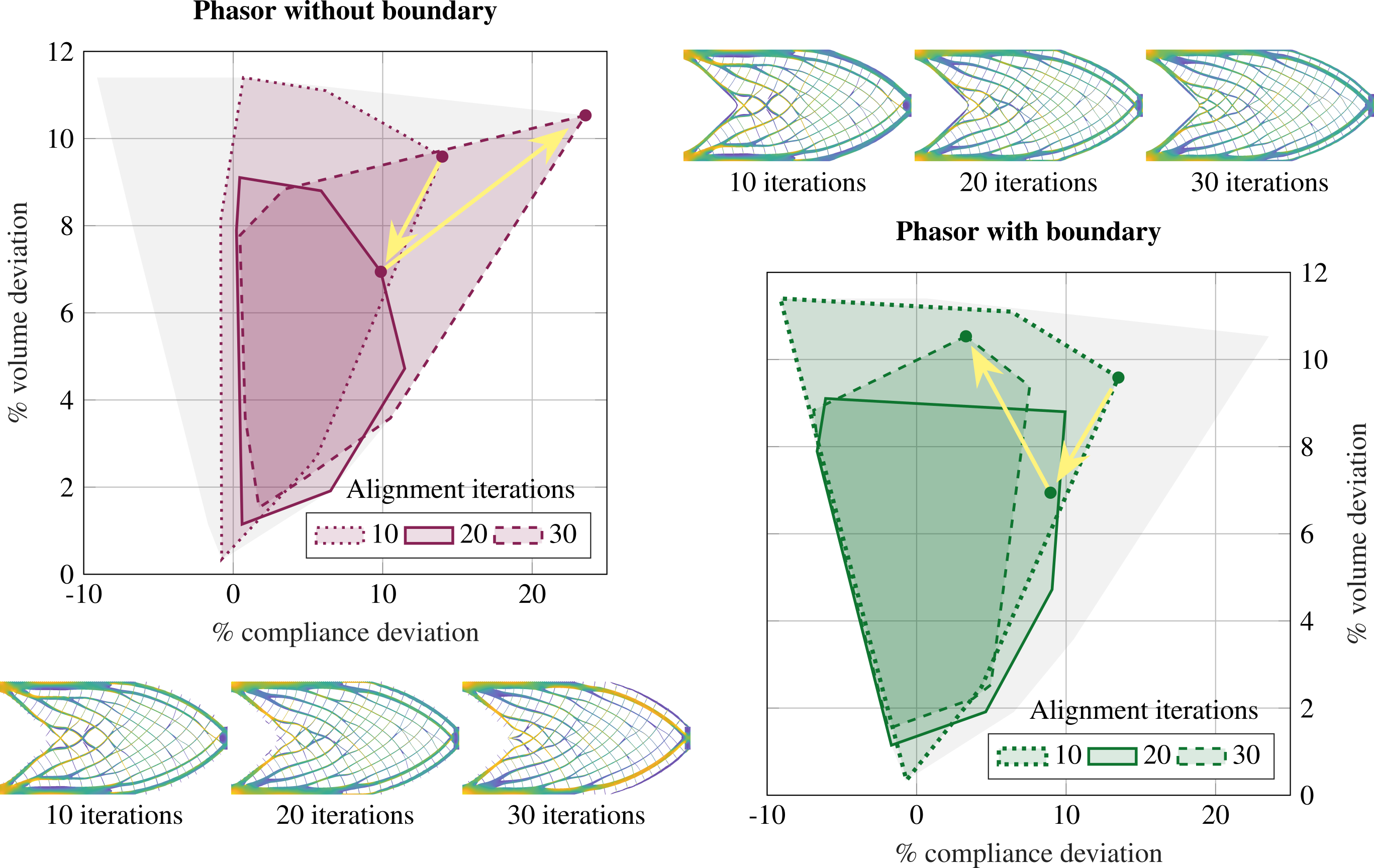}
    \caption{Illustrating the effect of the structural boundary for varying alignment iterations in terms of volume fraction and compliance deviation relative to the homogenised solution. The area plots indicate where in the compliance-volume deviation space for the cantilever instances in \autoref{tab:cantilever_elingaard_instances} the solutions obtained by the phasor approach (with or without boundary) for 10, 20 and 230 alignment iterations are located. The versions with and without added boundary are illustrated separately with the alignment iteration distinction represented by patches. The underlying grey patch illustrates the entire space covered by the phasor method for all boundary and alignment iteration options tested. The plot marks indicate the location of test instance A1 for each of these combinations, and the corresponding structures are illustrated top right (without boundary) and bottom left (with boundary). This is the instance with the largest variance in structural performance when the number of alignment iterations are increased. }
    \label{fig:wwo_boundary_alignment_iter}
\end{figure}

\autoref{fig:wwo_boundary_alignment_iter} illustrates these mentioned effects of varying the alignment iterations and adding the structural boundary. It can be seen that for both the cases with and without the added structural boundary, the variance in compliance and volume fraction deviation for the considered cantilever instances (\autoref{tab:cantilever_elingaard_instances}) is reduced when increasing from 10 to 20 alignment iterations. With the further increase to 30 alignment iterations it is evident that the variance in solution quality for the cases without added boundary increases again, while the same trend is not observed for the cases with this boundary. The A1 test instance is included for each dehomogenisation version to exemplify these trends. One can observe that for only 10 alignment iterations the underlying phasor field does not sufficiently adhere to the periodicity and underlying lamination orientations, reflected by the branch connection procedure failing to ensure closure for two of the branches. After 20 iterations connection is ensured and one can observe that the outer bars of the structures are aligned well with the location of the structural boundary, explaining why the performance is similar for both cases with and without the added boundary. After 30 iterations the phasor fields have been subjected to a global phase shift moving these outer bars into the void domain, such that their carrying capacity is diminished in the case without the added structural boundary. These results thus motivate running at least 20 alignment iterations, indicate that increasing the number of alignment iterations is not guaranteed to improve the quality of the solution, and that the structural boundary decreases the effect of the number of alignment iterations after a stable combination of local phase shifts is found. The impact of the added boundary is largest for structures with lower periodicity, many holes or vary narrow structural members. Adding the structural boundary constitutes about 1-3\% of the running time for the presented instances. Due to the fact that information from the boundary construction process is utilised in the phase-alignment procedure and is crucial to the presented performance of the method, the cost adding of the boundary itself is very limited. It should also be noted that the variance in the solution qualities is diminished with increasing periodicity, an effect which will be further investigated in \autoref{sec:periodicity_convergence}.\\

\begin{table}[htb!]
\centering
\caption{Comparing dehomogenisation time and solution quality of the proposed phasor-based method with added structural boundary for different number of phase alignment iterations for the Cantilever \citealt{Elingaardetal2022}. $V^t_{phasor}$, $C^t_{phasor}$, $\mathcal{\bar{R}}^t_{phasor}$ and $T^t_{phasor}$ denote the volume fraction, compliance, volume-compliance ratio to the CNN-based solution and dehomogenisation time for the phasor-based method using $t$ iterations of phase alignment, respectively.}
\label{tab:comp_cantilever_Elin_maxit_boundary}
\resizebox{0.95\textwidth}{!}{
\begin{tabular}{c|cccc|cccc|cccc}
\toprule
ID & $V_{phasor}^{10}$ & $C_{phasor}^{10}$ & $\mathcal{\bar{R}}_{phasor}^{10}$ & $T_{phasor}^{10}$ & $V_{phasor}^{20}$ & $C_{phasor}^{20}$ & $\mathcal{\bar{R}}_{phasor}^{20}$ & $T_{phasor}^{20}$ & $V_{phasor}^{30}$ & $C_{phasor}^{30}$ & $\mathcal{\bar{R}}_{phasor}^{30}$ & $T_{phasor}^{30}$ \\ \midrule
A1 & 0.2778            & 120.55            & 0.8867                            & 1.82              & 0.2711            & 115.72            & 0.8305                            & 1.56              & 0.2802            & 109.71            & \textbf{0.8137}  & 1.69              \\
A2 & 0.4312            & 72.11             & 0.9545                            & 1.39              & 0.4214            & 74.80             & 0.9675                            & 1.41              & 0.4269            & 71.77             & \textbf{0.9405}  & 1.47              \\
A3 & 0.2853            & 120.81            & 0.7741                            & 1.04              & 0.2794            & 124.90            & 0.7837                            & 1.10              & 0.2809            & 122.20            & \textbf{0.7710}  & 1.19              \\
A4 & 0.4397            & 69.92             & 0.9498                            & 1.00              & 0.4289            & 70.86             & \textbf{0.9389}  & 1.05              & 0.4301            & 70.66             & \textbf{0.9389}  & 1.15              \\
A5 & 0.2912            & 111.66            & \textbf{0.8619}  & 0.60              & 0.2852            & 115.38            & 0.8722                            & 0.65              & 0.2845            & 114.40            & 0.8625                            & 0.66              \\
A6 & 0.4468            & 72.51             & 0.9578                            & 0.88              & 0.4455            & 72.52             & 0.9552                            & 0.95              & 0.4473            & 71.97             & \textbf{0.9516}  & 1.01              \\ \midrule
B1 & 0.2687            & 108.62            & 0.7873                            & 3.22              & 0.2615            & 110.67            & \textbf{0.7805}  & 3.30              & 0.2678            & 109.24            & 0.7890                            & 3.33              \\
B2 & 0.4124            & 71.56             & 0.8916                            & 2.87              & 0.4101            & 71.75             & \textbf{0.8890}  & 2.97              & 0.4126            & 71.99             & 0.8974                            & 3.06              \\
B3 & 0.2664            & 118.27            & 0.8201                            & 2.09              & 0.2683            & 116.94            & \textbf{0.8168}  & 2.16              & 0.266             & 119.18            & 0.8252                            & 2.21              \\
B4 & 0.419             & 70.95             & 0.9365                            & 2.00              & 0.4171            & 70.90             & \textbf{0.9316}  & 2.07              & 0.4206            & 70.53             & 0.9344                            & 2.15              \\
B5 & 0.2623            & 122.02            & 0.8160                            & 1.27              & 0.2644            & 120.79            & \textbf{0.8143}  & 1.31              & 0.2654            & 120.53            & 0.8156                            & 1.37              \\
B6 & 0.4253            & 74.37             & 0.9455                            & 1.67              & 0.4262            & 74.16             & \textbf{0.9448}  & 1.76              & 0.4267            & 74.43             & 0.9493                            & 1.81              \\ \midrule
C1 & 0.2659            & 110.66            & 0.8723                            & 25.26             & 0.2632            & 110.53            & 0.8624                            & 26.48             & 0.2629            & 110.60            & \textbf{0.8618}  & 27.22             \\
C2 & 0.4146            & 71.20             & 0.9465                            & 24.11             & 0.4114            & 71.33             & 0.9409                            & 25.16             & 0.4112            & 71.18             & \textbf{0.9387}  & 26.05             \\
C3 & 0.2718            & 116.64            & 0.9515                            & 19.65             & 0.2692            & 114.32            & 0.9236                            & 20.81             & 0.2685            & 114.26            & \textbf{0.9208}  & 22.26             \\
C4 & 0.4224            & 70.01             & 0.9660                            & 19.74             & 0.4194            & 70.24             & 0.9623                            & 20.90             & 0.4186            & 70.18             & \textbf{0.9597}  & 22.51             \\
C5 & 0.2781            & 105.71            & 0.9396                            & 8.97              & 0.2775            & 105.66            & \textbf{0.9371}  & 9.16              & 0.2772            & 105.82            & 0.9377                            & 9.49              \\
C6 & 0.4343            & 70.83             & 0.9652                            & 16.53             & 0.4309            & 71.16             & \textbf{0.9622}  & 17.29             & 0.4303            & 71.30             & 0.9627                            & 18.51             \\ \bottomrule
\end{tabular}}
\end{table}

\begin{figure}[htb!]
\begin{minipage}{0.495\textwidth}
\begin{table}[H]
\centering
\caption{Comparing dehomogenisation time and solution quality of the proposed phasor-based method with and without added structural boundary the best performing number of alignment iterations from \autoref{tab:comp_cantilever_Elin_maxit_boundary}. $V^{no}_{phasor}$, $C^{no}_{phasor}$, and  $\mathcal{\bar{R}}^{no}_{phasor}$ denote the volume fraction, compliance and volume-compliance ratio to the CNN-based solution  for the phasor-based method without added boundary. The last column measures the improvements achieved by adding the structural boundary in terms of the relative relationship between the compliance-volume ratio to the homogenised solution for the two alternatives. The lower relative ratio is the larger the gain from adding the structural boundary is. }
\label{tab:comp_cantilever_Elin_maxit_noboundary}
\resizebox{\textwidth}{!}{
\begin{tabular}{cc|ccc|c}
\toprule
ID & Iter. & $C_{phasor}^{no}$ & $V_{phasor}^{no}$ & $\mathcal{\bar{R}}_{phasor}^{no}$ & $\mathcal{{R}}_{phasor}^{yes}/\mathcal{{R}}_{phasor}^{no}$ \\ \midrule
A1 & 30    & 131.24            & 0.2693            & 0.9358   & {0.8696}\\
A2 & 30    & 77.62             & 0.4139            & 0.9862                            & 0.9537   \\
A3 & 30    & 132.79            & 0.2754            & 0.8215                            & 0.9386\\
A4 & 20    & 71.99             & 0.4280            & 0.9520                            & 0.9862  \\
A5 & 10    & 123.70            & 0.2900            & 0.9507                            & 0.9065  \\
A6 & 30    & 76.50             & 0.4341            & 0.9818                            & 0.9694   \\ \midrule
B1 & 20    & 111.01            & 0.2653            & 0.7943                            & 0.9826  \\
B2 & 20    & 73.05             & 0.4113            & 0.9076                            & 0.9795 \\
B3 & 20    & 121.01            & 0.2680            & 0.8442                            & 0.9675   \\
B4 & 20    & 71.94             & 0.4169            & 0.9448                            & 0.9860  \\
B5 & 20    & 123.59            & 0.2802            & 0.8830                            & 0.9222   \\
B6 & 20    & 75.74             & 0.4299            & 0.9732                            & 0.9708   \\ \midrule
C1 & 30    & 117.12            & 0.2577            & 0.8949                            & 0.9630   \\
C2 & 30    & 71.92             & 0.4071            & 0.9389                            & 0.9997 \\
C3 & 30    & 117.14            & 0.2648            & 0.9310                            & 0.9891  \\
C4 & 30    & 71.12             & 0.4137            & 0.9611                            & 0.9986    \\
C5 & 20    & 113.44            & 0.2698            & 0.9784                            & {0.9578}   \\
C6 & 20    & 72.40             & 0.4296            & 0.9760                            & 0.9859   \\ \bottomrule
\end{tabular}}
\end{table}
\end{minipage}
\hfill
\begin{minipage}{0.495\textwidth}
\vspace{10mm}
    \includegraphics[width=\linewidth]{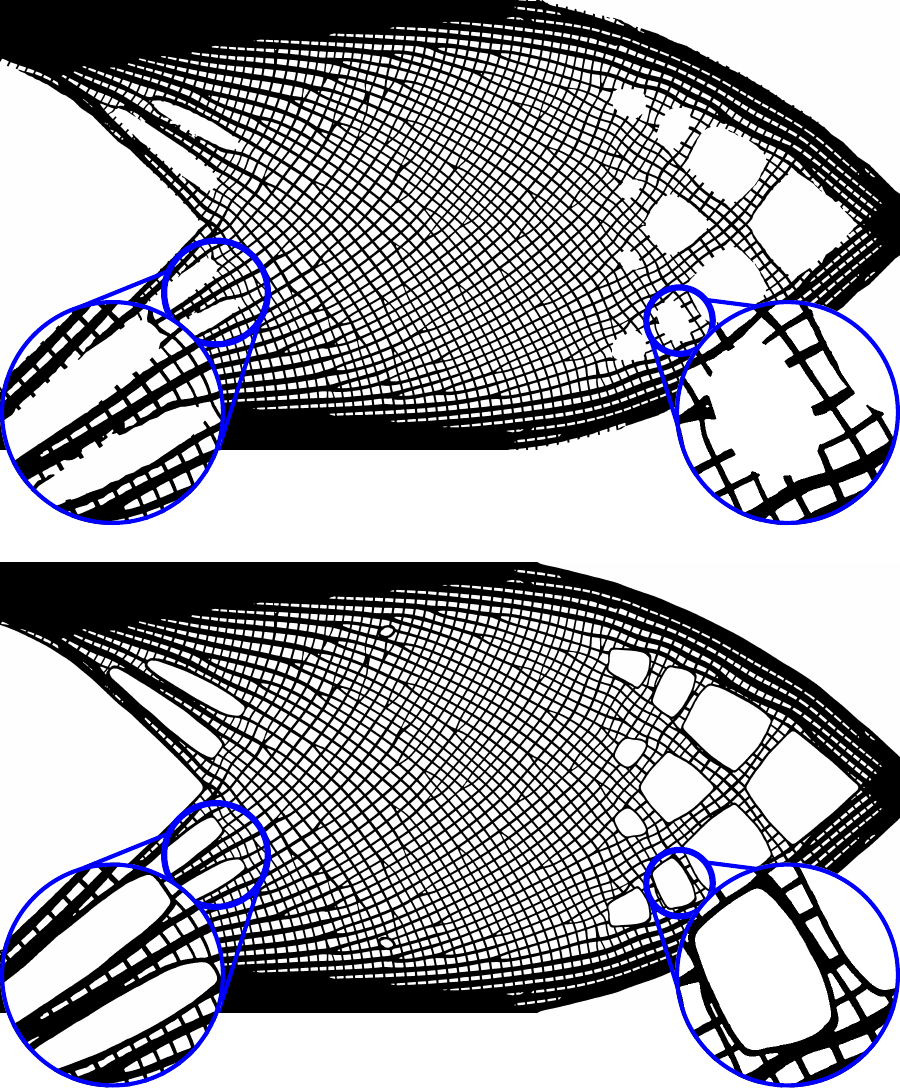}
    \caption{Illustrating the effect of adding the structural boundary to thin members.}
    \label{fig:boundary zoom}
\end{minipage}
\end{figure}
\FloatBarrier
\section{Computational time} \label{app:time_test}
\begin{figure}[!htb]
        \centering
\includegraphics[width=0.8\linewidth]{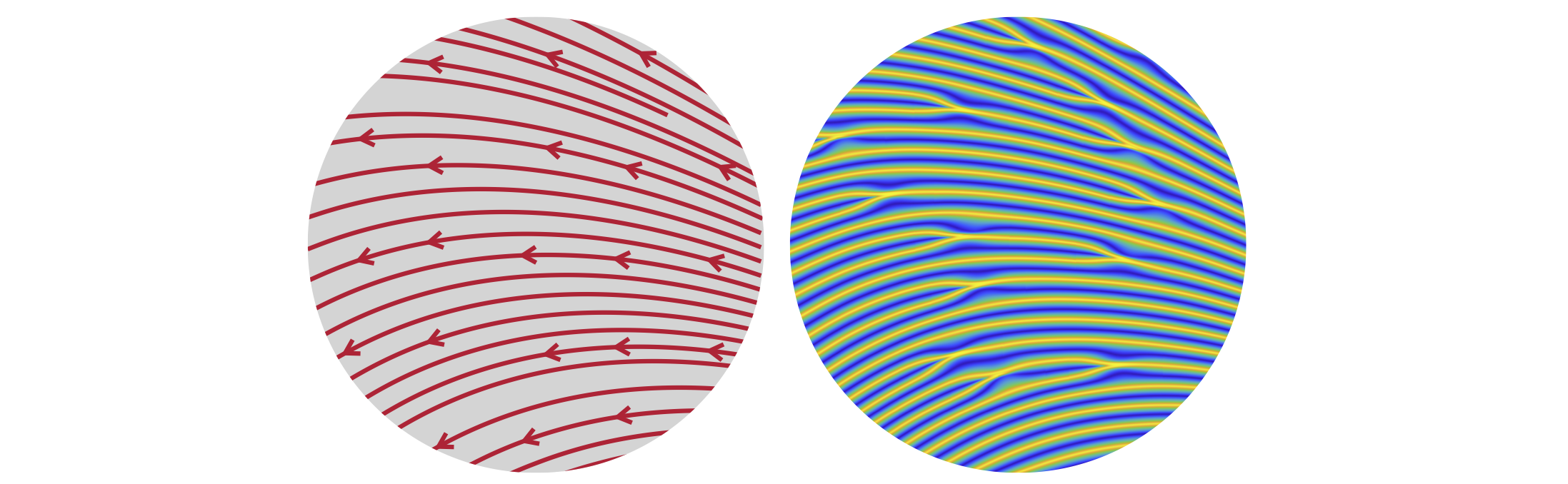}
   \caption{Illustrating the dehomogenisation problem considered for testing the computational complexity of the phasor-based method and its subprocesses (left). The problem instance is described by a Rank-1 laminate on a square domain with uniform relative thickness within the circular indicator field shown in grey and void regions outside this circle are blank. The synthetic definition allows for varying the size of the coarse mesh, the minimum relative thickness and the periodicity. The dehomogenised triangular field with connected branches of an example instance is shown to the right.}
    \label{fig:time_circle_test}
\end{figure}

A more detailed test of the dehomogenisation time, considering also the individual computation times of the different sub-procedures and how they contribute to the overall running time, is conducted considering a circular structural domain as an extension to the square test problem in \autoref{sec:time_test}. The upscaling factors from \autoref{eq:intermediate_resolution_limits} with equality and $h_{min}=3$ are considered for each dehomogenised instance. The coarse mesh resolution is varied such that $n_c\in \{[20,\;80]\;\land \mathbb{Z}\}$ and the circular structural domain is filled with uniform thickness equal to the minimal relative thickness $\mu_{min}{\in} ]0.05,\;0.20[$. For each instance defined by these two metrics, the procedure is applied for prescribed periodicities in the range $\omega\in[30,\; 80]$. The circular indicator field is utilised, rather then the full square, to ensure more representative boundary generation time.
\autoref{fig:time_circle_test} illustrates the nature of the problem being dehomogenised in the considered test cases in terms of the orientation field (left) and an example dehomogenised patch (right).

\begin{figure}[!htb]
    \centering
    \includegraphics[width=0.9\linewidth]{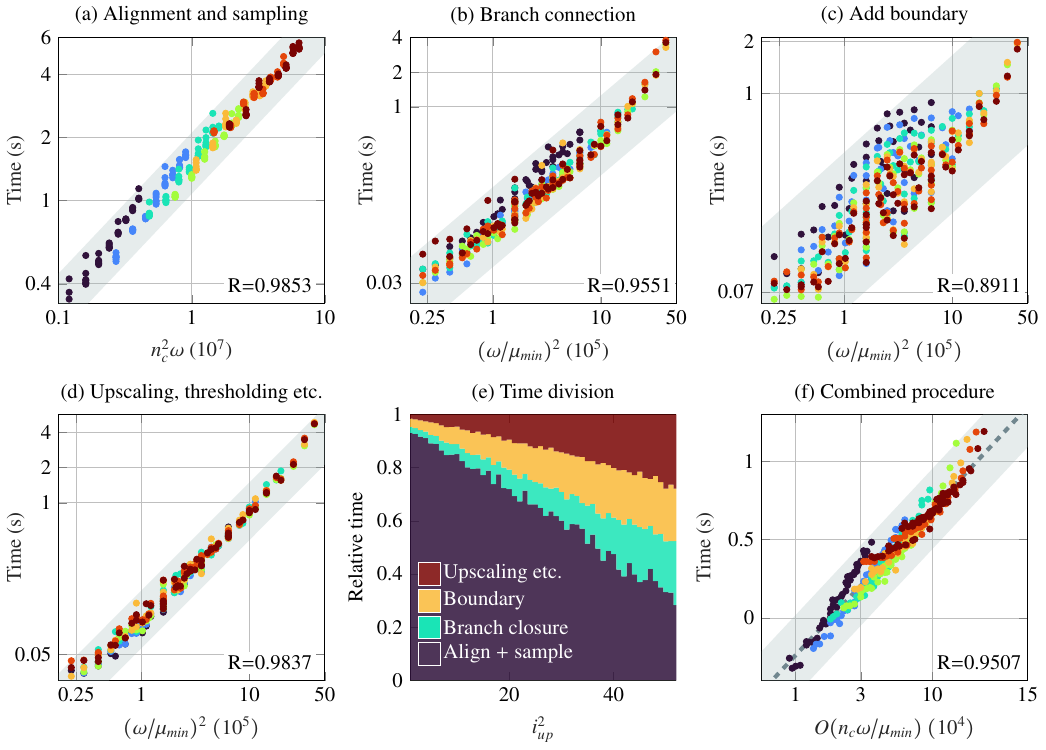}
     \caption{Illustrating the computational complexity for the different subprocesses involved in the phasor-based dehomogenisation by comparing the determining sizes established in \autoref{app:computational_complexity} to the running time and evaluating the strength of their linear relationship. Scatter points are coloured based on the resolution of the underlying coarse mesh homogenised solution being dehomogenised.}
    \label{fig:time_subprocesses_separate_scatter}
\end{figure}

\autoref{fig:time_subprocesses_separate_scatter} (a)-(d) shows the computational time of the different sub-procedures as a function of the sizes determining the computational complexity, discussed in \autoref{app:computational_complexity}. For the phase alingment and latter phasor field sampling processes the major computational expense is determined by the sampling procedure, which is confirmed by the clear linear relation between the theoretical complexity of the sampling procedure and the measured time of the combined procedure. The linear trend of the branch connection procedure is almost as convincing, even when considering the conservative assumption about maximal number of branching points utilised in the derivation of the theoretical complexity. In most cases the number of branching points will be much smaller than the utilised measure, but as the same orientation field is considered throughout the tests here, a proportionality is expected between the number of branches and the prescribed periodicity. Adding the structural boundary presents largest variation from the linear trend. This is expected because the procedure combines its own versions of constructing, aligning, sampling, upscaling and thresholding a phasor field, thus containing more subprocedures affecting the running time trend. The mesh construction, upscaling and transformation overhead is highly linear in the number of fine resolution elements, as expected. 

The bar plot in \autoref{fig:time_subprocesses_separate_scatter} (e) illustrates the average running time distribution among the subprocedure as a function of the second intermediate upscaling factor applied. For lower upscaling factors the phasor field sampling is by far the most computationally demanding part of the overall dehomogenisation procedure, but as the upscaling factor increases, so does the relative time spent on each of the remaining subprocedures. The complexity of the sampling procedure is highly dependant upon the resolutions of the coarse mesh and the first intermediate upscaling factor, while the finer resolutions have no influence. The finer resolutions are, however, the main contributors to the computational expense of the remaining procedures. This explains why the cost of these procedures increase when $i_{up}^2$ increases, while the time of the sampling procedure is not directly affected.

\begin{figure}[!htb]
    \centering
    \includegraphics[width=0.75\linewidth]{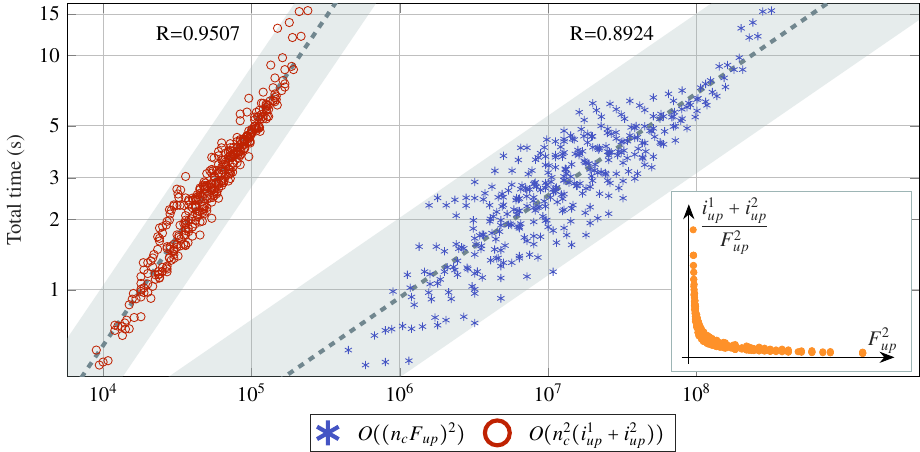}
    \caption{Illustrating how the localised implementation of the different subprocedures in the dehomogenisation procedure allows for obtaining a computational complexity which is sublinear in the number of elements in the intermediate and fine-resolution number of elements.}
    \label{fig:total_time_order_comparison}
\end{figure}

\autoref{fig:time_subprocesses_separate_scatter} (f) illustrates the linear trend of the computational time as a function of the number of coarse scale elements, the desired periodicity and the minimal relative thickness, as for the square structural domain instances in \autoref{fig:time_square_test}(c). It is evident that the strong linear trend in the complexity $O(n_c\omega/\mu_{min})$ remains. The main takeaway from this relation is further illustrated in \autoref{fig:total_time_order_comparison}, which shows how the computational time has a stronger dependency upon the size-measure $n_c^2(i_{up}^1+i_{up}^2)=O(n_c\omega/\mu_{min})$ than the number of fine resolution elements $(n_cF_{up})^2$. This means that the computational complexity is sublinear in the number of fine-resolution elements, even considering that the computational complexity of the upscaling and thresholding operators depend upon this finest resolution. Interestingly, the found complexity is sublinear in the number of intermediate mesh elements as well. The benefit from this observation is that it proves very promising scalability of the phasor-based dehomogenisation method, such that the computational cost will not be subjected to sudden and unrealistic increases when the problem size grows. These results further indicate that the length-scale of the dehomogenised design is a much more important measure of the complexity of the final design, than the mesh resolution in itself.

\end{document}